%% file: positronium_gate.tex
\documentclass{iopjournal}
\input{config}

\begin{document}

\articletype{Paper} %	 e.g. Paper, Letter, Topical Review...

\title{A Monte Carlo positronium decay source model with multiple annihilation channels in GATE}

\author{Wojciech Krzemien$^{1,*}$\orcid{0000-0002-9546-358X}, Mateusz Bala$^2$\orcid{0000-0002-2505-7949}, Kamil Dulski$^{3}$\orcid{0000-0002-4093-8162}, Wojciech Zdeb$^2$\orcid{0009-0000-1485-251X}, Aurélien Coussat$^4$\orcid{0000-0002-8336-7056}, Beatrix C. Hiesmayr$^{5,6}$\orcid{0000-0001-9062-6039}, Konrad Klimaszewski$^2$\orcid{},Michał Obara$^2$\orcid{0009-0006-4779-6757},L. Raczyński$^2$\orcid{0000-0002-7039-2084}, Roman Y. Shopa$^2$\orcid{0000-0002-1089-5050}}

\affil{$^1$National Centre for Nuclear Research, Department of High Energy Physics, 05-400 Otwock, Poland}\\
\affil{$^2$National Centre for Nuclear Research, Department of Complex Systems, 05-400 Otwock, Poland}\\
\affil{$^3$Faculty of Physics, Astronomy and Applied Computer Science, Jagiellonian University, 30-348 Kraków, Poland}\\
\affil{$^4$INSA-Lyon, Université Claude Bernard Lyon 1, CNRS, Inserm, CREATIS UMR 5220, U1294, F-69373, Lyon, France}\\
\affil{$^5$IT:U Interdisciplinary Transformation University, Freistädter Strasse 400, 4040 Linz, Austria}\\
\affil{$^6$University of Vienna, Faculty of Physics, Währingerstrasse 17, 1090 Vienna, Austria}

\email{$^*$wojciech.krzemien@ncbj.gov.pl}

\keywords{positronium, Monte Carlo, Computer Simulations}

\begin{abstract}
{\it Objective}.
Positronium-based imaging, including \ac{PLI} and multi-photon \ac{PET}, requires realistic modelling of \ac{Ps} decay in matter. Existing \ac{MC} implementations in \ac{GATE} are limited to simplified single- or two-channel descriptions, preventing accurate representation of the multi-component lifetime structure observed in biological and material systems. The objective of this work is to develop and validate a flexible \ac{MC} \ac{Ps} decay source model supporting multiple annihilation channels within the \ac{GATE} framework.

{\it Approach}.
We introduce a modular \ac{Ps} decay model implemented in \ac{GATE}~9.4 and \ac{GATE}~10, enabling the definition of an arbitrary number of decay channels characterised by lifetime, branching fraction, annihilation multiplicity (\twogamma/\threegamma), and optional prompt photon emission. The model is validated through analytical and numerical benchmarks, including lifetime distributions, branching fraction consistency, photon kinematics, and prompt photon emission. Its practical applicability is demonstrated using simulations of mixed annihilation scenarios and the NEMA IEC phantom with a large field-of-view \ac{PET} system.

{\it Results}.
The proposed model accurately reproduces input lifetime distributions as weighted sums of exponential components and correctly samples decay channel fractions. Simulated two- and three-photon annihilation kinematics are consistent with theoretical expectations. Complex mixtures of decay channels, including varying \threetotwo ratios and multi-component \ac{o-Ps} lifetimes, are correctly modelled, with observable signatures reflected in both temporal and energy distributions. Phantom simulations demonstrate the capability to generate realistic positronium-sensitive datasets.

{\it Significance}.
This work provides the first general-purpose, multi-channel positronium decay model integrated into \ac{GATE}, enabling realistic simulations of positronium behaviour in complex media. The model supports the development and optimisation of positronium-based imaging techniques, including \ac{PLI} and multi-photon \ac{PET}, and applies to medical imaging, industrial tomography, and fundamental physics studies. Its public availability and compatibility with standard \ac{GATE} workflows make it a valuable tool for the broader research community.

\end{abstract}
\section{Introduction}
\label{sec:Introduction}

Positronium imaging has recently emerged as a promising extension of \acf{PET}, offering access to novel biomarkers related to the microscopic properties of matter~\citep{ hourlierExperimentalUsesPositronium2024, tashimaThreeGammaImagingNuclear2024a}. In particular, \ac{PLI} has demonstrated sensitivity to tissue composition and microstructure, with recent studies establishing its feasibility in both preclinical and clinical settings, including phantom experiments, in vivo human brain imaging, and tumour hypoxia sensing~\citep{shibuyaOxygenSensingAbility2020a,mercolliVivoPositroniumLifetime2024,steinbergerPositroniumLifetimeValidation2024a, mercolliFirstPositroniumLifetime2025a, mercolliPhantomImagingDemonstration2025, mercolliVivoVoxelwisePositronium2026, moskalPositroniumImageHuman2024a}.
These developments have been accompanied by rapid progress in reconstruction algorithms for \ac{Ps}-based imaging~\citep{qiPositroniumLifetimeImage2022, huangSPLITStatisticalPositronium2024, huangHighresolutionPositroniumLifetime2024, huangFastHighresolutionLifetime2025}.
Measurements of multi-photon annihilation fractions have similarly been proposed as complementary imaging markers~\citep{kacperskiThreegammaAnnihilationImaging2004,kacperskiPerformanceThreephotonPET2005,fujimotoAdvancingPETDirect2025}.

\ac{MC} simulations play an essential role in \ac{PET} tomography, supporting the design of imaging modalities, the development and validation of image reconstruction and correction algorithms, and the optimisation of acquisition protocols. Several \ac{MC} packages cover different aspects of \ac{PET}~\citep{arce2014,faddegon2020}.
Among them, \ac{GATE}~\citep{janGATESimulationToolkit2004,sarrutAdvancedMonteCarlo2021a,sarrutOpenGATEEcosystemMonte2022b} is an open-source, community-driven simulation framework built on top of Geant4~\citep{allisonRecentDevelopmentsGeant42016} and specifically designed for nuclear medicine and medical physics applications. \ac{GATE} enables detailed modelling of imaging systems and radioactive sources, including time-dependent phenomena such as radioactive decay, detector motion, and acquisition dynamics, while remaining accessible through a high-level user interface. As a result, \ac{GATE} has become a widely adopted reference platform for \ac{PET}, SPECT, CT, and related imaging and dosimetry studies.
The recent release of \ac{GATE}~10~\citep{sarrutGATE10Monte2026, krahGATE10Monte2026} introduced substantial architectural improvements, including a Python-based user interface replacing the macro scripting system, further consolidating its role as a reference simulation environment.

Despite \ac{GATE}'s wide adoption, modelling of \ac{Ps} decay remains limited to at most two predefined channels, corresponding to \ac{p-Ps} and \ac{o-Ps} decay \citep{sarrutAdvancedMonteCarlo2021a}, which is insufficient for the multi-component lifetime structure observed in real materials (see \cref{sec:Physical_model}). Dedicated Geant4-based applications such as the J-PET simulation package \citep{moskalFeasibilityStudyPositronium2019c} support multi-channel Ps decay modelling, but are tailored to a specific detector geometry and are not integrated into a general-purpose framework. A detailed comparison with these and other existing approaches is given in \cref{subsec:discussion_comparison}.

In this work, we introduce a \ac{MC} \ac{Ps} decay source model with multiple annihilation channels implemented in both \ac{GATE}~9.4 and \ac{GATE}~10. The model provides a flexible and modular description of positronium decay, allowing the user to define an arbitrary number of decay channels, each characterised by annihilation multiplicity, lifetime, branching fraction, and optional prompt photon emission. This enables a realistic simulation of the full complexity of \ac{Ps} decay in matter. The proposed model is particularly relevant for \ac{PLI}, for which a multi-component description of \ac{Ps} lifetimes has not previously been available in a general-purpose simulation framework.
The model is validated against selected benchmarks, demonstrating correct reproduction of lifetime distributions, branching fractions, and photon kinematics for both two- and three-photon annihilation modes. Its utility is further illustrated through simulations of the NEMA IEC phantom filled with the \ac{Ps} sources. The implementation is fully compatible with standard \ac{GATE} workflows and is publicly available, providing the community with a validated and extensible simulation tool for positronium-based imaging research.

The rest of the article is organised as follows:
\Cref{sec:Physical_model} presents the physical model of the \ac{Ps} multi-channel decays.
\Cref{sec:Implementation} describes the implementation details.
\Cref{sec:Materials_and_methods} presents the validation methodology and describes selected use cases used as simulation benchmarks.
\Cref{sec:Results} presents the results of the performed simulations, followed by a discussion in \cref{sec:Discussion}. Conclusions are given in \cref{sec:Conclusions}.

\section{Physical Model}
\label{sec:Physical_model}
Positronium is a bound state of an electron and a positron, formed after the thermalisation of the positron in matter~\citep{harpenPositroniumReviewSymmetry2003}. Depending on the relative spin configuration of the electron–positron pair, positronium exists in two distinct states: \ac{p-Ps} with antiparallel spins, and \ac{o-Ps} with parallel spins. These states differ both in their lifetimes and decay modes.
In a vacuum, \ac{p-Ps} decays into two back-to-back photons, each with an energy of 511~keV, with a mean lifetime of approximately 125~ps. \Ac{o-Ps} decays into three photons with a continuous energy spectrum constrained by energy and momentum conservation, with a vacuum lifetime of approximately 142~ns. In matter, these lifetimes are substantially modified by interactions with the surrounding medium. The dominant quenching mechanism for \ac{o-Ps} in biological matter is pick-off annihilation, in which the positron of the \ac{o-Ps} pair annihilates with an electron from a neighbouring molecule rather than its bound partner, producing a two-photon final state with a lifetime that depends on the local electron density. Spin-exchange processes, in which the spin state of \ac{o-Ps} is converted to \ac{p-Ps} through interaction with unpaired electrons in the medium, provide an additional quenching channel. As a result, the effective \ac{o-Ps} lifetime in matter, typically in the range of 1–4~ns in biological tissue~\citep{mercolliVivoVoxelwisePositronium2026, moskalPositroniumImageHuman2024a}, is sensitive to the local chemical environment and the free-volume microstructure of the medium (called {\it voids}), which is the physical basis of \ac{PLI}.

Although \ac{Ps} can also decay into higher photon multiplicities, such as \ac{p-Ps} into four photons or \ac{o-Ps} into five photons, these channels are strongly suppressed \citep{ADKINS2002136,PhysRevA.68.032512}, with branching ratios of the order of $10^{-6}$. Consequently, they can be neglected for virtually all practical applications.

\Ac{Ps} is an unstable system with a finite decay width, which corresponds to a constant probability of decay per unit time. The probability differs for \ac{p-Ps} and \ac{o-Ps}, and may also depend on the surrounding environments.  Under the assumption of a constant decay rate, the number of decays in a given time interval follows a Poisson process, implying an exponential distribution of lifetimes. Regarding environmental effects, such as pick-off and conversion processes, it is assumed that the corresponding decay rates remain unchanged over the timescale relevant for the positronium lifetime. In the presence of multiple decay channels, the standard description is a sum of exponential components, each associated with a different decay rate.

The lifetime spectrum of a given medium is not predicted by theory. The \ac{Ps} formation probability and the \ac{o-Ps} quenching rates depend on the local electron density, the chemical composition and the free-volume structure of the medium, and in biological systems additionally on factors such as oxygenation. They are established experimentally by positron annihilation lifetime spectroscopy, whose analysis decomposes the measured spectrum into several exponential components characterised by their mean lifetimes $\tau_i$  and intensities $I_i$. This decomposition is the standard characterisation of a material's positronium response, and it constitutes the physical input required to describe \ac{Ps} decay in that medium. The number of components and their intensities are themselves the outcome of a fitting procedure, so this description is empirical rather than derived. Given such a decomposition, the division of each component into two- and three-photon channels is, however, fixed: by \cref{eq:intensity_o_ps} for the o-Ps components and \cref{eq:intensity_direct} for direct annihilation. The description assumes that the decay rates are constant over the timescale of the positronium lifetime and that the composition is homogeneous within the region considered; media with spatially varying composition are described by a separate set of components for each region. 

In addition to annihilation photons, positronium formation can be accompanied by the emission of a prompt photon arising from nuclear de-excitation correlated with positron emission. The energy and emission probability of the prompt photon depend on the radionuclide.
The temporal relationship between the prompt photon and the annihilation photons provides the measurement principle underlying \ac{PLI}: the prompt photon marks the moment of positronium formation, while the annihilation photons mark the moment of decay. The interval between these two events, measurable in coincidence, is the positronium lifetime. By reconstructing this interval on an event-by-event basis and accumulating statistics over many annihilations, a voxel-wise distribution of \ac{o-Ps} lifetimes can be obtained, from which the mean lifetime, and hence information about the local environment, can be extracted for each image voxel.

\begin{figure}
 \centering
 \includegraphics[width=0.95\textwidth]{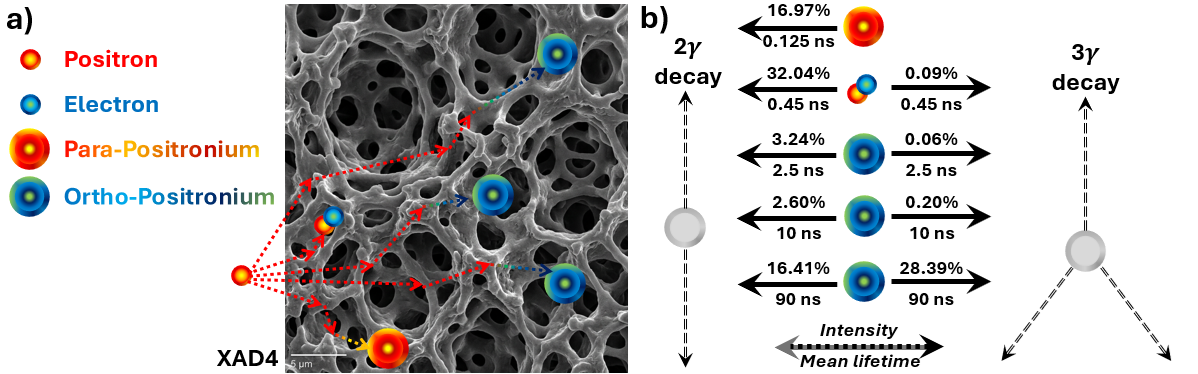}
 \caption{a) A positron introduced into the porous XAD4 polymer can annihilate either directly with an electron or first to form positronium, which, depending on the total spin, forms \ac{p-Ps} or \ac{o-Ps}. Due to the diverse structure of the polymer, three possible annihilation modes are experimentally observed after the formation of nanostructure-sensitive \ac{o-Ps}, which can be interpreted as three types of pores inside the sample. b) Decay scheme of positron components in XAD4, illustrating the multiple lifetime components required for a realistic simulation with a chance (intensity) of each type of decay occurring (\twogamma/\threegamma): \ac{p-Ps} decay, direct annihilation decay, and three \ac{o-Ps} pick-off components with lifetimes ranging from 2.5~ns to 90~ns. This example motivates the need for an arbitrary multi-channel decay source as implemented in the present work.}
 \label{fig:xad4}
\end{figure}

The complexity of positronium decay in real materials makes the simultaneous modelling of multiple lifetime components essential. \Cref{fig:xad4} shows the decay scheme of positronium in XAD4 resin, a material commonly used as a reference in positron annihilation lifetime spectroscopy. The full description of this system requires at least five distinct components \citep{jasinskaDetermination3gammaFraction2016}: a \ac{p-Ps} two-photon decay channel with a mean lifetime of 0.125~ns, a direct annihilation decay channel with a mean lifetime of 0.45~ns, and three additional \ac{o-Ps} components with mean lifetimes ranging from 2.5~ns to 90~ns, each with distinct branching fractions. Additionally, it should be noted that while the \ac{p-Ps} component can be represented by a decay into two photons (\twogamma decay), the rest of the components should be divided into contributions of two-photon and three-photon decays (\threegamma decay). The relative chances (intensities $I$) of a given component, for given decay type (\twogamma/\threegamma) in \cref{fig:xad4} can be determined based on the mean lifetime ($\tau$) of the \ac{o-Ps} components. In matter, pick-off annihilation quenches \ac{o-Ps} into \twogamma final states, reducing the surviving \threegamma fraction relative to vacuum. Since this quenching shortens the effective lifetime proportionally, the \threegamma intensity scales as the ratio of the measured lifetime to the vacuum value of 142 ns:
\begin{equation}
\label{eq:intensity_o_ps}
I_{o-Ps, 3\gamma} = I_{o-Ps} \frac{\tau _{o-Ps}}{142 \text{ ns}}
\end{equation}
or from the theory for the direct annihilation component~\citep{oreThreePhotonAnnihilationElectronPositron1949, pevovarRatioPositronAnnihilation2007}:
\begin{equation}
\label{eq:intensity_direct}
I_{direct, 3\gamma} = \frac{I_{direct}}{372}, 
\end{equation}
where the factor 372 reflects the ratio of the two-photon to three-photon cross-sections for direct positron-electron annihilation.

No simple two-channel model can capture this complexity, and any simulation restricted to two channels will yield an incorrect lifetime distribution and branching-fraction mixture, potentially biasing the results of \ac{PLI} studies that rely on such simulations for system optimisation or reconstruction validation. Therefore, the proposed component structure is the closest representation of positronium annihilation in XAD4 that we can use in simulations.

\section{Implementation in GATE}
\label{sec:Implementation}

\subsection{Overview and design philosophy}

The positronium decay model was implemented in \ac{GATE} as an \emph{effective gamma-emission source}, fully compatible with the standard \ac{GATE} source mechanism. Internally, it encapsulates the physics of positronium formation, decay, and associated photon emissions, while remaining transparent to the user.
The positronium decay is represented by a set of decay channels characterised by annihilation multiplicity (\twogamma or \threegamma decays), mean lifetime, and an optional prompt photon emission probability. This representation supports an arbitrary number of simultaneously defined decay channels, enabling composite source configurations of arbitrary complexity.
The decay model is isolated from the rest of the \ac{GATE} system, through three well-defined interfaces: (1) loading user-defined parameters at initialisation, via a dedicated messenger class in \ac{GATE}~9 or the corresponding Python interface in \ac{GATE}~10; (2) receiving the decay position and event time from the \ac{GATE} source framework; and (3) returning the generated gamma vectors to the Geant4 event. The internal decomposition into modular components is described in \cref{subsec:architecture}.
Helper functions are additionally provided to simplify the definition and extension of decay channels and their associated emission components.

\subsection{Software architecture}
\label{subsec:architecture}

The positronium decay model is decomposed into the four logical components, each reflecting an element of the physical process described in \cref{sec:Physical_model}:
\begin{itemize}
  \item \textbf{Decay channel definition:} each channel is characterised by an annihilation multiplicity (\twogamma or \threegamma), a mean lifetime, and an optional prompt photon with a specified emission probability.

  \item \textbf{Lifetime sampling:} for each event, the positronium lifetime is sampled from an exponential distribution parameterised by the channel's mean lifetime.

  \item \textbf{Annihilation kinematics sampling:} dedicated procedures generate photon energies and emission directions for two- and three-photon final states consistent with the selected decay channel.

  \item \textbf{Prompt photon emission:} an optional additional photon is generated with a channel-dependent probability and energy.
\end{itemize}

\subsubsection{Mapping of physics concepts to software classes}
\label{subsubsec:classes}

\begin{figure}
\centering
\begin{tikzpicture}[
  pinkbox/.style={draw, rounded corners=8pt, fill=red!15, align=center,
                  minimum width=5cm, minimum height=1cm, font=\small},
  bluebox/.style={draw, rounded corners=8pt, fill=blue!15, align=center,
                  minimum width=3.8cm, minimum height=1cm, font=\small},
  blueoval/.style={draw, ellipse, fill=blue!20, align=center,
                   minimum width=3.2cm, minimum height=1.6cm, font=\small},
  note/.style={align=left, font=\scriptsize},
  arr/.style={->, >=stealth, thick}
]

%% ── NODES ──────────────────────────────────────────────
%% Left column x=0, Right column x=8

\node[blueoval] (cloud) at (8, 0) {Input params\\via macro\\commands};
\node[pinkbox] (messenger)at (8, -3.5) {GatePositroniumSourceMessenger};
\node[pinkbox] (source) at (0, -3.5) {Positronium\\Source};
\node[pinkbox] (model) at (0, -8) {GatePositroniumDecayModel};
\node[pinkbox] (paramgen) at (8, -7) {GatePositroniumDecay\\ParamGenerator};
\node[bluebox] (params) at (8, -9.2) {GatePositronium\\DecayParams};
\node[pinkbox] (ps) at (0, -14) {GatePositronium};
\node[pinkbox] (channel) at (8, -14) {GatePositroniumDecayChannel};

%% ── ANNOTATION NODES (free-standing, never on arrows) ──

%% label 1 — right of cloud->messenger arrow
\node[note, anchor=west] at (8.4, -1.8)
  {1. Gathering user parameters};

%% label 3 — above messenger->source arrow
\node[note, anchor=south, align=center] at (4, -2.8)
  {\scriptsize 3. Positronium decay params\\
   \scriptsize passed to create the decay model};

%% label 2 — right of messenger->paramgen arrow
\node[note, anchor=west] at (8.4, -5.2)
  {\scriptsize 2. Validation and recalculation\\
   \scriptsize of input parameters into a\\
   \scriptsize standardised format};

%% label 4 — right of source->model arrow
\node[note, anchor=west] at (0.4, -5.8)
  {\scriptsize 4. Creation of a dedicated\\
   \scriptsize Ps model made of set\\
   \scriptsize of positronium entities\\
   \scriptsize with decay channels attached};

%% label 5 — right of model->ps arrow
\node[note, anchor=west] at (0.4, -10)
  {\scriptsize 5. Creation of a vector of\\
   \scriptsize Positronium entities with\\
   \scriptsize relative probabilities and\\
   \scriptsize associated prompt photons};

%% assigning label — above ps->channel arrow
\node[note, anchor=south, align=center] at (4, -13.4)
  {\scriptsize Assigning GatePositroniumDecayChannel\\
   \scriptsize to each GatePositronium entity};

%% ── PARAMS EXAMPLE BLOCK ───────────────────────────────
\node[note, anchor=north west,
  draw=blue!60,
  rounded corners=2pt,
  fill=blue!5,
  inner sep=4pt] at (5.8, -10.2)
  {\begin{tabular}{@{}ll@{}}
   Prob: & 0.4,\ 0.1,\ 0.5 \\[2pt]
   Decay: & \twogamma,\ \threegamma,\ \twogamma \\[2pt]
   Lifetime: & 0.5 ns,\ 2 ns,\ 1 ns \\[2pt]
   E Prompt: & 1.22 MeV \\[2pt]
   Prob Prompt: & 1,\ 0.9,\ 0.2 \\
   \end{tabular}};

%% ── ARROWS (clean, no labels) ──────────────────────────

\draw[arr] (cloud) -- (messenger);
\draw[arr] (messenger) -- (source);
\draw[arr] (messenger) -- (paramgen);
\draw[arr] (source) -- (model);
\draw[arr] (model) -- (ps);
\draw[arr] (ps) -- (channel);

\end{tikzpicture}
\caption{Schematic overview of the positronium decay model in \ac{GATE}, illustrating
the mapping between physical decay channels, logical components, and software classes. Steps indicate the processing sequence from parameter loading to event generation.}
\label{fig:ps_architecture}
\end{figure}
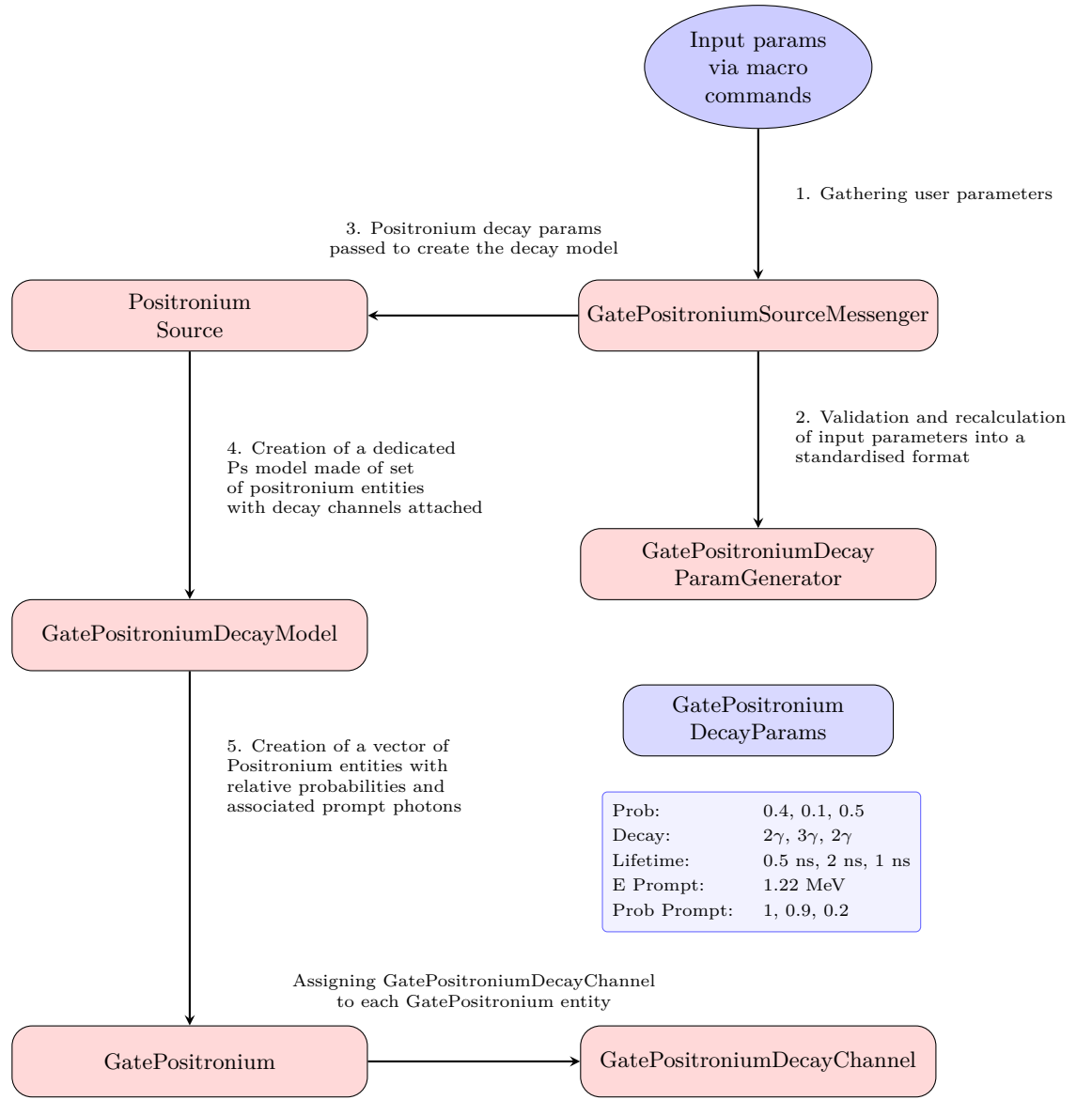

\begin{figure}
\centering
\begin{tikzpicture}[
  box/.style={draw, rounded corners, align=center,
              minimum width=5.5cm, minimum height=1cm}
]

\node[box] (input) at (0, 0) {Event\\time + position};
\node[box] (source) at (7, 0) {GatePositroniumSource\\calls model};
\node[box] (model) at (7, -2) {Select decay channel};
\node[box] (prompt) at (0, -2) {Generate prompt photons};
\node[box] (lifetime) at (0, -4) {Sample lifetime};
\node[box] (range) at (7, -4) {Add positron range (optional)};
\node[box] (decay) at (7, -6) {Generate \twogamma / \threegamma photons};
\node[box] (output) at (0, -6) {Photons returned\\to event};

\draw[->] (input) -- (source);
\draw[->] (source) -- (model);
\draw[->] (model) -- (prompt);
\draw[->] (prompt) -- (lifetime);
\draw[->] (lifetime) -- (range);
\draw[->] (range) -- (decay);
\draw[->] (decay) -- (output);

\end{tikzpicture}
\caption{Event generation flow in the positronium decay model. For each event,
a decay channel is selected according to the user-defined branching fractions, the corresponding lifetime is sampled, and photon
kinematics are generated. If a non-zero positron range is configured, the annihilation position is displaced from the emission point before photon generation.}
\label{fig:ps_flow}
\end{figure}
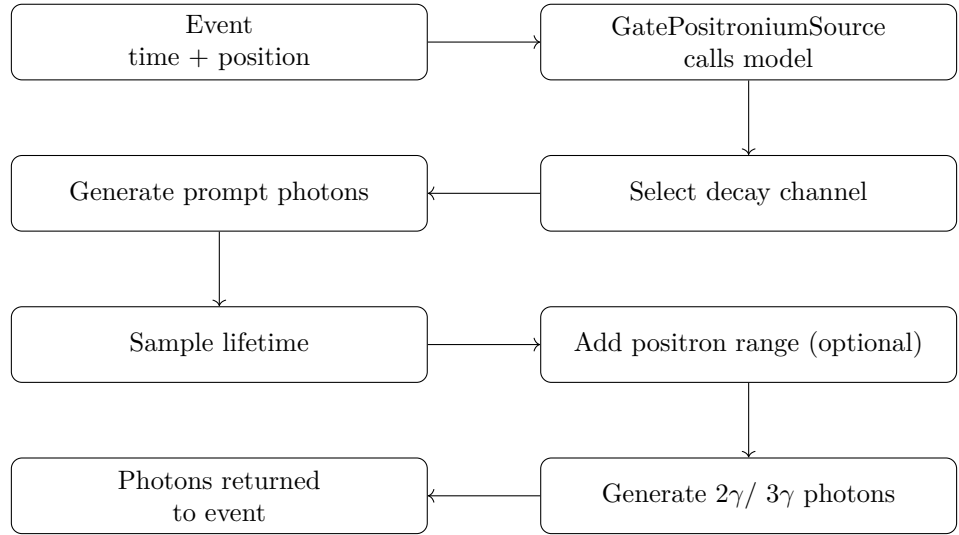

The core interface between \ac{GATE} and the positronium decay model is the \texttt{GatePositroniumSource} class, which serves a twofold role.
First, \texttt{GatePositroniumSource} manages all user-defined parameters. In \ac{GATE}~9, this is handled through the standard messenger mechanism. User inputs are validated and encapsulated in a \texttt{PositroniumDecayModelParams} data structure, which is passed to the decay model during initialisation.
Second, \texttt{GatePositroniumSource} acts as the entry point for primary particle generation, delegating spatial and temporal sampling to the standard \ac{GATE} source mechanism.
The positronium decay logic is implemented in \texttt{PositroniumDecayModel}, which defines the available decay channels and handles the probabilistic channel selection.
For each event, \texttt{PositroniumDecayModel} selects a decay channel according
to user-defined branching fractions, samples the corresponding positronium lifetime, and triggers the generation of annihilation photons and, when applicable, prompt photons.
Each decay channel is represented by the \texttt{GatePositronium} class, which encapsulates the channel’s physical properties and provides the corresponding decay products.
\Cref{fig:ps_architecture} summarises the overall processing flow and the mapping between physical concepts and software components. The event generation sequence is illustrated in \cref{fig:ps_flow}.

\clearpage

\subsubsection{Decay channel parametrization}
\label{subsub:decay_channel_parameterization}

A \texttt{GatePositronium} instance corresponds to a specific physical decay mode and encapsulates its annihilation characteristics.
The detailed kinematics are handled by a dedicated \texttt{PositroniumDecayChannel} object, which generates two- or three-photon final states for each event.
From the user's perspective, the positronium source was designed to be configured via a set of \ac{GATE}~9 macro commands that define its decay-channel structure. Each command accepts a list of values, one per decay channel, allowing an arbitrary number of channels to be specified simultaneously. \Cref{lst:macro}
illustrates the definition of a four-component source.

\begin{lstlisting}[caption={Example macro definition of a four-component positronium source. The configuration is not intended to represent a physical scenario, but serves only to illustrate the data structure and connections between parameter sets.}, label={lst:macro}, basicstyle=\footnotesize]
/gate/source/addSource ps PositroniumSource
/gate/source/ps/setType Ps
/gate/source/ps/setPositroniumFractions         0.4     0.3     0.2     0.1
/gate/source/ps/setPositroniumLifetimes         0.125   6.0     2.0     3.0   ns
/gate/source/ps/setPromptPhotonProbabilites    1.0     0.3     0.7     0.9
/gate/source/ps/setPromptPhotonEnergies         1.274   1.274   1.274   1.274 MeV
/gate/source/ps/setMeanPositronRange            0.20    0.05    0.09    0.4   mm
/gate/source/ps/setElectronCaptureProbabilities 0.7     0.7     0.7     0.7
/gate/source/ps/setDecayKinds                   k2Gamma k3Gamma k2Gamma k2Gamma
\end{lstlisting}

The \texttt{setType Ps} command declares the source as a positronium decay source. The remaining commands each accept a list of values, one per decay channel, defining the physical properties of all channels simultaneously.
The \texttt{setPositroniumFractions} command specifies the relative occurrence probability of each decay channel. In the example above, the four channels are assigned fractions of 0.4, 0.3, 0.2, and 0.1, respectively. The model normalises these values internally so that they need not sum to unity, providing flexibility in source definition.
The \texttt{setPositroniumLifetimes} command sets the mean lifetime of each channel, from which the positronium lifetime for each event is sampled from an exponential distribution. Lifetimes are specified with an explicit unit, here nanoseconds.
The \texttt{setDecayKinds} command specifies the annihilation multiplicity of each channel, using the keywords \texttt{k2Gamma} for two-photon decay and \texttt{k3Gamma} for three-photon decay. In the example, the first, third, and fourth channels produce two-photon final states, while the second channel produces a three-photon final state.
\texttt{setPromptPhotonProbabilites} and \texttt{setPromptPhotonEnergies} together define the optional prompt photon emission for each channel. A probability of 1 indicates that a prompt photon is emitted with certainty for every event in that channel, while a probability of 0 suppresses prompt photon emission entirely. In this example, all four channels emit a prompt photon of 1.274~MeV, consistent with the \nucNa decay scheme.
The \texttt{setMeanPositronRange} sets the effective positron range, modelling the displacement between the positron emission point and annihilation site as a Gaussian smearing with the specified mean. This is an effective parameterisation rather than a full transport simulation of the positron, consistent with the overall design philosophy of the model.
The \texttt{setElectronCaptureProbabilities} command specifies the probability of the decay via electron capture rather than positron emission and therefore no annihilation photon emission.

\subsubsection{From a measured lifetime spectrum to a source definition}
The decay-channel structure is supplied by the user rather than derived internally from the material definition, reflecting the empirical character of the input described in \cref{sec:Physical_model}. Since no material model predicts the lifetime spectrum, embedding a fixed mapping between the material and lifetime would restrict the source to the measurements available at the time of implementation. The parameter lists were instead chosen to mirror the output of a PALS decomposition directly. \Cref{lst:xad4} shows the correspondence explicitly for a measured spectrum.
The intended workflow is the following:
(i) obtain the lifetime spectrum of the medium of interest, i.e. the set of ($\tau_i,I_i$) from a measurement or from the literature, (ii) split each component into its \twogamma and \threegamma contributions using \cref{eq:intensity_o_ps} and \cref{eq:intensity_direct}. 
Finally, (iii) declare the resulting channels: lifetime, fraction, multiplicity, and optional prompt photon properties as source parameters.
Step (ii) is purely arithmetic and is additionally available as a helper function in the implementation so that measured ($\tau_i, I_i$) values may be passed directly. Its use is a convenience and does not affect the model. The helper function is useful when a source is specified by its physical components, the direct, \ac{p-Ps} and \ac{o-Ps} intensities and lifetimes, and the photon multiplicity channel list is to be generated automatically.  
The scenarios of \cref{subsec:3to2} are constructed in the same way.

\subsection{Integration with \ac{GATE}~10}

The positronium decay model was integrated into \ac{GATE}~10 while preserving the core C++ physics and event generation logic. The classes responsible for decay channel definition, lifetime sampling, and photon generation are identical to those used in the \ac{GATE}~9 implementation.
In \ac{GATE}~10, sources are configured using Python dictionaries, which are passed to the simulation kernel.
The positronium source follows the standard \ac{GATE}~10 source pattern, with user-defined parameters, such as lifetimes, branching fractions, decay modes, and prompt photon properties, extracted from the dictionary during initialisation and used to construct the decay model.
The use of Python dictionaries allows direct mapping between user input and internal data structures, removing the need for string-based macro parsing and simplifying parameter validation and extension.
During initialisation, the extracted parameters are encapsulated into \texttt{PositroniumDecayModelParams} and used to construct the decay model instance. This separation between the user interface layer and the physics model means that the \ac{GATE}~10 integration required only minor modifications to the core C++ physics classes.
Event generation follows the standard \ac{GATE}~10 source workflow, with spatial and temporal event vertices received from the base source infrastructure. The downstream physics, such as decay channel selection, lifetime sampling, and photon generation, proceed identically to the \ac{GATE}~9 implementation, with the core C++ classes shared between both versions.

From the user's perspective, the transition from macro-based configuration to Python significantly improves flexibility and usability. Complex multi-channel positronium sources can be defined using standard Python data structures, enabling easier scripting, parameter scanning, and integration with external workflows. At the same time, the underlying physics behaviour and configuration concepts remain identical to those introduced for \ac{GATE}~9, ensuring continuity and reproducibility of simulation studies.
For illustration, the same \ac{GATE}~9 example given in \cref{lst:macro} is reproduced for \ac{GATE}~10 in \cref{lst:macro_python}.

\begin{lstlisting}[caption={Example of \ac{GATE}~10 definition of a four-component positronium source. The configuration is not intended to represent a physical scenario, but serves only to illustrate the data structure and connections between parameter sets.}, label={lst:macro_python}, basicstyle=\footnotesize]
source = sim.add_source("PositroniumSource", "source")
source.channels_from_fractions.fractions = [0.4, 0.3, 0.2, 0.1]
source.channels_from_fractions.decay_kinds = ["k2Gamma", "k3Gamma", "k2Gamma", "k2Gamma"]
source.positronium_lifetimes = [0.125 * ns, 6. * ns, 2. * ns, 3. * ns]
source.prompt_gamma_probabilities = [1., .3, .7, 0.9]
source.prompt_gamma_energies = [1.274 * MeV, 1.274 * MeV, 1.274 * MeV, 1.274 * MeV]
source.mean_positron_range = [.2 * mm, .05 * mm, .09 * mm, .4 * mm]
source.electron_capture_probabilities = [.7, .7, .7, .7]
\end{lstlisting}

The implementation was merged into the official \ac{GATE}~9 and is available on \href{https://github.com/OpenGATE/Gate}{OpenGATE GitHub repository}. 

\section{Materials and methods}
\label{sec:Materials_and_methods}

\subsection{Validation methodology}
Validation and testing of the model were performed at several levels.
First, code correctness was verified through a suite of unit tests targeting individual class methods.
The model was then validated through a series of analytical and numerical benchmarks targeting four key components: lifetime distributions, branching fraction consistency, photon kinematics, and prompt photon emission.

\paragraph{Lifetime distributions}
Simulated lifetime distributions were compared with the corresponding fitted exponential functions for each single-channel configuration. For multi-channel sources, the total lifetime spectrum was compared against the expected weighted sum of exponential components, with weights equal to the specified branching fractions.
For the single-channel validation, three simulations of \threegamma annihilation were performed with mean lifetimes of 1, 2, and 5~ns.
The lifetime sampling mechanism is common to both \twogamma and \threegamma channels, as it depends only on the user-defined mean lifetime and not on the annihilation multiplicity. Validation of \threegamma channels, therefore, implicitly validates the sampling for \twogamma channels as well.

\paragraph{Photon kinematics}
The kinematics of annihilation photons were validated for three-photon decay modes.
In particular, the energy spectrum and the joint angular distribution were compared against the theoretical predictions of \citet{oreThreePhotonAnnihilationElectronPositron1949,raczynskiVertex2026}.

\paragraph{Prompt photon emission benchmarks}
Prompt photon emission was validated for three clinically relevant radionuclides: \nucSc,
\nucGa, and \nucI, whose decay schemes~\citep{LNHB_database} are summarised in \cref{tab:beta_and_ec_sim}. 
For all radionuclides, only transitions involving the $\beta^+$ decay mode were considered, while transitions proceeding exclusively via electron capture, which by definition do not generate annihilation photons, were excluded. In the case of \nucI, which contains a complicated decay scheme with many low-probability transition lines, the analysis was further restricted to transitions with a $\beta^+$ branching ratio exceeding $0.25\%$, which allows for constraining the number of transitions to a reasonable amount while ensuring high statistical accuracy. Those selection choices do not affect the benchmark results, and it should be noted that the used approximation might not be sufficient for realistic \ac{PET} simulations: depending on the simulation goal, a complete decay scheme may be required.

For each source, the simulated prompt photon rate and energy were compared against the configured values.

\begin{table}[htbp]
\centering
\caption{Summary of decay modes modelled in simulations for prompt photon emission benchmarks. Here, $\beta^+_{i,j}$ represents a positron-emission transition and $\epsilon_{i,j}$ an electron-capture transition. The indices $(i,j)$ denote the initial and final nuclear energy levels, respectively.}
\label{tab:beta_and_ec_sim}
    \begin{tabular}{ccccc}
        \toprule
        \textbf{Transition} & \textbf{Prompt energy} & \textbf{Emissions per} & \textbf{Relative} & \textbf{Defined} \\
        \textbf{} & \textbf{[MeV]} & \textbf{100 disint.} & \textbf{probability} & \textbf{probability (approx.)} \\
        \midrule
        \multicolumn{5}{c}{\nucSc} \\
        \midrule
        $\beta^+_{0, 1}$ & 1.157 & 94.27 & 100 & 95.25\% \\
        $\epsilon_{0, 1}$ & 1.157 & 4.7 & 4.99 & 4.75\% \\
        \midrule
        \multicolumn{5}{c}{\nucGa} \\
        \midrule
        $\beta^+_{0, 0}$ & — & 87.68 & 100 & 96.66\% \\
        $\beta^+_{0, 1}$ & 1.077 & 1.2 & 1.379 & 1.32\% \\
        $\epsilon_{0, 1}$ & 1.077 & 1.8 & 2.053 & 1.99\% \\
        $\beta^+_{0, 2}$ & 1.656 & 0.00026 & 0.000297 & 0.00029\% \\
        $\epsilon_{0, 2}$ & 1.656 & 0.0335 & 0.0382 & 0.037\% \\
        \midrule
        \multicolumn{5}{c}{\nucI} \\
        \midrule
        $\beta^+_{0, 0}$ & — & 10.32 & 100 & 19.46\% \\
        $\beta^+_{0, 1}$ & 0.603 & 11.45 & 111 & 21.59\% \\
        $\epsilon_{0, 1}$ & 0.603 & 25.55 & 248 & 48.17\% \\
        $\beta^+_{0, 3}$ & 1.326 & 0.287 & 2.78 & 0.54\% \\
        $\epsilon_{0, 3}$ & 1.326 & 5.43 & 52.6 & 10.24\% \\
        \bottomrule
    \end{tabular}
\end{table}

\paragraph{Branching fractions and channel selection.}
Simulated channel populations were compared with the input branching fractions for representative multi-channel configurations, confirming convergence to the expected values within statistical uncertainty. Branching fractions were further tested for two use cases described in
\cref{subsec:3to2} and \cref{subsec:NEMA_2}: simulations of varying \threetotwo ratios with a point source, and simulations of the NEMA IEC phantom with tissue-mimicking positronium sources (\cref{fig:nemaIEC}).

\subsection{Simulation setup}
The Biograph Vision Quadra \ac{PET}/CT scanner \citep{SiemensBiographVision2022}, a large field-of-view system, was used as the reference detector for all \ac{MC} simulations. Only the geometry and physical parameters of the \ac{PET} detection system were modelled.

The detector has a cylindrical geometry with an inner radius of \( R_{\mathrm{in}} = 409\ \mathrm{mm} \) and an outer radius of \( R_{\mathrm{out}} = 430\ \mathrm{mm} \). The axial field of view (aFOV) is \( 1060\ \mathrm{mm} \). The detection structure consists of four coaxial cylindrical segments arranged along the scanner axis (Z-axis), each with a length of \( 263.2\ \mathrm{mm} \). These segments are axially shifted relative to each other by \( 265.6\ \mathrm{mm} \).

Each cylinder contains 38 linear detector modules uniformly distributed around the system axis. The dimensions of a single module are \( 263.2\ \mathrm{mm} \times 64.6\ \mathrm{mm} \times 20.0\ \mathrm{mm} \), and its geometric center is located at a radial distance of \( 420\ \mathrm{mm} \) from the cylinder axis. The modules are oriented with their longest dimension along the Z-axis.

Each linear detector module is composed of eight detector blocks arranged along the Z-axis. The dimensions of a single block are \( 32.2\ \mathrm{mm} \times 64.6\ \mathrm{mm} \times 20.0\ \mathrm{mm} \). Each block consists of a \( 2 \times 4 \) matrix of submodules with dimensions of \( 16\ \mathrm{mm} \times 16\ \mathrm{mm} \times 20\ \mathrm{mm} \).

Each submodule contains a \( 5 \times 5 \) array of LSO (\( \mathrm{Lu_2SiO_5:Ce} \)) scintillation crystals with individual dimensions of \( 3.2\ \mathrm{mm} \times 3.2\ \mathrm{mm} \times 20\ \mathrm{mm} \).

\subsubsection{Event selection}
\label{subsec:event_selection}
The multiphoton coincidence formation and event selection were performed during the postprocessing stage using a dedicated sorter. In addition, selected counters and statistics were collected during the simulations with the \texttt{GateMultiPhotonAnalysis} module, implemented as a replacement for the standard \texttt{GateAnalysis} module and distributed as part of the developed software. The extension was required because the standard module contains hardcoded limitations that restrict event aggregation to two-hit events only.
As the performed studies were intended primarily for self-consistency checks rather than for a fully realistic reproduction of the data acquisition process, several simplifications were introduced. First, only genuine coincidences were considered; thus, neither scatter nor random events were included in the analysis. Second, the true interaction points were used as the registered hit positions within the detector.
No dedicated energy window selection was applied. In addition, benchmark studies involving prompt photon emission were performed using raw data without energy smearing. For simulations including energy smearing, an energy resolution of \( \Delta E / E = 10.1\% \) at 511 keV was assumed.
Since the emission vertices were reconstructed from simulated data under idealised conditions, the reconstructed positions coincide with the Monte Carlo truth.
\subsubsection{Exemplary use case: \texorpdfstring{\threetotwo}{3-to-2 gamma} ratio}
\label{subsec:3to2}
The study focused on scenarios combining multiple positronium lifetime components and competing annihilation channels, demonstrating independent control over the \threetotwo ratio within a single source definition.

Each simulation employed a composite source consisting of four decay channels: direct positron annihilation, \ac{p-Ps} decay and \ac{o-Ps} decay via \twogamma and \threegamma modes. The simulations were performed for an acquisition time of 10~s. The direct annihilation component was assigned a lifetime of 0.4~ns, corresponding to a fraction of 68\% and an activity of 340~kBq, while the \ac{p-Ps} component was assigned a lifetime of 0.125~ns, with a fraction of 8\% and an activity of 40~kBq. Both components were kept fixed across all three simulations.

These simulations were defined by varying the \ac{o-Ps} lifetime with values: 2~ns, 40~ns, and 100~ns. For each simulation, both \twogamma and \threegamma decay channels were included simultaneously, with the total \ac{o-Ps} fraction set to 24\% and total activity of 120~kBq. The \twogamma/\threegamma split was determined by the lifetime-dependent formula \cref{eq:intensity_o_ps,eq:intensity_direct} introduced in \cref{sec:Physical_model}. A prompt photon with an energy of 1.157~MeV, consistent with the energy of \nucSc, was included in all channels.
The \threegamma component of the direct annihilation channel was not included, as its expected fraction, approximately $0.18\%$, estimated from \cref{eq:intensity_direct}, falls below the statistical precision of the present simulations, and is therefore negligible in this context.  

A summary of the simulation parameters is provided in \cref{tab:3to2_sim}.

\begin{table}[htbp]
\centering
\caption{Summary of \threetotwo ratio simulation parameters.}
\label{tab:3to2_sim}
    \begin{tabular}{ccccc}
        \toprule
        \textbf{Decay channel} & \textbf{Lifetime [ns]} & \textbf{Decay mode} & \textbf{Fraction [\%]} & \textbf{Activity[kBq]} \\
        \midrule
        \multicolumn{5}{c}{Simulation 1} \\
        \midrule
        direct & 0.4 & \twogamma & 68 & 340 \\
        \ac{p-Ps} & 0.125 & \twogamma & 8 & 40 \\
        \ac{o-Ps} & 2 & \threegamma & 0.338 & 1.692 \\
        \ac{o-Ps} & 2 & \twogamma & 23.662 & 118.308 \\
        \midrule
        \multicolumn{5}{c}{Simulation 2} \\
        \midrule
        direct & 0.4 & \twogamma & 68 & 340 \\
        \ac{p-Ps} & 0.125 & \twogamma & 8 & 40 \\
        \ac{o-Ps} & 40 & \threegamma & 6.76 & 33.804 \\
        \ac{o-Ps} & 40 & \twogamma & 17.24 & 86.196 \\
        \midrule
        \multicolumn{5}{c}{Simulation 3} \\
        \midrule
        direct & 0.4 & \twogamma & 68 & 340 \\
        \ac{p-Ps} & 0.125 & \twogamma & 8 & 40 \\
        \ac{o-Ps} & 100 & \threegamma & 16.9 & 84.504 \\
        \ac{o-Ps} & 100 & \twogamma & 7.1 & 35.496 \\
        \bottomrule
    \end{tabular}
\end{table}

\subsubsection{Exemplary use case: simulations of the NEMA IEC phantom -- \texorpdfstring{\nucSc}{Sc-44} with three decay modes}
\label{subsec:NEMA_1}
Simulations were performed with a model of a standard NEMA IEC phantom  \citep{NEMA:2018} filled with the \nucSc. The setup was loosely inspired by the experimental investigation presented in~\citep{mercolliPhantomImagingDemonstration2025}. 

The phantom has an active length of 22~cm and contains six spheres filled with activity 40.68 kBq/ml, and background (including central cylinder) with activity 3.9 kBq/ml.
The phantom was positioned isocentrically within the scanner.
Hot spheres with diameters of 10, 13, 22, and 28~mm and cold spheres with diameters of 17 and 37~mm were simulated; spheres are referred to by their diameter in millimetres.

For this simulation, only \twogamma decay channels were included. The \threegamma components of both the direct annihilation channel and the \ac{o-Ps} channel were omitted.
Both fractions are small in absolute terms, and this simulation was designed as a simplified three-component benchmark to validate lifetime fitting performance across sphere sizes rather than to achieve full physical completeness of the annihilation channel mixture. The impact of these omissions on the lifetime fitting results is expected to be negligible.

The detailed decay configuration is given in \cref{tab:experiment_scandium_44}.

\begin{table}[htbp]
\centering
\caption{Simulation parameters for \nucSc with three decay modes.}
\label{tab:experiment_scandium_44}
    \begin{tabular}{cccc}
        \toprule
        \textbf{Decay channel} & \textbf{Lifetime [ns]} & \textbf{Decay mode} & \textbf{Fraction [\%]} \\
        \midrule
        direct    & 0.388 & \twogamma & 60 \\
        \ac{p-Ps} & 0.125 & \twogamma & 7 \\
        \ac{o-Ps} & 1.78  & \twogamma & 33 \\
        \bottomrule
    \end{tabular}
\end{table}

\subsubsection{Exemplary use case: simulations of the NEMA IEC phantom with tissue-mimicking sources}
\label{subsec:NEMA_2}
To further illustrate the model's capabilities, simulations of the NEMA IEC Image Quality phantom were performed with tissue-mimicking positronium sources.

\begin{figure}
 \centering
 \includegraphics[width=0.95\textwidth]{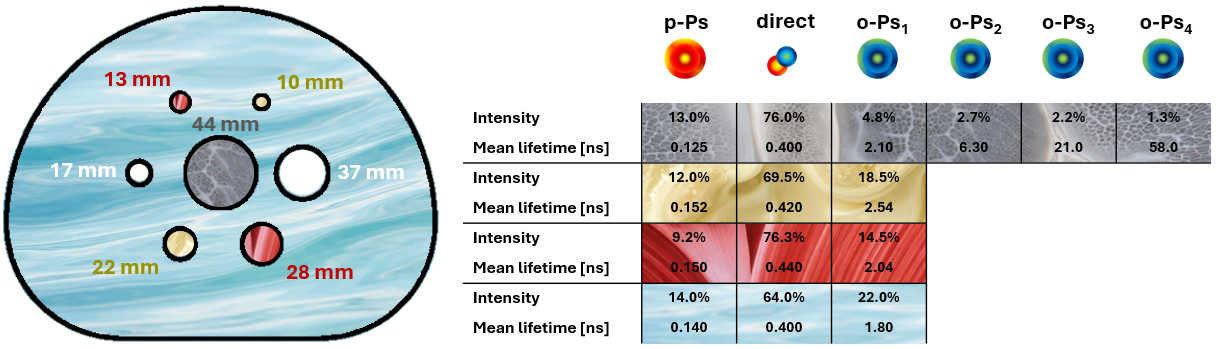}
 \caption{NEMA IEC phantom configuration used for the tissue-mimicking simulations in \cref{subsec:NEMA_2}. The diagram shows the diameters of the individual elements of the phantom. Four tissue types have been predefined: bone (grey), fat/adipose tissue (yellow), muscle tissue (red), and water (blue), with their respective positronium lifetime components shown in the table on the right based on \cite{ahnHierarchicalNatureNanoscale2021}(bone), \cite{avachatOrthopositroniumLifetimeSofttissue2024a}(muscle and adipose tissues) and \cite{koteraOrthopositroniumLifetimeWater2005a}(water). Water is filled with activity 1.3 kBq/ml, while the central cylinder with bone is filled with activity 0.5 kBq/ml. Each non-empty sphere (10, 13, 22 and 28 mm) is filled with activity 6.4 kBq/ml. Cold spheres (17 and 37~mm) are filled with water. They contain no activity and are shown in white.}
 \label{fig:nemaIEC}
\end{figure}

Spheres 10 and 22 were assigned lifetime parameters characteristic of fat/adipose tissue, spheres 13 and 28 of muscle tissue \citep{avachatOrthopositroniumLifetimeSofttissue2024a}, the central cylinder of bone \citep{ahnHierarchicalNatureNanoscale2021}, and the water-filled background of water \citep{koteraOrthopositroniumLifetimeWater2005a}, as summarised in \cref{fig:nemaIEC}. In each source, a prompt photon of 1.274~MeV was included, consistent with the \nucNa decay scheme and providing the temporal start information required for positronium lifetime measurement.
The direct annihilation and \ac{p-Ps} intensities and lifetime components were assigned individually for different tissue types (see \cref{fig:nemaIEC} for details). The numerical values and their sources are collected in \cref{tab:tissues}.

%NEMA IEC.

\section{Results}
\label{sec:Results}

\subsection{Lifetime distributions.}
\label{subsec:results_lifetime_dist}
\begin{figure}[h]
 \centering
 \includegraphics[width=0.75\textwidth]{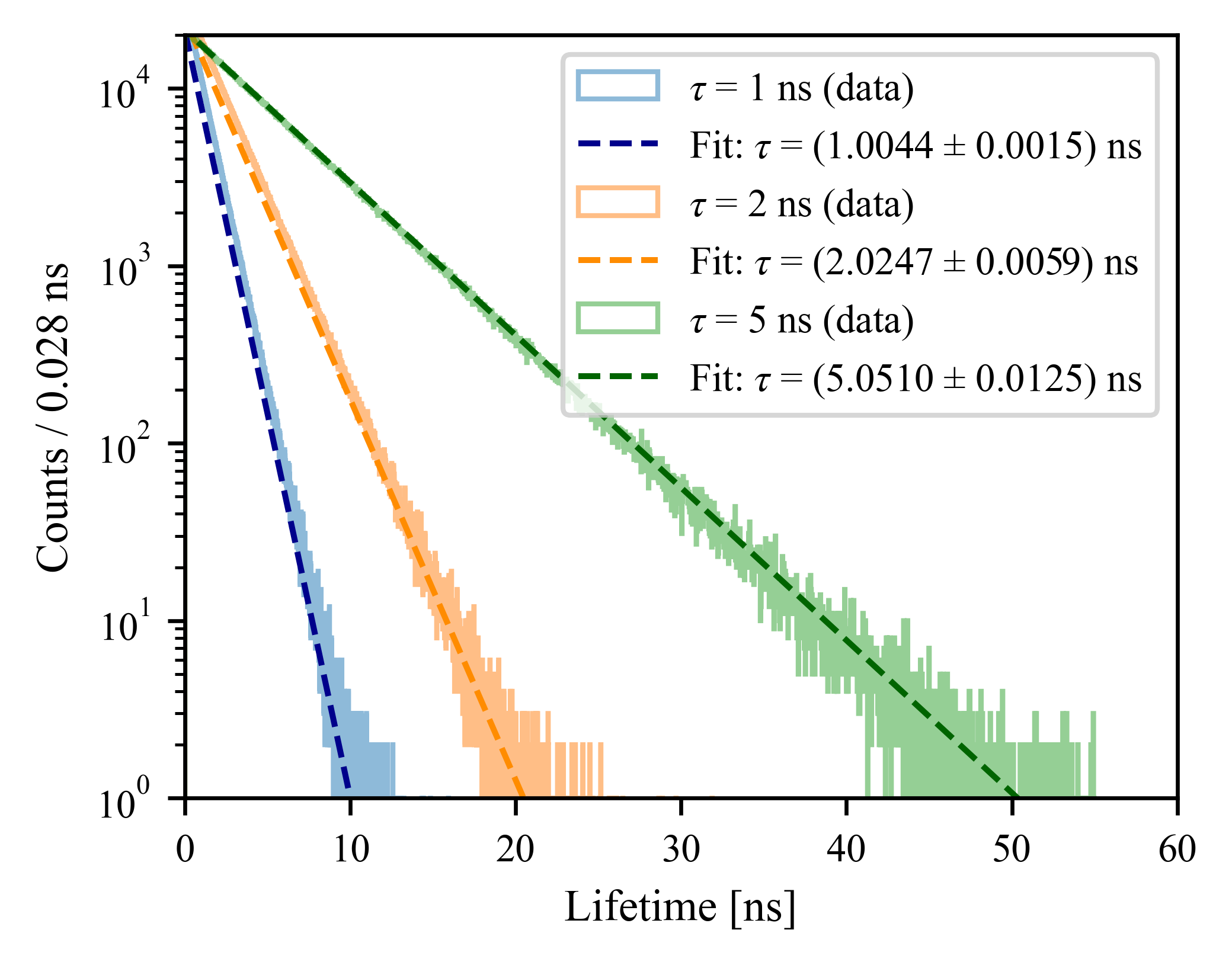}
 \caption{Lifetime distributions for three \threegamma simulations with mean lifetimes of 1, 2, and 5~ns. An exponential function $y = A\cdot\exp(-t/\tau)$ was fitted to each distribution. The fitted parameter values of $\tau$ are shown in the legend.}
 \label{fig:lifetimes_with_fits}
\end{figure}

Lifetime distributions for \threegamma decays, calculated from the simulation data, are shown in \cref{fig:lifetimes_with_fits}. The binning scheme was based on the 5 ns simulation dataset, as its range encompassed the ranges of the other two datasets. The number of bins was calculated as the square root of the total number of entries in the dataset, and the bin width was calculated by taking the difference between the maximum and minimum values in the dataset and dividing it by the bin count. The achieved bin width and edges were then uniformly applied to all datasets presented in the histogram in question, allowing for a consistent visual comparison between the datasets.
Lifetime distributions were fitted with an exponential decay function $y = A \cdot \exp(-t/\tau)$. Only bins with more than 25 counts were included in the fit, corresponding to a relative uncertainty below 20\%. The fitted values of $\tau$ are $(1.0044 \pm 0.0015)$~ns, $(2.0247 \pm 0.0059)$~ns, and $(5.051 \pm 0.013)$~ns, consistent with the input lifetimes with an error of at most 1.3\%. The $R^2$ values of 99.666\%, 99.184\%, and 99.311\% confirm that the simulated distributions follow the expected exponential form.

\subsection{Photon kinematics}
\begin{figure}[h]
    \centering

    \subfloat[]{
        \includegraphics[width=0.47\textwidth]{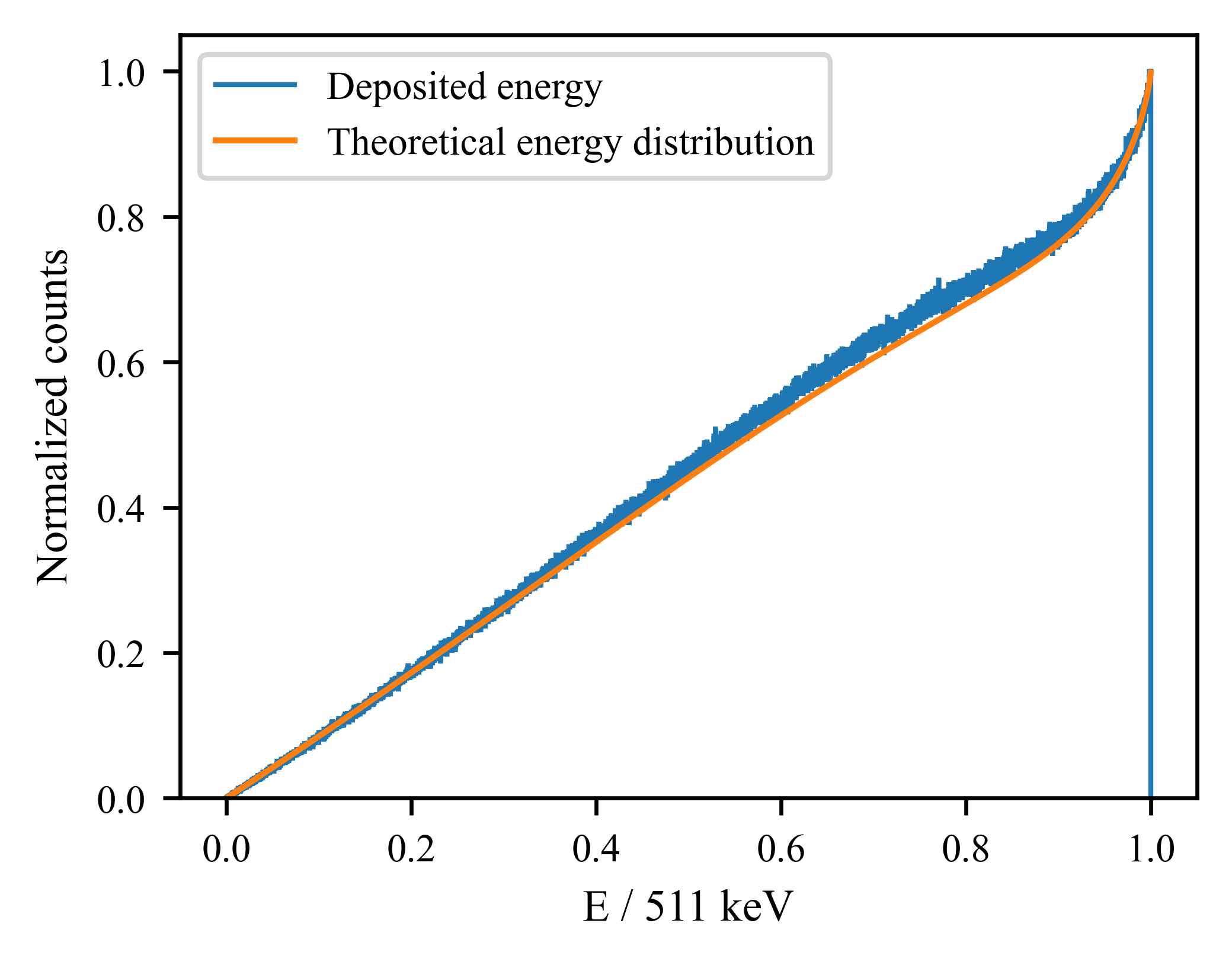}
        \label{fig:energy_spectrum}
    }
    \subfloat[]{
        \includegraphics[width=0.47\textwidth]{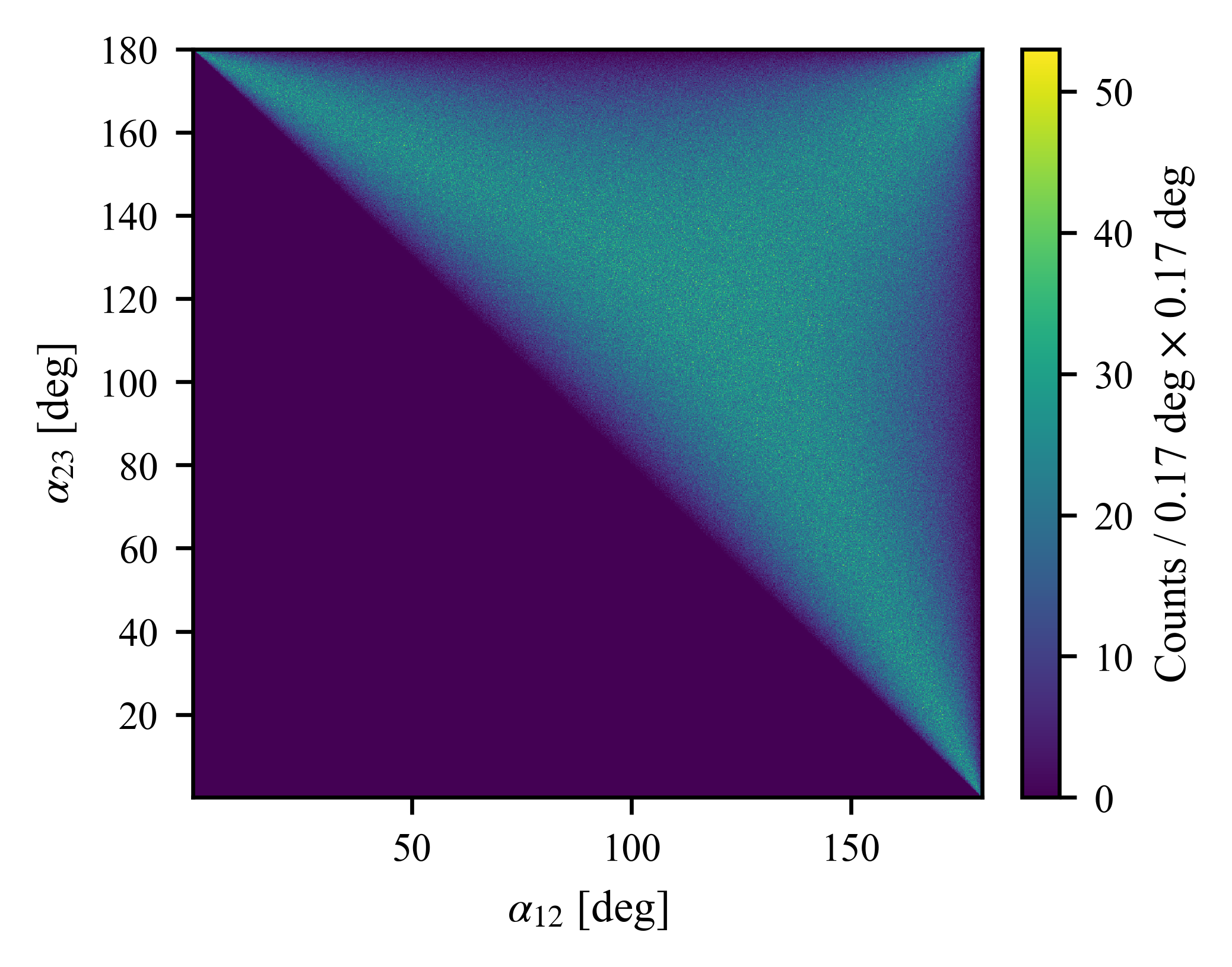}
        \label{fig:angle_heatmap}
    }

    \caption{
    (a) Energy spectrum of photons from simulated \threegamma annihilation events, compared with the theoretical distribution from \citep{oreThreePhotonAnnihilationElectronPositron1949}. Energies are normalised to 511~keV.
    (b) Joint distribution of angles between photon pairs ($\alpha_{12}$ and $\alpha_{23}$) in simulated \threegamma annihilation events. The populated triangular region corresponds to the kinematically allowed phase space. The non-uniformity of the distribution inside the triangle is the consequence of the modelled decay dynamics.
    }

\end{figure}

Simulated three-photon decay kinematics were found to be in good agreement with theoretical predictions, as described below.
The resulting histogram of the energy spectrum, together with the theoretical distribution~\citep{oreThreePhotonAnnihilationElectronPositron1949}, is shown in \cref{fig:energy_spectrum}.
The joint angular distribution is shown in \cref{fig:angle_heatmap} and is in good agreement with the kinematic constraints from three-photon annihilation~\citep{raczynskiVertex2026}.

\subsection{Three-to-two ratio}
Simulated channel fractions, lifetime distributions, and energy deposition distributions are shown for the three \threetotwo ratio scenarios defined in \cref{subsec:3to2}.
Obtained channel fractions (\cref{tab:3to2_sim_fractions}) are consistent with the input branching fractions.

\begin{figure}
    \centering
    \subfloat[Lifetime: 2 ns]{%
        \includegraphics[width=0.45\textwidth]{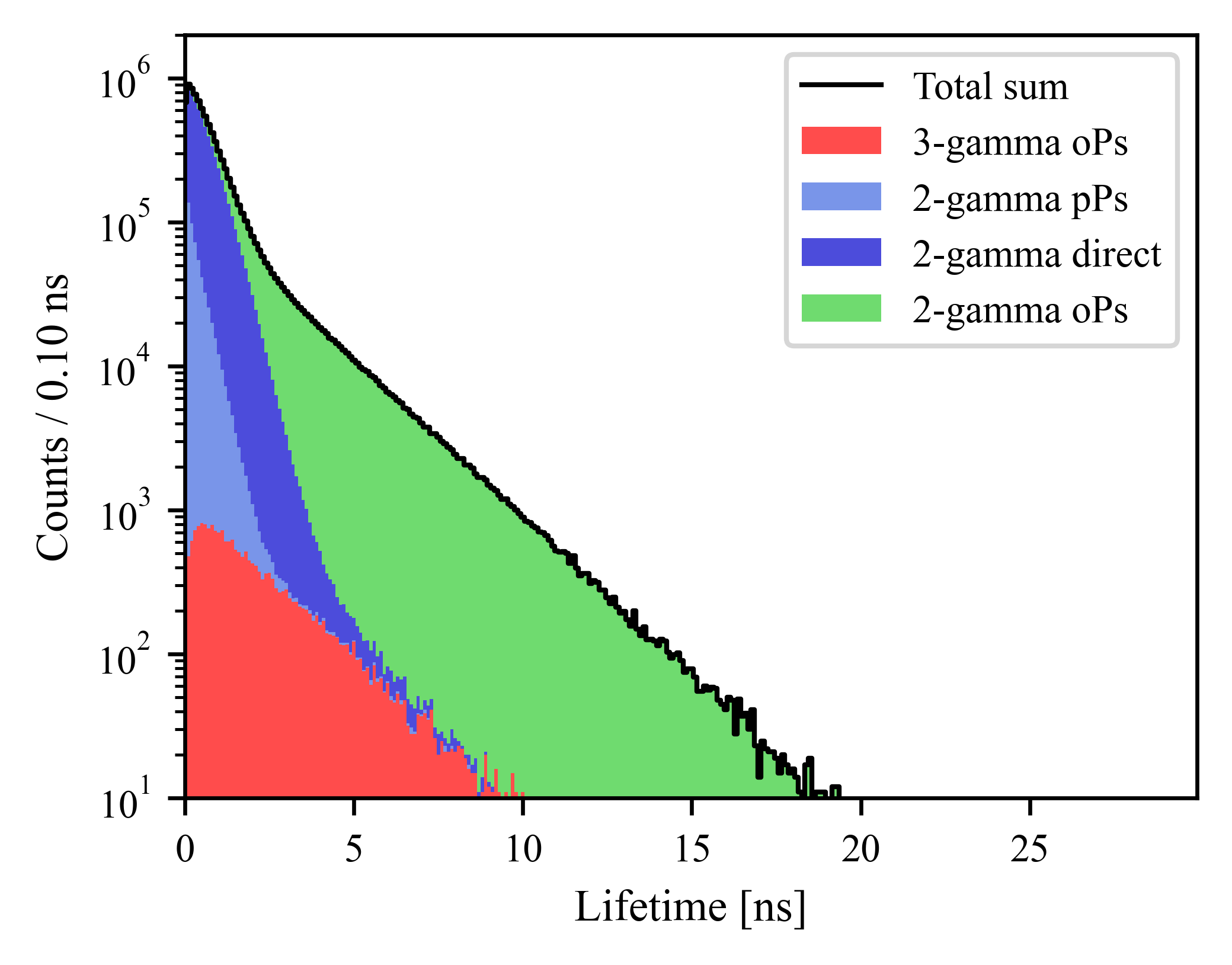}
    }
    \hfill
    \subfloat[Lifetime: 40 ns]{%
        \includegraphics[width=0.45\textwidth]{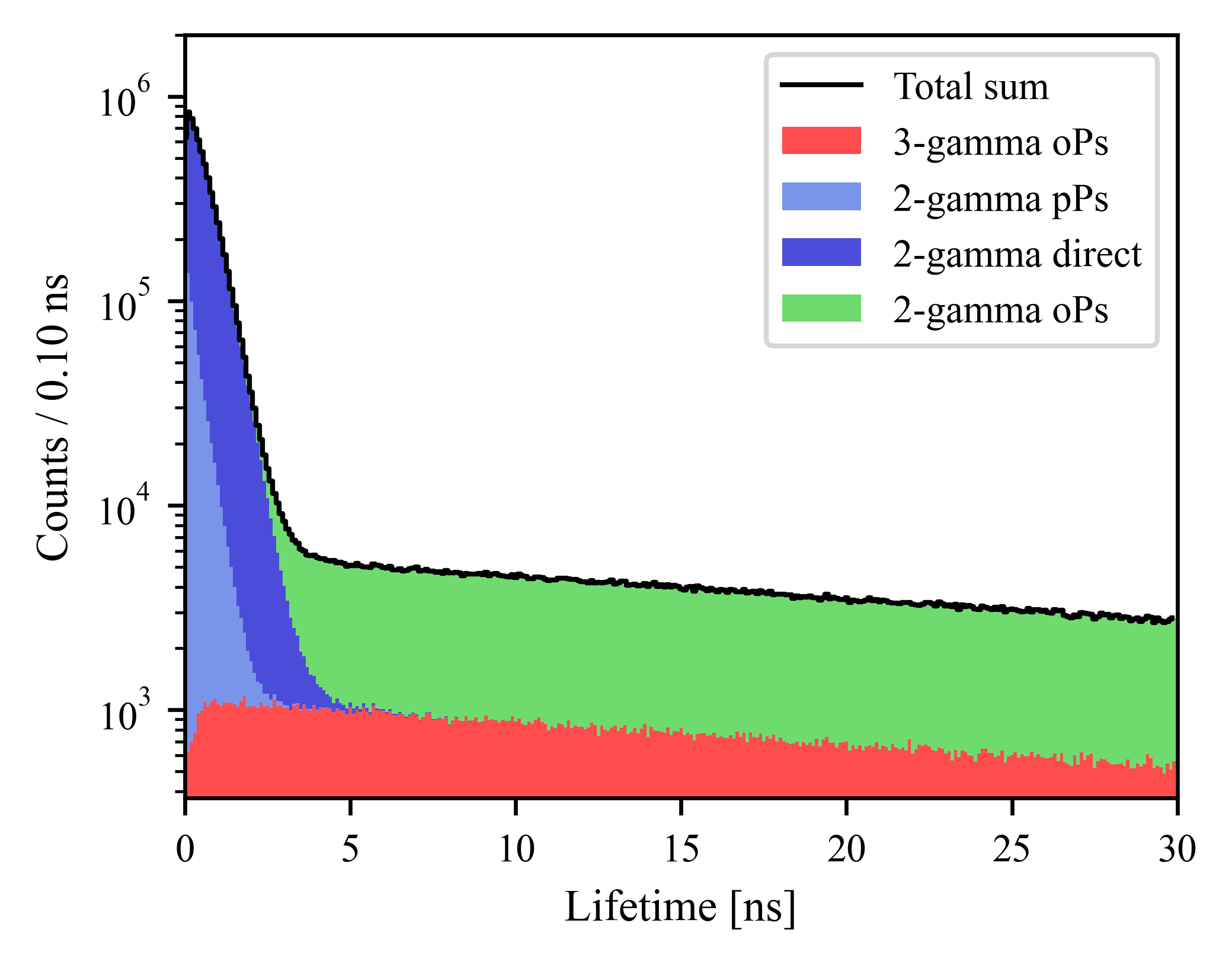}
    }

    \vspace{0.2em}

    % Bottom row (same size)
    \subfloat[Lifetime: 100 ns]{%
        \includegraphics[width=0.45\textwidth]{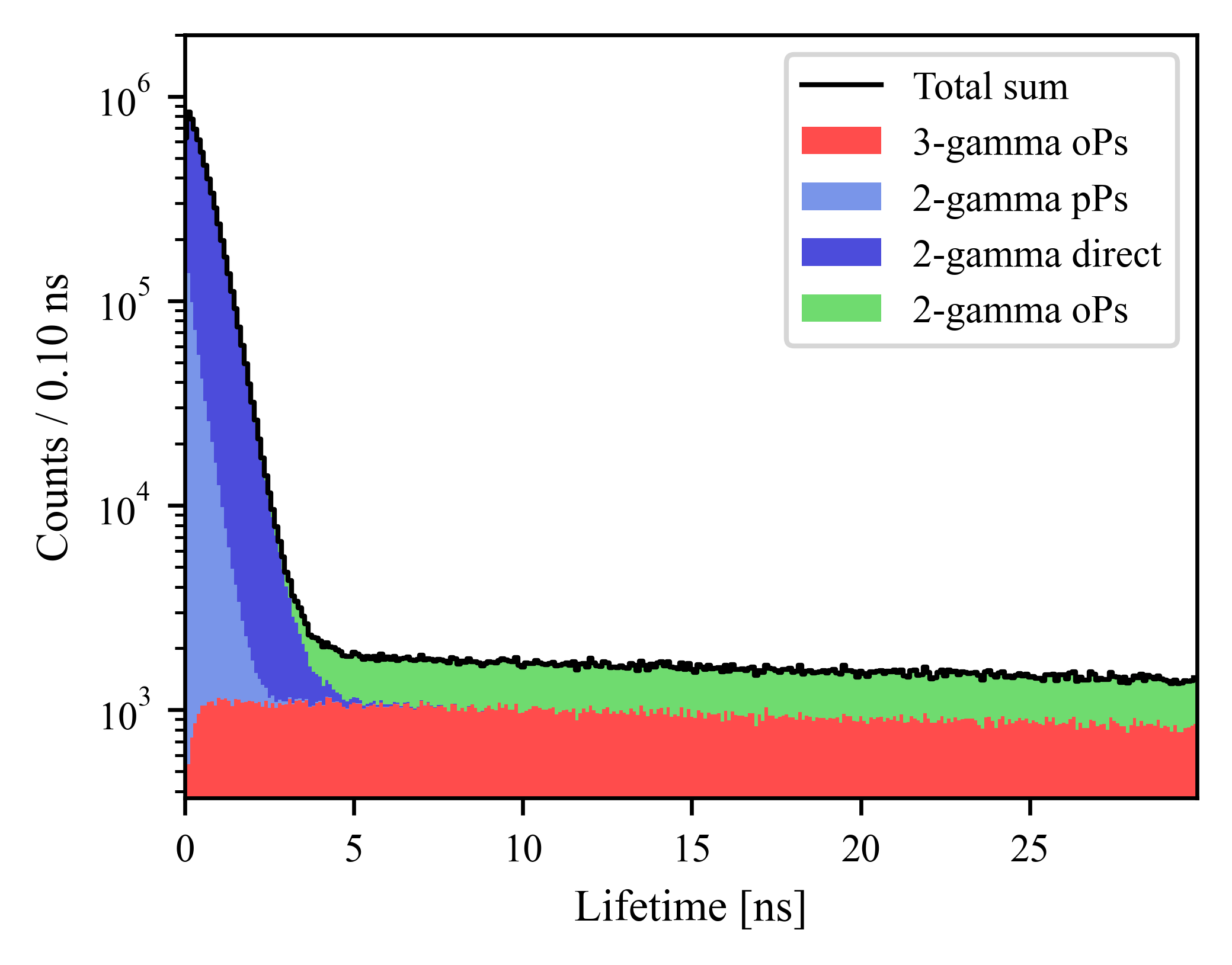}
    }

    \caption{Stacked lifetime distributions for the three \threetotwo simulations defined in \cref{tab:3to2_sim}, corresponding to \ac{o-Ps} lifetimes of (a) 2~ns, (b) 40~ns, and (c) 100~ns. Each histogram shows the contributions of the individual decay channels to the total lifetime spectrum.}
    \label{fig:3_to_2_lifetime}
\end{figure}

As the \ac{o-Ps} mean lifetime increases from 2~ns to 40~ns and 100~ns, the lifetime distributions extend and progressively broaden, with the long-lived tail becoming dominant for both \twogamma and \threegamma channels, as seen in \cref{fig:3_to_2_lifetime}.

\begin{table}[h]
\centering
\caption{Comparison of achieved fractions for each decay channel.}
\label{tab:3to2_sim_fractions}
    \begin{tabular}{ccccc}
        \toprule
        \textbf{\ac{o-Ps} lifetime [ns]} &
        \textbf{direct [\%]} & \textbf{\ac{p-Ps} [\%]} & \textbf{\ac{o-Ps} (\threegamma) [\%]} & \textbf{\ac{o-Ps} (\twogamma) [\%]} \\
        \midrule
           2 & 68.03 & 8.01 & 0.27 & 23.69 \\
          40 & 68.94 & 8.13 & 5.44 & 17.49 \\
         100 & 70.48 & 8.30 & 13.85 & 7.37 \\
        \bottomrule
    \end{tabular}
\end{table}

Consistently, the energy deposition spectra vary systematically with the \threegamma fraction (see \cref{fig:3_to_2_energy} in the appendix). For the 2~ns lifetime, the \ac{o-Ps} contribution is dominated by the \twogamma channel. For the 40~ns and 100~ns lifetimes, the growing \threegamma fraction produces a more pronounced low-energy continuum, while the relative contribution of the \twogamma channel decreases.

\subsection{Prompt photon emission benchmarks}

The deposited energy distributions from simulations of the dominant decay modes of \nucSc, \nucGa, and \nucI are shown in the appendix in \cref{fig:deposited_energy_44Sc}, \cref{fig:deposited_energy_68Ga} and \cref{fig:deposited_energy_124I}, respectively.

In each case, photoelectric peaks corresponding to all modelled decay modes were identified, with relative intensities consistent with matching the theoretical ratios. The Compton distributions associated with each decay mode were also reproduced correctly.

The overall ratios expressed as $\beta^+$ to electron capture ($\beta^+$-to-EC) of emitted prompt gammas are also in agreement with their expected values for all three radionuclides: $19.87\!:\!1$ vs.\ $20.06\!:\!1$ (expected) for \nucSc, $1\!:\!1.53$ vs.\ $1\!:\!1.53$ (expected) for \nucGa, and $1\!:\!2.6$ vs.\ $1\!:\!2.64$ (expected) for \nucI.

\subsection{Results of simulations of the NEMA IEC phantom -- \texorpdfstring{\nucSc}{Sc-44} with three decay modes}
\begin{figure}
    \centering
    \subfloat[Projection X-Y]{%
        \includegraphics[width=0.45\textwidth]{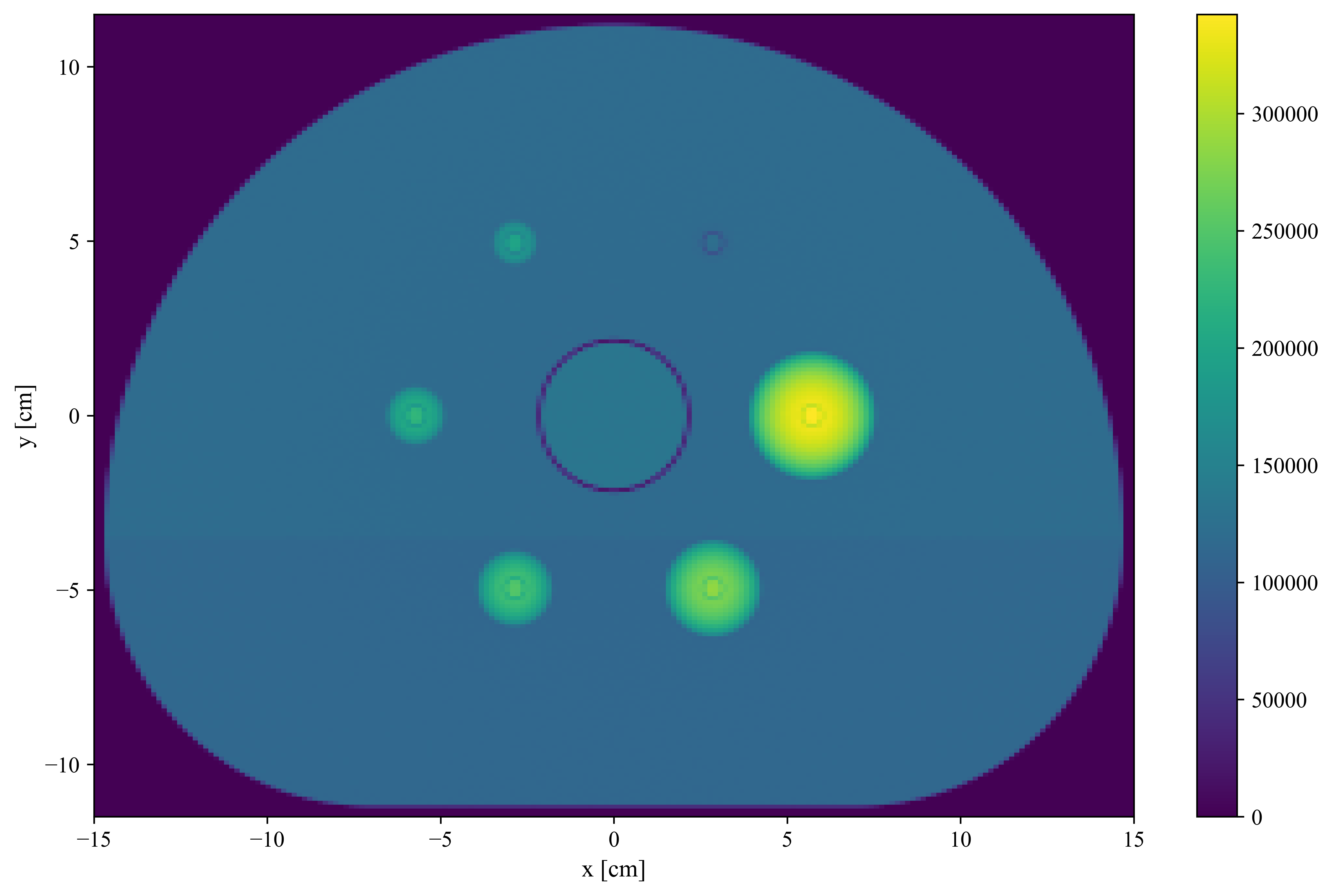}
    }
    \hfill
    \subfloat[Projection X-Z]{%
        \includegraphics[width=0.45\textwidth]{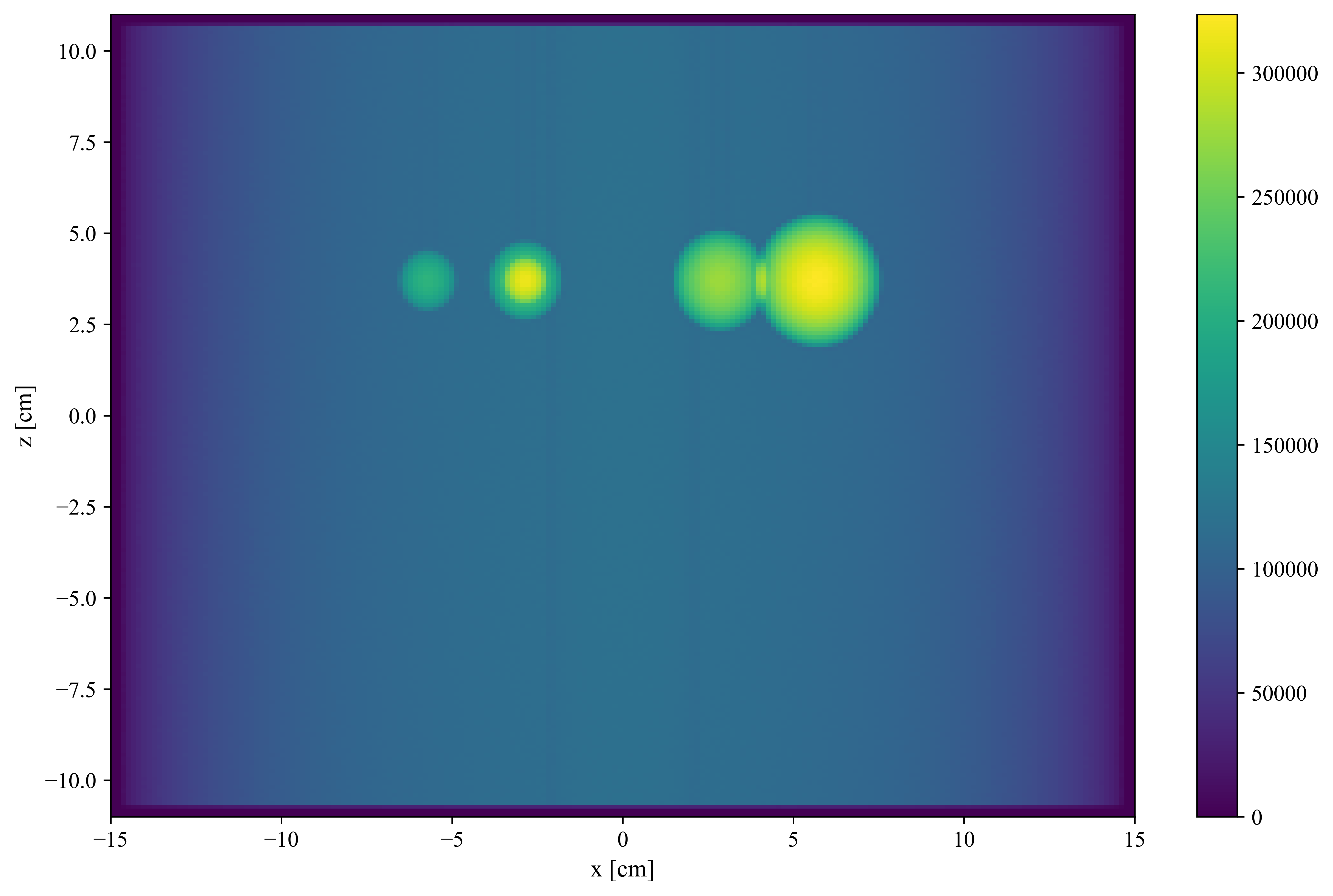}
    }

    \vspace{0.2em}

    % Bottom row (same size)
    \subfloat[Projection Y-Z]{%
        \includegraphics[width=0.45\textwidth]{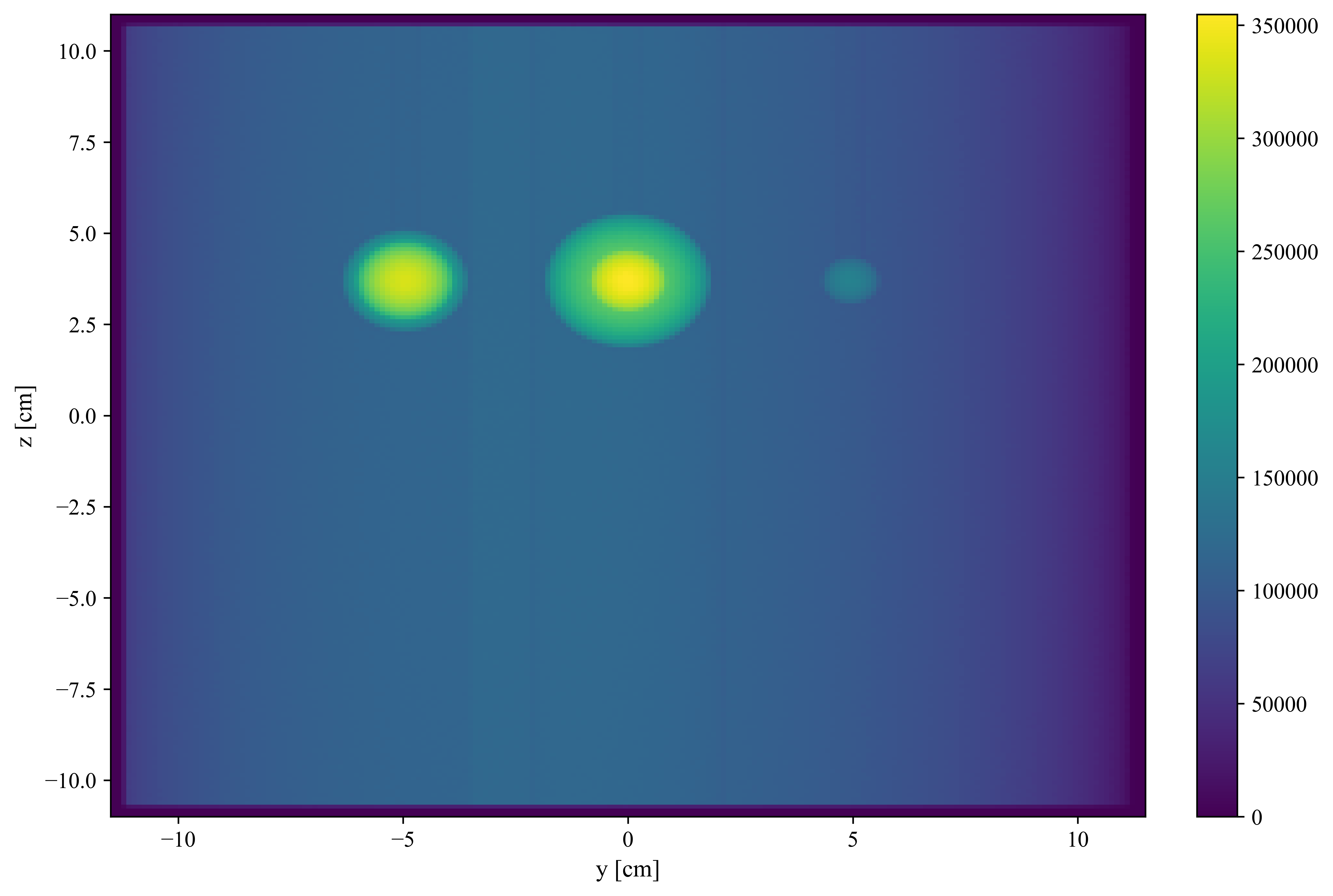}
    }

    \caption{Spatial distribution of emission positions of photons detected in the \nucSc NEMA IEC phantom simulation, shown as two-dimensional projections onto the (a) X-Y, (b) X-Z, and (c) Y-Z planes. The individual hot spheres are visible as localised regions of high activity against the lower-activity background.}
    \label{fig:scandium44_experiment_source_projections}
\end{figure}

The emission source distribution of the registered photons is presented in \cref{fig:scandium44_experiment_source_projections}.

\begin{figure}
    \subfloat[sphere 10 mm]{%
        \includegraphics[width=0.45\textwidth]{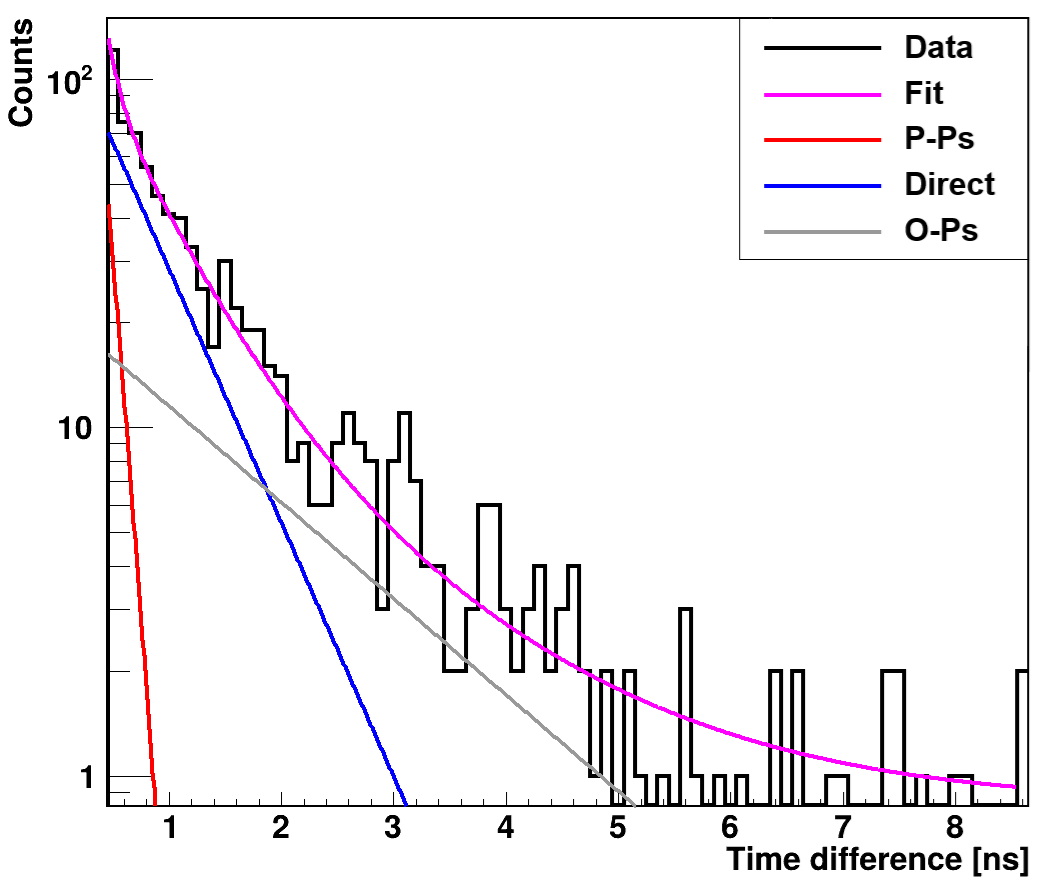}
    }
    \hfill
    \subfloat[sphere 13 mm]{%
        \includegraphics[width=0.45\textwidth]{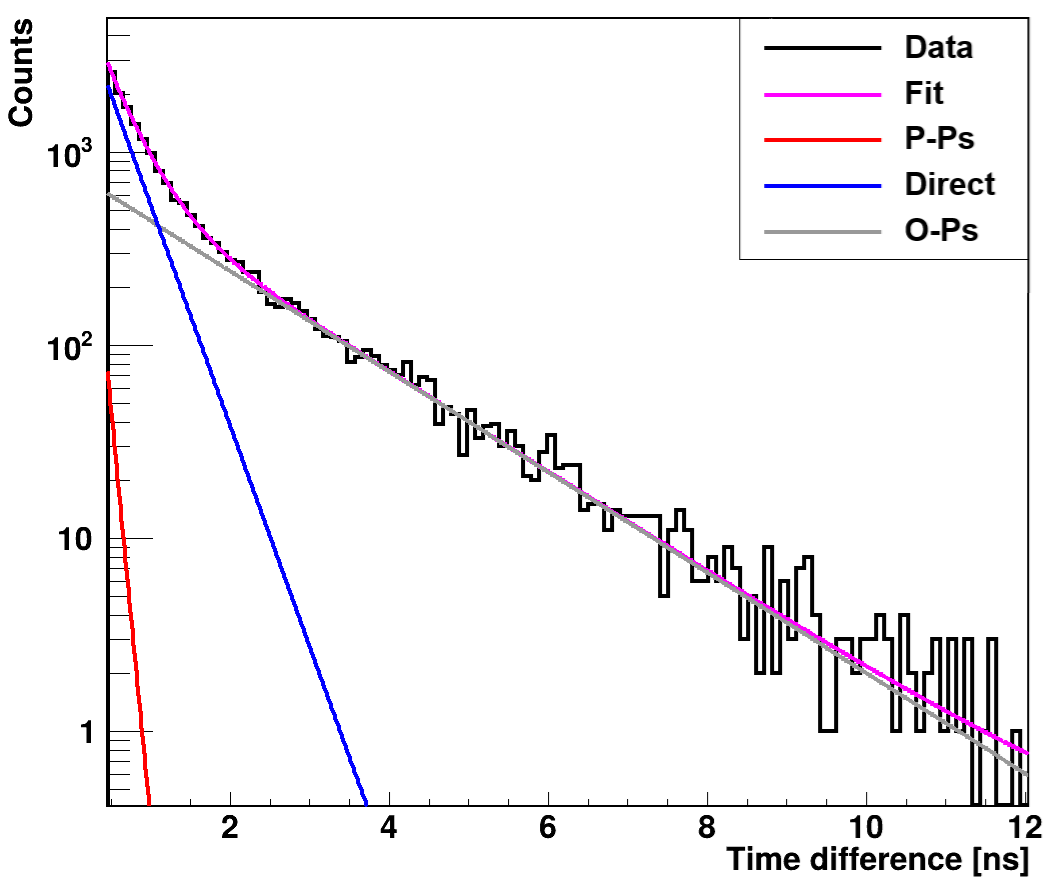}
    }

    \vspace{0.5em}

    \subfloat[sphere 17 mm]{%
        \includegraphics[width=0.45\textwidth]{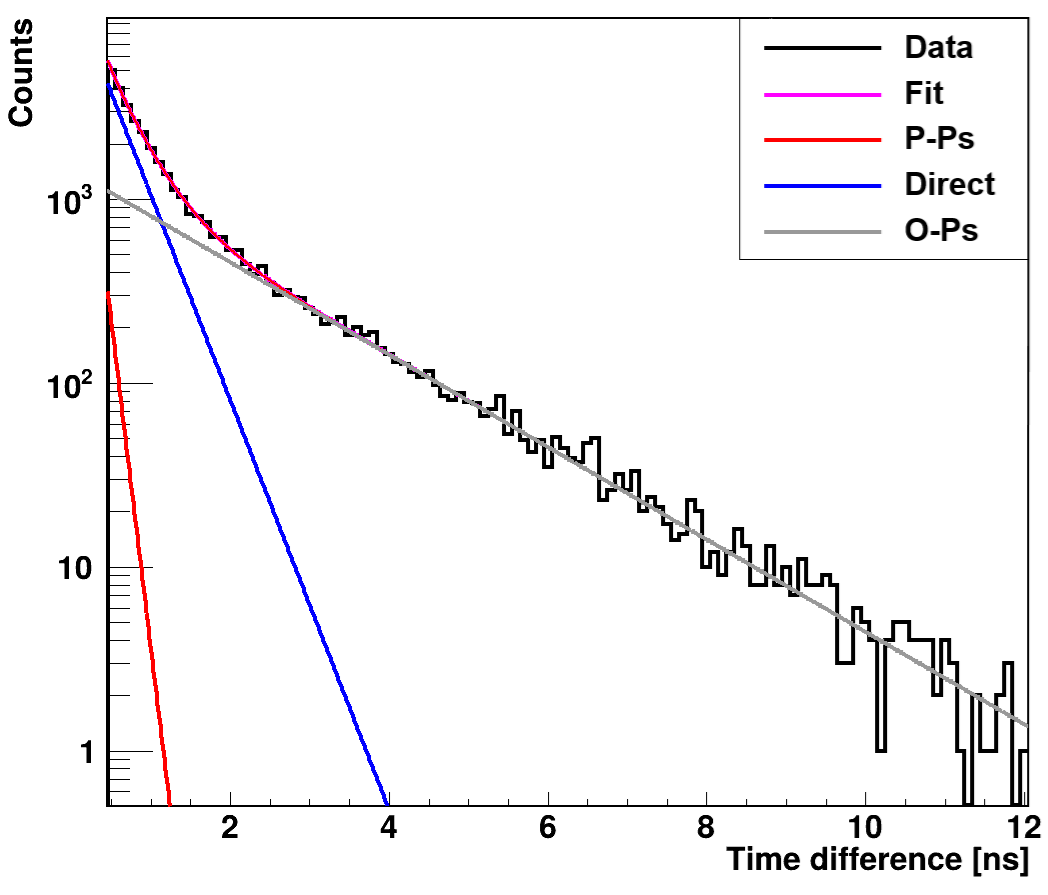}
    }
    \hfill
    \subfloat[sphere 22 mm]{%
        \includegraphics[width=0.45\textwidth]{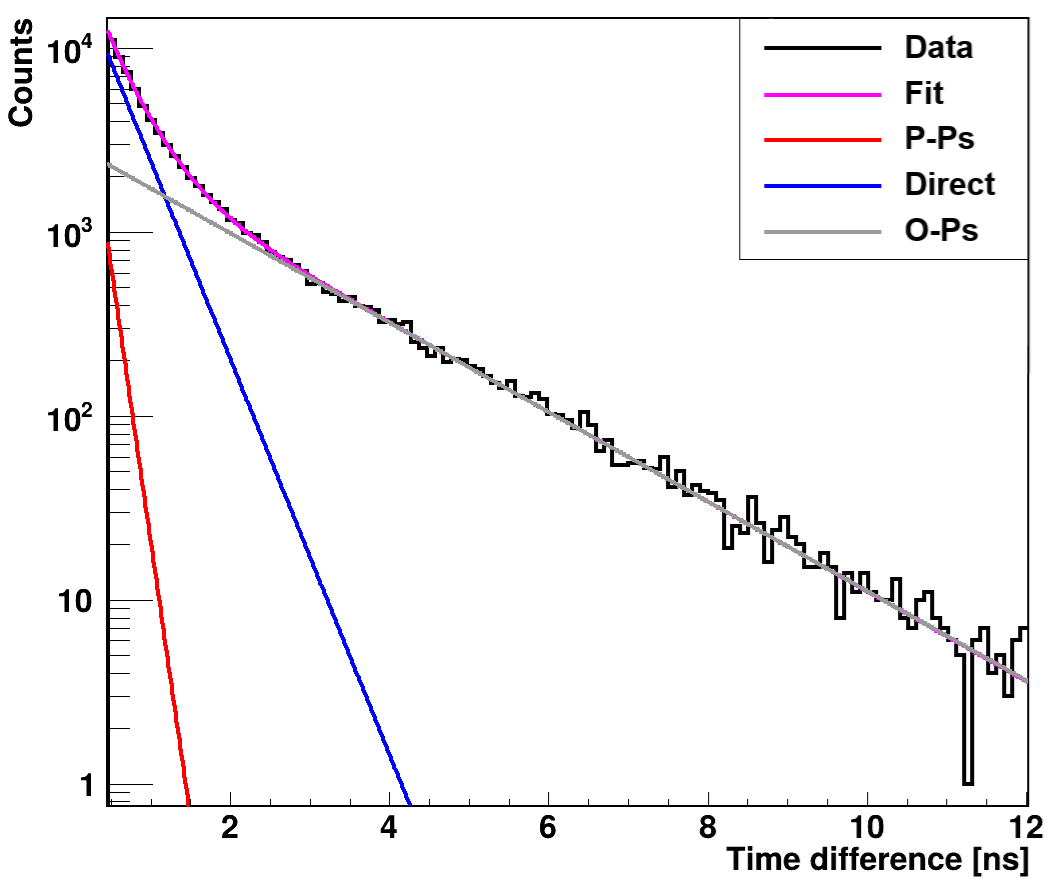}
    }

    \vspace{0.5em}

    \subfloat[sphere 28 mm]{%
        \includegraphics[width=0.45\textwidth]{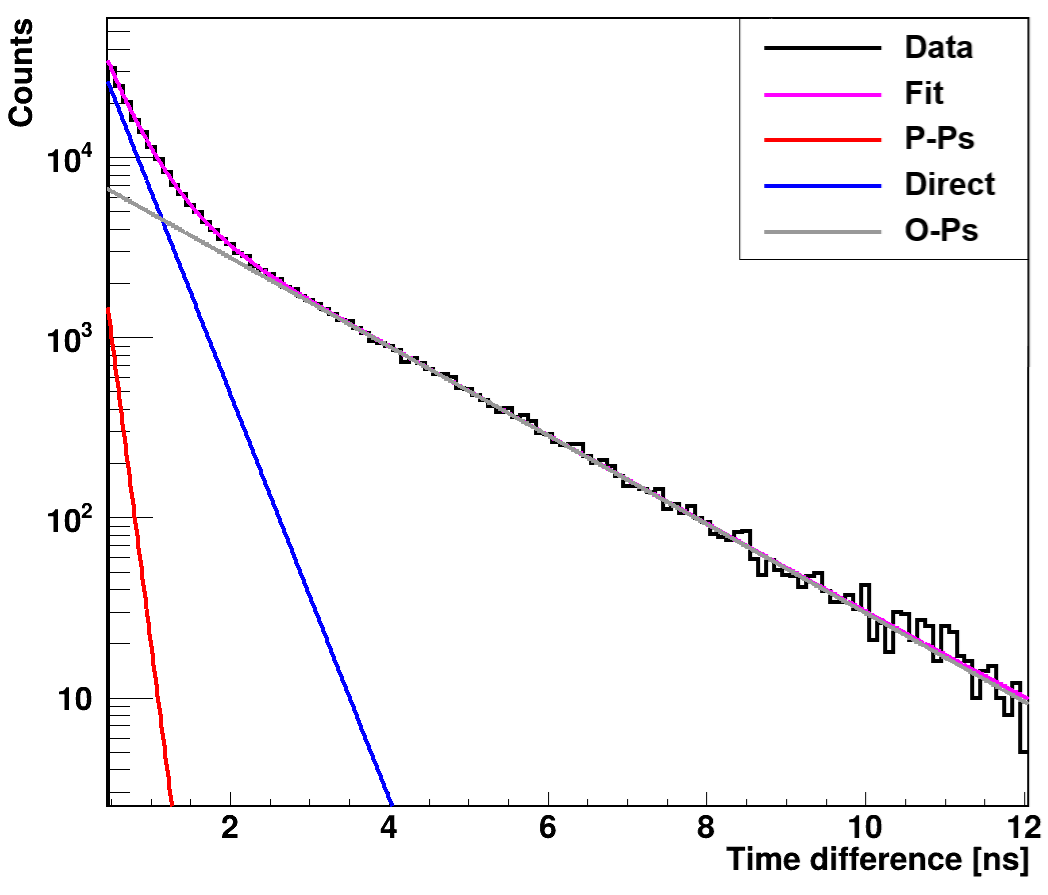}
    }
    \hfill
    \subfloat[sphere 37 mm]{%
        \includegraphics[width=0.45\textwidth]{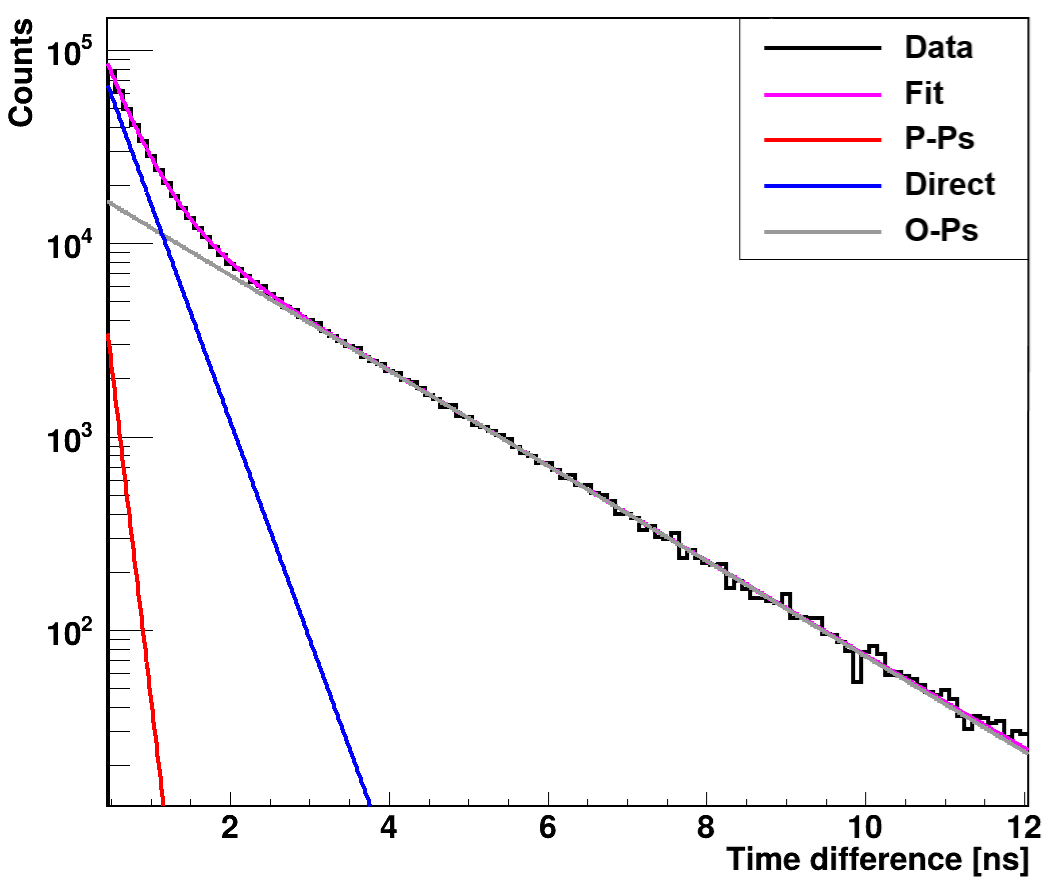}
    }

    \caption{Fitted lifetime histograms for the \nucSc NEMA IEC phantom simulation, for each of the six spheres. Fits were performed using PALS Avalanche, 
    with three decay components (\ac{p-Ps}, direct annihilation, and \ac{o-Ps}). The fitted parameter values are reported in \cref{tab:spheres_fit_scandium44}.}
    \label{fig:scandium44_experiment_spheres_fit}
\end{figure}

Lifetime distributions were 
fitted using PALS Avalanche \citep{PALSAV1, PALSAV2}. The results of the fit are reported in \cref{tab:spheres_fit_scandium44} and illustrated in \cref{fig:scandium44_experiment_spheres_fit}.

\begin{table}[htbp]
\centering
\caption{Results from fitting lifetime spectra for each sphere.}
\label{tab:spheres_fit_scandium44}
    \begin{tabular}{cccc}
        \toprule
        \textbf{Sphere} & \textbf{\ac{p-Ps}} & \textbf{direct annihilation} & \textbf{\ac{o-Ps}} \\
        \textbf{diameter} & \textbf{lifetime [ns]} & \textbf{lifetime [ns]} & \textbf{lifetime [ns]} \\
        \textbf{[mm]} & \textbf{/ intensity [\%]} & \textbf{/ intensity [\%]} & \textbf{/ intensity [\%]} \\
        \midrule
        10 & 0.103 (11) / 35.1 (2.2) & 0.596 (39) / 44.90 (10) & 1.57 (18) / 20.0 (2.2) \\
        13 & 0.101 (07) / 6.48 (02) & 0.380 (04) / 58.92 (52) & 1.67 (02) / 34.60 (52) \\
        17 & 0.122 (05) / 9.13 (33) & 0.390 (03) / 57.65 (30) & 1.73 (01) / 33.22 (30) \\
        22 & 0.145 (04) / 9.32 (20) & 0.404 (02) / 58.04 (02) & 1.79 (01) / 32.64 (20) \\
        28 & 0.127 (02) / 7.05 (11) & 0.386 (01) / 59.59 (01) & 1.76 (01) / 33.36 (12) \\
        37 & 0.125 (09) / 6.77 (02) & 0.386 (04) / 59.86 (61) & 1.76 (01) / 33.37 (61) \\
        \bottomrule
    \end{tabular}
\end{table}

Large bin-to-bin fluctuations observed in the tail of \cref{fig:scandium44_experiment_spheres_fit} are expected statistical fluctuations arising from the low event counts, rather than an indication of instability in the fitting procedure or the estimated lifetime parameters.
The accuracy of the fitted lifetime and intensity estimates improves with sphere size, reflecting the increase in the number of collected events: for the largest spheres, the fitted parameters converge closely to the input values, while for the smallest sphere (10~mm), the discrepancies are largest.
For the smallest sphere, the fitted \ac{o-Ps} lifetime is systematically shorter than the input value. This behaviour is consistent with a well-known fitting bias under low-count conditions: when the number of events is insufficient to resolve the multi-exponential structure, the minority component can be pulled toward the timescale of the dominant component, resulting in an underestimate of the longer lifetime. The effect disappears as statistics increase, as confirmed by the convergence of the fitted values toward the input lifetimes for the larger spheres.
Components with low intensity are associated with larger relative uncertainties in the fitted values, as seen for the direct annihilation component in the smaller spheres.

\subsection{Results of simulations of the NEMA IEC phantom - tissues with different distribution of positronium lifetime components}

In the second simulation with the NEMA IEC phantom described in \cref{fig:nemaIEC}, different distributions of positronium lifetimes corresponding to four sample types were simulated: water (background), bone, muscle, and fat. For each element, the positronium lifetime distribution was plotted and fitted similarly to the previous phantom.

\begin{figure}
    \subfloat[background]{%
        \includegraphics[width=0.45\textwidth]{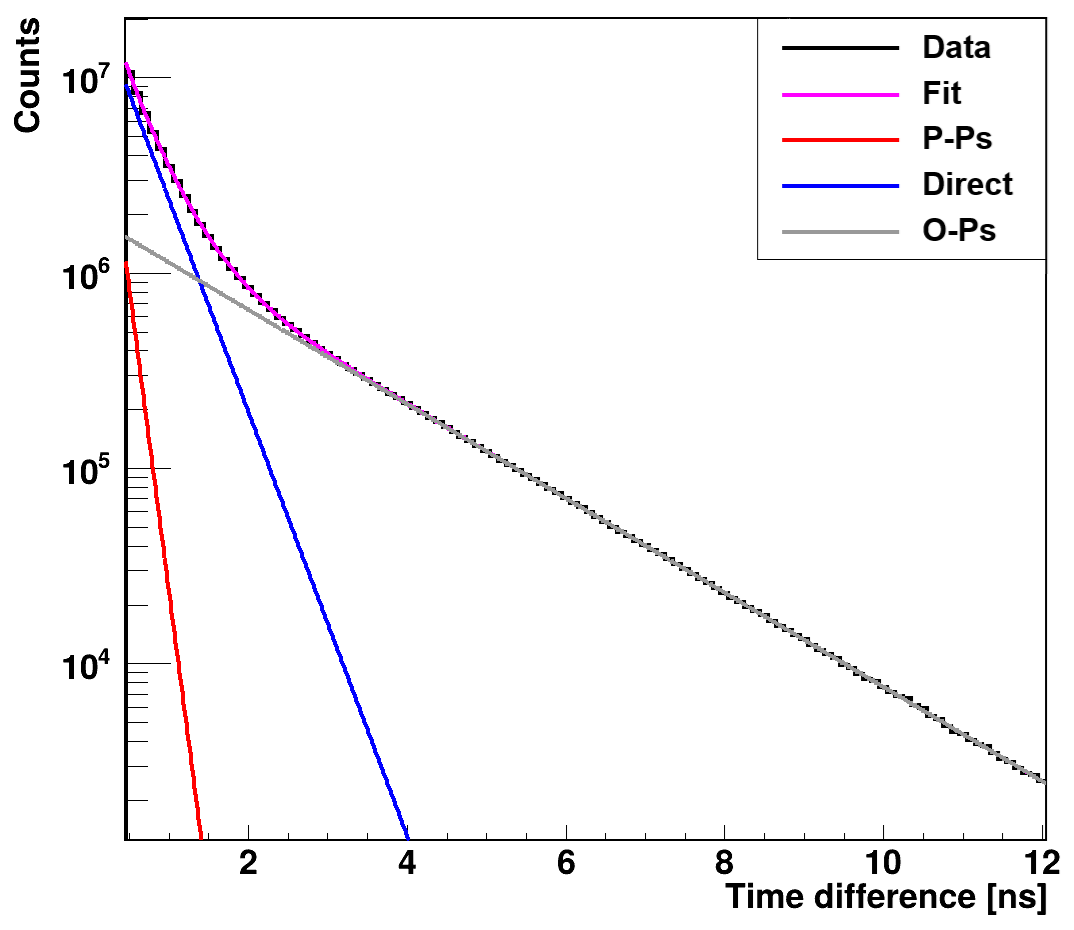}
    }
    \hfill
    \subfloat[sphere 10 mm]{%
        \includegraphics[width=0.45\textwidth]{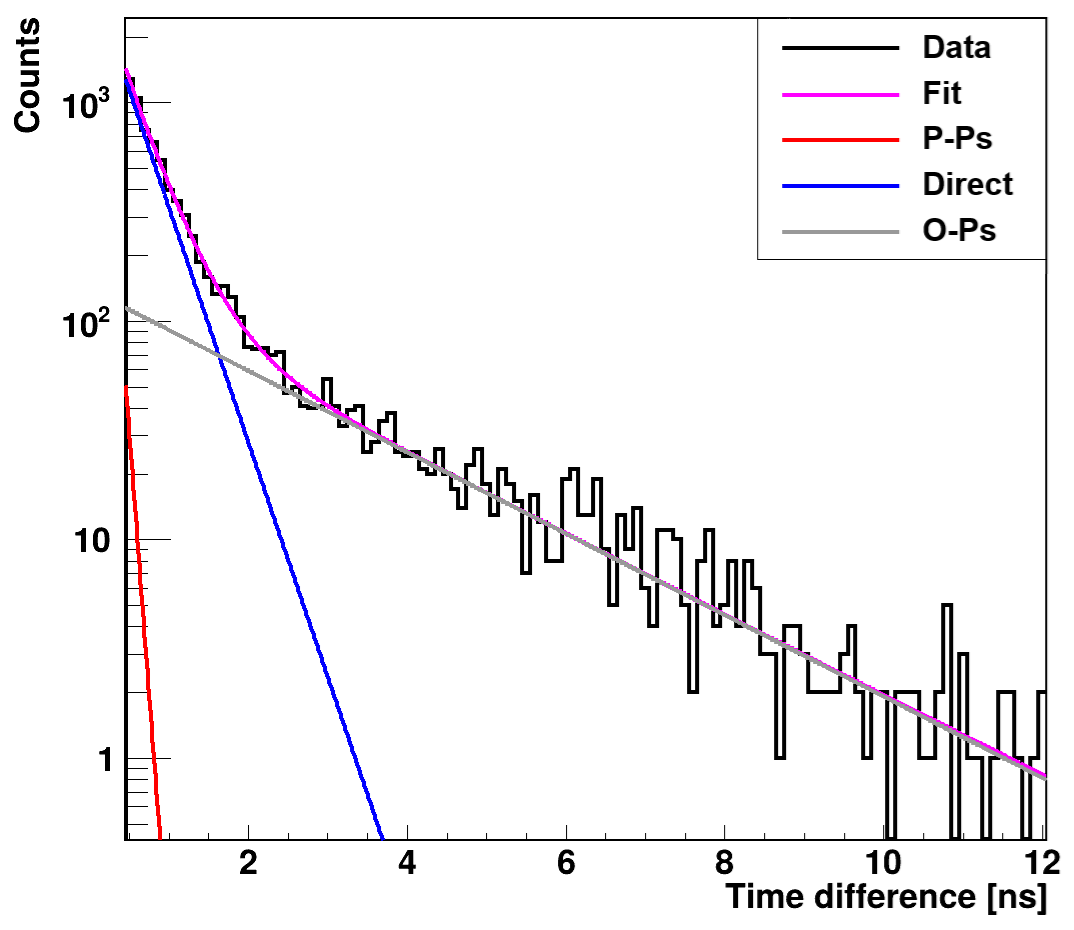}
    }

    \vspace{0.5em}

    \subfloat[sphere 13 mm]{%
        \includegraphics[width=0.45\textwidth]{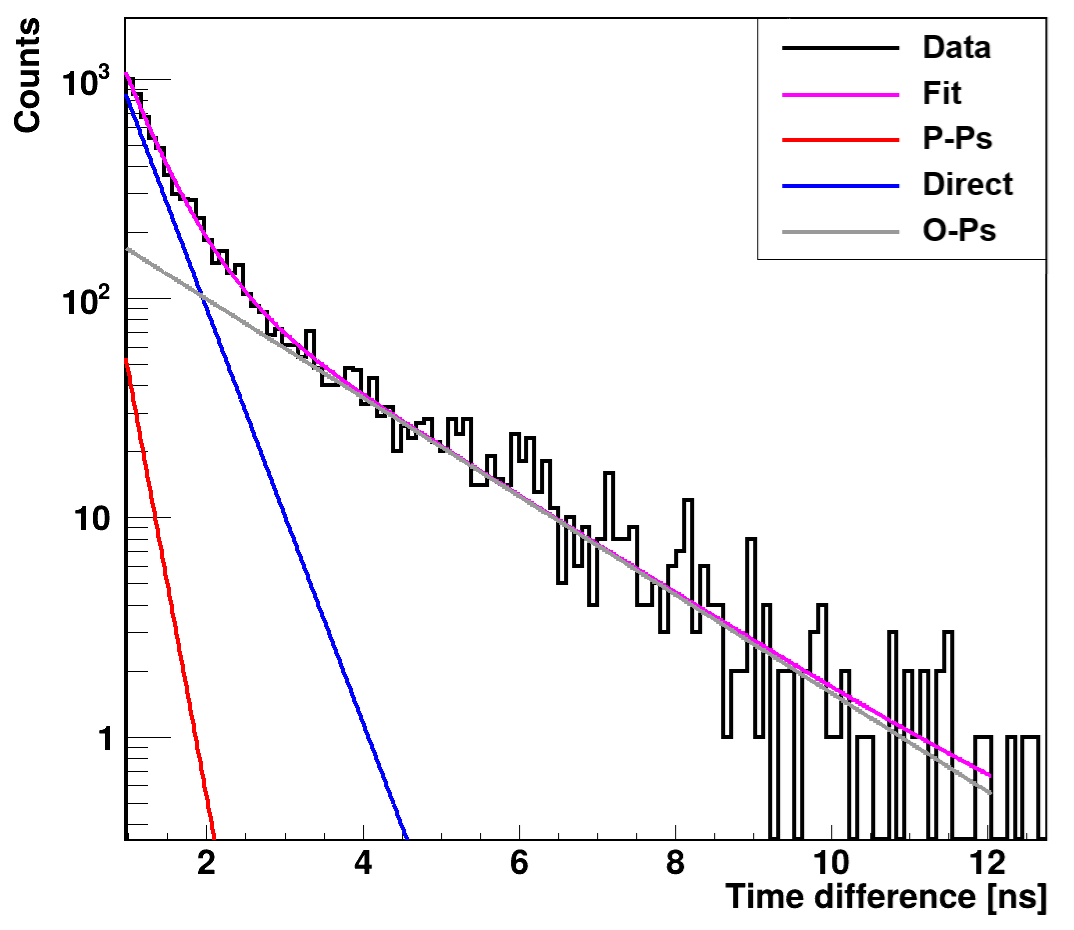}
    }
    \hfill
    \subfloat[sphere 22 mm]{%
        \includegraphics[width=0.45\textwidth]{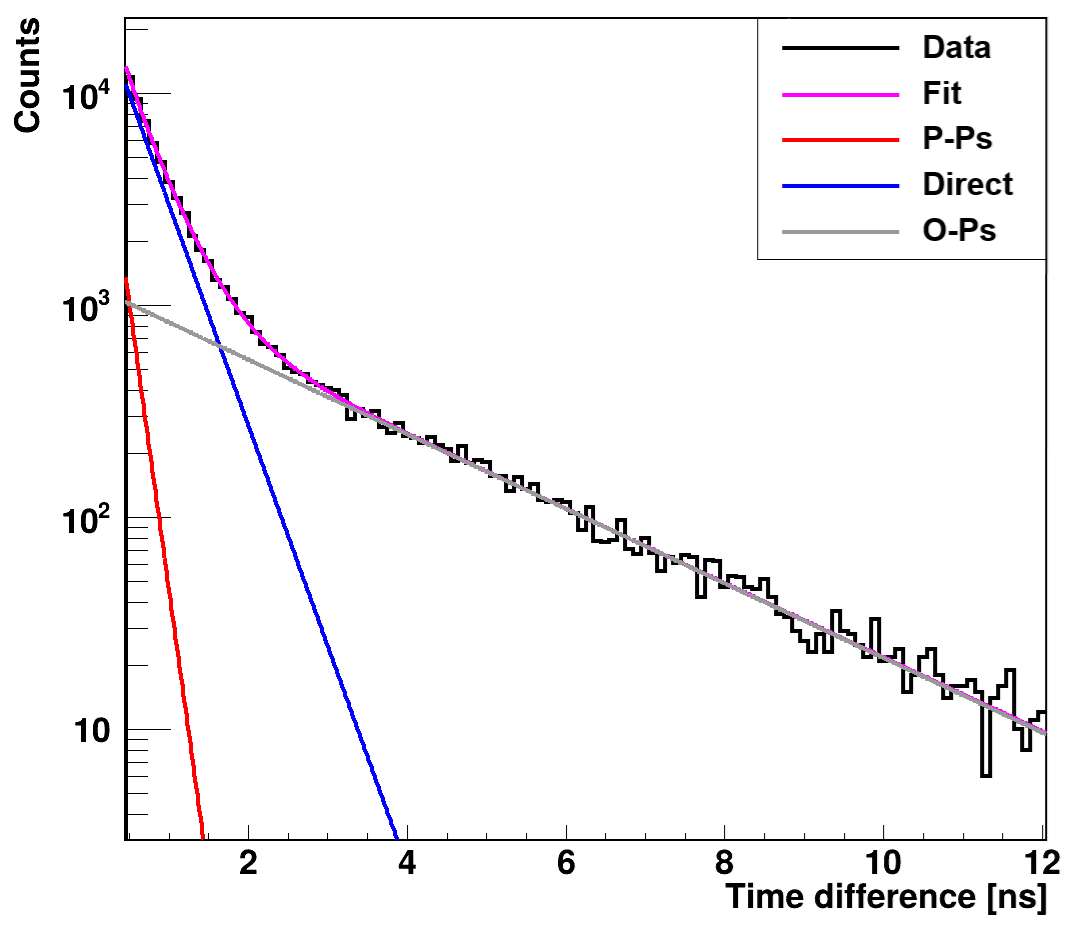}
    }

    \vspace{0.5em}

    \subfloat[sphere 28 mm]{%
        \includegraphics[width=0.45\textwidth]{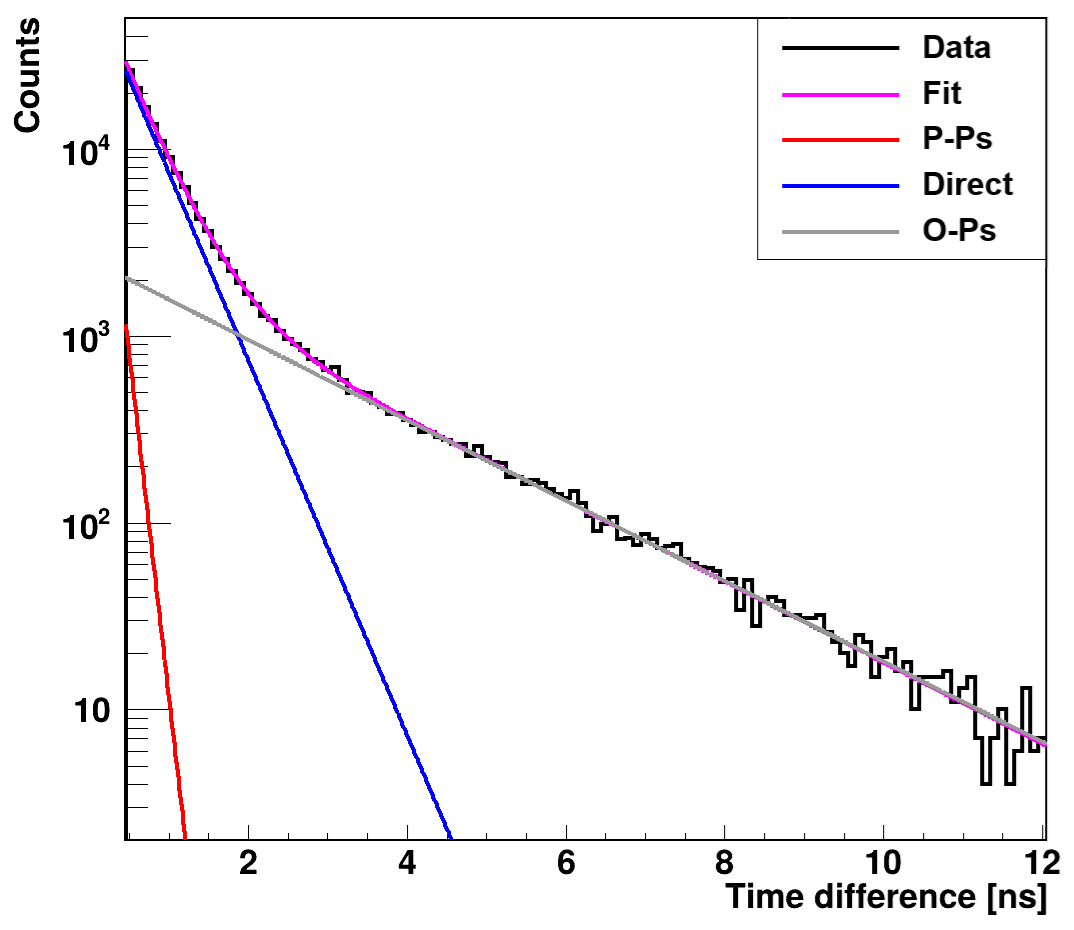}
    }
    \hfill
    \subfloat[central cylinder]{%
        \includegraphics[width=0.45\textwidth]{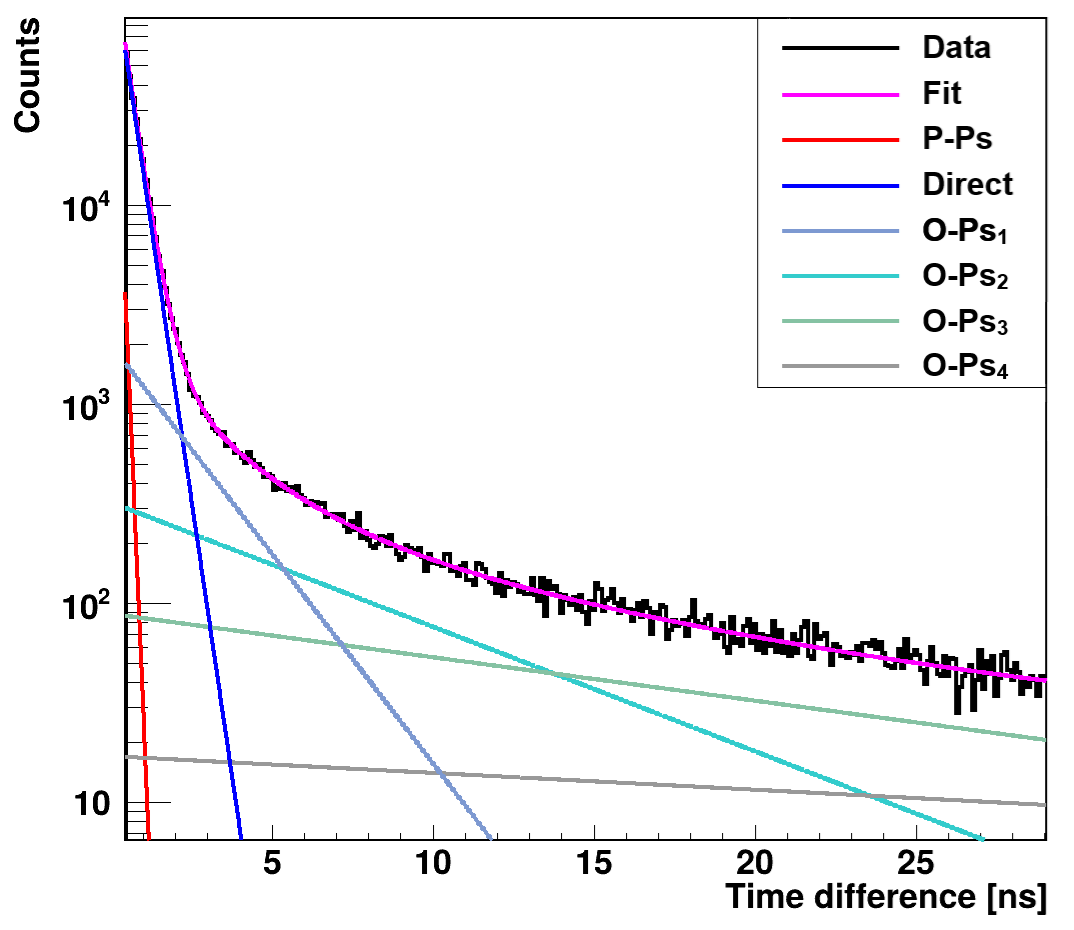}
    }

    \caption{Fitted lifetime histograms for the tissue-mimicking NEMA IEC phantom simulation, for the water background (a), the four hot spheres (b-e), and the central cylinder (f). Fits were performed using PALS Avalanche. The fitted parameter values are reported in \cref{tab:nema_fit_tissues}.}
    \label{fig:nemaIEC_experiment_tissues_fit}
\end{figure}

The fit results are shown in \cref{fig:nemaIEC_experiment_tissues_fit} and \cref{tab:nema_fit_tissues}. These results also show similar trends: the lower the statistics and the lower the intensity of a given component, the greater the error in estimating its mean lifetime, which is especially evident for the bone sample, whose distribution consisted of four \ac{Ps} components with decreasing intensity.

\begin{table}[htbp]
\centering
\caption{Results from fitting lifetime spectra for each part of the NEMA IEC phantom with tissues.}
\label{tab:nema_fit_tissues}
    \begin{tabular}{cccc}
        \toprule
        \textbf{Element} & \textbf{\ac{p-Ps}} & \textbf{direct annihilation} & \textbf{\ac{o-Ps}}$_1$ \\
        \textbf{of the} & \textbf{lifetime [ns]} & \textbf{lifetime [ns]} & \textbf{lifetime [ns]} \\
        \textbf{phantom} & \textbf{/ intensity [\%]} & \textbf{/ intensity [\%]} & \textbf{/ intensity [\%]} \\
        \midrule
        background & 0.140 (01) / 13.93 (01) & 0.400 (01) / 62.70 (01) & 1.80 (01) / 23.37 (01) \\
        sphere 10 mm & 0.091 (09) / 9.98 (45) & 0.404 (05) / 71.50 (01) & 2.33 (06) / 18.52 (45) \\
        sphere 13 mm & 0.225 (09) / 15.38 (72) & 0.460 (04) / 70.50 (70) & 1.93 (03) / 14.12 (02) \\
        sphere 22 mm & 0.161 (03) / 12.56 (01) & 0.417 (05) / 68.61 (15) & 2.47 (02) / 18.83 (15) \\
        sphere 28 mm & 0.118 (03) / 7.72 (10) & 0.433 (01) / 77.42 (01) & 2.02 (01) / 14.86 (10) \\
        central cylinder & 0.115 (01) / 11.57 (17) & 0.395 (01) / 76.83 (06) & 2.06 (02) / 5.21 (06) \\
        \midrule
         & \textbf{\ac{o-Ps}}$_2$ & \textbf{\ac{o-Ps}}$_3$ & \textbf{\ac{o-Ps}}$_4$ \\
        & \textbf{lifetime [ns]} & \textbf{lifetime [ns]} & \textbf{lifetime [ns]} \\
        & \textbf{/ intensity [\%]} & \textbf{/ intensity [\%]} & \textbf{/ intensity [\%]} \\
        \midrule
        central cylinder & 6.95 (60) / 2.91 (24) & 19.9 (1.9) / 2.32 (18) & 51.4 (5.1) / 1.16 (05) \\
        \bottomrule
    \end{tabular}
\end{table}

\section{Discussion}
\label{sec:Discussion}
\subsection{Validation results}
\label{subsec:discussion_validation}

\paragraph{Lifetime sampling.} 
The agreement between fitted and input lifetimes of around $1\%$, combined with coefficients of determination around $99\%$, confirms that the exponential sampling introduces no detectable bias, and reproduces the full temporal distribution, not only its mean. Since the same sampling mechanism is used for both the \twogamma and \threegamma channels, these results validate the model's temporal component across all annihilation modes.

\paragraph{Three-photon kinematics.}
The comparison between the simulated energy spectrum and the theoretical predictions shows that the model accurately reproduces the distribution of energy between emitted photons. The energy spectrum closely follows the expected continuous distribution, including the characteristic peak near 511~keV and the subsequent cutoff above that energy. Similarly, the joint angular distribution exhibits the expected theoretical results of energy and momentum conservation constraints. The observed triangular region matches the theoretically allowed phase space, with no unphysical configurations being produced by the model. The consistency between simulation and theory indicates that the implementation correctly models the underlying three-body decay phase space and respects the correlations between photon energies and emission angles.

\paragraph{Channel fraction reproduction.}
The multi-channel simulations demonstrate that the model reliably reproduces user-defined branching fractions, including scenarios with competing \twogamma and \threegamma processes. For the \twogamma channels, the achieved fractions are in close agreement with the input values across all three simulations. For the \threegamma channels, a consistent underestimation of approximately $19\%$ in relative terms is observed, corresponding to absolute deviations of 0.07, 1.32 and 3.05 $\%$ for the three simulations, respectively. This systematic trend is attributed to the difference in detection efficiency between \twogamma and \threegamma annihilation events. No efficiency corrections were applied in the present simulations as they were designed primarily as self-consistency checks. The observed deviations reflect known efficiency effects, rather than a deficiency of the source model.
Moreover, the results also show that changes in the energy spectra and lifetime distributions are consistent with changes in \threetotwo ratios.

Together, these results confirm that the model correctly describes multiple positronium decay channels, accurately reproducing the interplay between lifetimes, branching fractions, and their observable signatures.

Similar conclusions can be drawn from the simulation of a NEMA IEC phantom in which the full positronium lifetime distributions were simulated. The results from fitting the obtained positronium lifetime distributions were consistent with the simulated distributions for each phantom element in terms of statistical uncertainty. Additionally, it was shown that assuming a different distribution for a given phantom element significantly affects the obtained distribution of positronium lifetimes, which confirms the correct operation of the introduced functionalities. Simulations with NEMA IEC phantoms confirmed that the precision of the positronium lifetime parameters is primarily influenced by the collected statistics for a given lifetime component. Simulations of this type can significantly contribute to the development of appropriate measurement procedures for verifying the positronium lifetime imaging performance of \ac{PET} systems.

\subsection{Comparison with existing approaches}
\label{subsec:discussion_comparison}
The proposed model advances beyond existing positronium approaches in several respects.
In standard Geant4, positron annihilation is treated within electromagnetic processes without an explicit representation of positronium as a physical particle or bound state. The commonly used \texttt{G4eplusAnnihilation} process models positron annihilation as a direct interaction with an atomic electron, resulting in two-photon emission under the assumption of a free electron at rest.
The formation and decay of positronium, and the associated lifetime distributions, are not explicitly described. Recent developments (Geant4 11.4) introduced the possibility of assigning an effective positronium lifetime during the annihilation process; however, this approach phenomenologically assigns a single lifetime parameter and does not provide a multi-channel description of \ac{Ps} decay.

Tashima et al.~\citep{tashimaModelingPositroniumLifetime2023} implemented \ac{Ps} lifetime
modelling as a Geant4-based application by supplying a spatially varying lifetime image as input, with the annihilation time sampled via Poisson fluctuations around a mean determined by the \ac{o-Ps} branching
ratio. Two separate simulations were performed, each assuming
a single uniform \ac{o-Ps} lifetime (1.62~ns and 2.10~ns, respectively), within a geometry specific to Whole Gamma Imaging. The approach does not support multi-component lifetime
structures or variable annihilation multiplicity.

\ac{GATE} inherits these limitations from Geant4, although the \texttt{ExtendedVSource} class (introduced in \ac{GATE}~9.3) partially addresses them by enabling the simulation of \ac{p-Ps} and \ac{o-Ps} decay channels within a source-based framework.
The present model removes its two-channel restriction and, as demonstrated by the XAD4 example in \cref{fig:xad4}, supports the multi-component lifetime structures observed in realistic biological and material systems. To verify backward compatibility, the present model was additionally benchmarked against \ac{GATE}~9.3 \texttt{ExtendedVSource} implementation for the two-channel case supported by both models. The results were found to be consistent within statistical uncertainties, confirming that the new implementation correctly reproduces the behaviour of the established tool.

Alternative approaches, such as the J-PET Geant4 package~\citep{moskalFeasibilityStudyPositronium2019c}, provide the capability for effective modelling of multi-channel positronium decay; however, the package is tailored to the specific geometry and software requirements of the J-PET system and is not integrated into \ac{GATE}.

\subsection{Scope of applicability}
\label{subsec:discussion_scope}

The validated model is applicable across a range of research contexts, as outlined below.

The NEMA phantom simulations presented in \cref{subsec:NEMA_1} and \cref{subsec:NEMA_2} demonstrate the model's ability to generate positronium-sensitive datasets with tissue-specific lifetime signatures, directly supporting the development and validation of \ac{PLI} reconstruction algorithms.

Beyond medical imaging, the model could support feasibility studies of positronium-sensitive imaging for industrial tomography and non-destructive testing, though such applications were not investigated in the present work.

Finally, the model may be used for fundamental physics studies involving positronium, where accurate modelling of lifetime distributions and branching fractions between two- and three-photon annihilation channels is required for detector optimisation and background estimation.

The incorporation of full positronium lifetime distributions enables more realistic simulations of positronium behaviour in matter, improving the reliability and interpretability of \ac{MC} studies across these application domains.

\subsection{Limitations and future directions}
\label{subsec:discussion_limitations}

The present model treats positronium decay as an effective gamma-emission source, with positronium formation represented implicitly through the user-defined lifetime and branching
fraction parameters rather than through physical simulation of positron thermalisation and Ps formation. A deeper integration
with Geant4 electromagnetic transport processes, starting from positron emission and tracking the particle through
thermalisation to positronium formation, would provide a more complete physical description and is identified as a possible direction
for future development. At the same time, a full formation-level simulation will introduce additional processing time cost; therefore, for many applications, it might be more beneficial to stay at the effective simulation level.

The introduced positron range model should be treated rather as an exemplary approximation: the displacement between the positron emission point and the annihilation site is modelled as a Gaussian smearing with a user-defined mean, rather than through a full positron transport simulation. This approximation is computationally efficient and adequate for source-level simulations, but may introduce bias
in applications where the spatial distribution of annihilation sites is critical.

Finally, the validation presented in this work is based entirely on analytical and numerical benchmarks. Validation against experimental positronium lifetime data is planned as a subsequent step and will be necessary before the model can be used for quantitative clinical or experimental studies.

\section{Conclusions}
\label{sec:Conclusions}

We presented a \ac{MC} model of positronium decay with multiple annihilation channels, implemented as a modular, user-configurable source within the \ac{GATE}~9.4 and \ac{GATE}~10 simulation frameworks. To the best of our knowledge, this is the first such general-purpose, multi-channel positronium decay model integrated into a widely adopted simulation framework, validated through the analytical and numerical benchmarks described in \cref{sec:Materials_and_methods,sec:Results,sec:Discussion}.

The model reproduces lifetime distributions, branching fractions and photon kinematics.
Prompt photon emission was validated for three clinically relevant radionuclides (\nucSc, \nucGa, and \nucI).
The practical utility of the model was demonstrated through simulations of multi-channel annihilation scenarios and the NEMA IEC phantom filled with tissue-mimicking positronium sources, confirming its applicability to \ac{PLI} and multi-photon \ac{PET} for industrial tomography and fundamental physics studies (\cref{subsec:discussion_scope}).

The current effective gamma-source implementation leaves several aspects open for future development, most notably a deeper integration with Geant4 electromagnetic transport, potentially enabling simulations that start from positron emission and track the particle through thermalisation to \ac{Ps} formation explicitly, and validation against experimental positronium lifetime data.

\section*{Acknowledgments}
W.K. thanks Nils Krah and Thomas Baudier for their help with porting the code into \ac{GATE}~10.
The authors acknowledge the support by the Foundation for Polish Science through the FIRST TEAM FENG.02.02-IP.05-0152/23 programme co-financed by the European Union under the European Funds for Smart Economy 2021-2027 (FENG). The work is co-financed by the Polish National Agency for Academic Exchange in the frame of the project BPN/BFR/2025/1/00036/U/00001 and BPN/BAT/2025/1/00009 and co-supported by the Austrian Ministry for Women, Science and Research (project WTZ PL 11/2026).
This work was partially supported by the “PHC POLONIUM” program (project number: 55194PB), funded by the French Ministry for Europe and Foreign Affairs, the French Ministry for Higher Education, Research and Space, and the Polish NAWA.
This work was completed with resources provided by the Świerk Computing Centre at the National Centre for Nuclear Research.

\bibliography{references}

% Appendix
\newpage
\appendix

\setcounter{figure}{0}
\setcounter{table}{0}
\setcounter{lstlisting}{0}
\renewcommand{\thefigure}{A.\arabic{figure}}
\renewcommand{\thetable}{A.\arabic{table}}
\renewcommand{\thelstlisting}{A.\arabic{lstlisting}}
\renewcommand{\theHfigure}{A.\arabic{figure}}
\renewcommand{\theHtable}{A.\arabic{table}}
\renewcommand{\theHlstlisting}{A.\arabic{lstlisting}}

\newpage
\section{Supplementary material for \texorpdfstring{\cref{sec:Results} (Results)}{results section}}
\label{app:validation_results}

% Energy spectra for 3-to-2 gamma
\begin{figure}[H]
    \centering

    \subfloat[Lifetime: 2 ns]{%
        \includegraphics[width=0.49\textwidth]{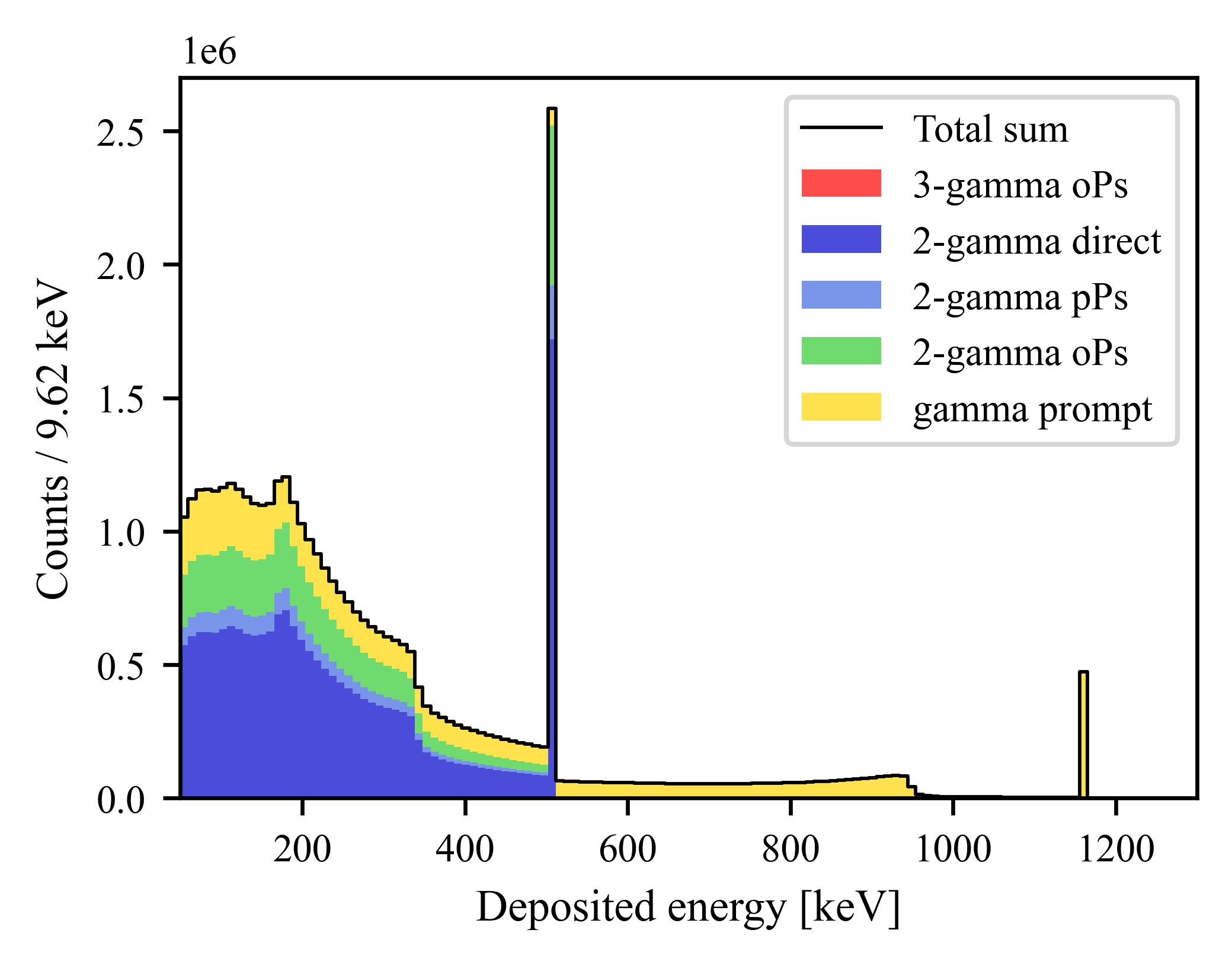}
        \includegraphics[width=0.49\textwidth]{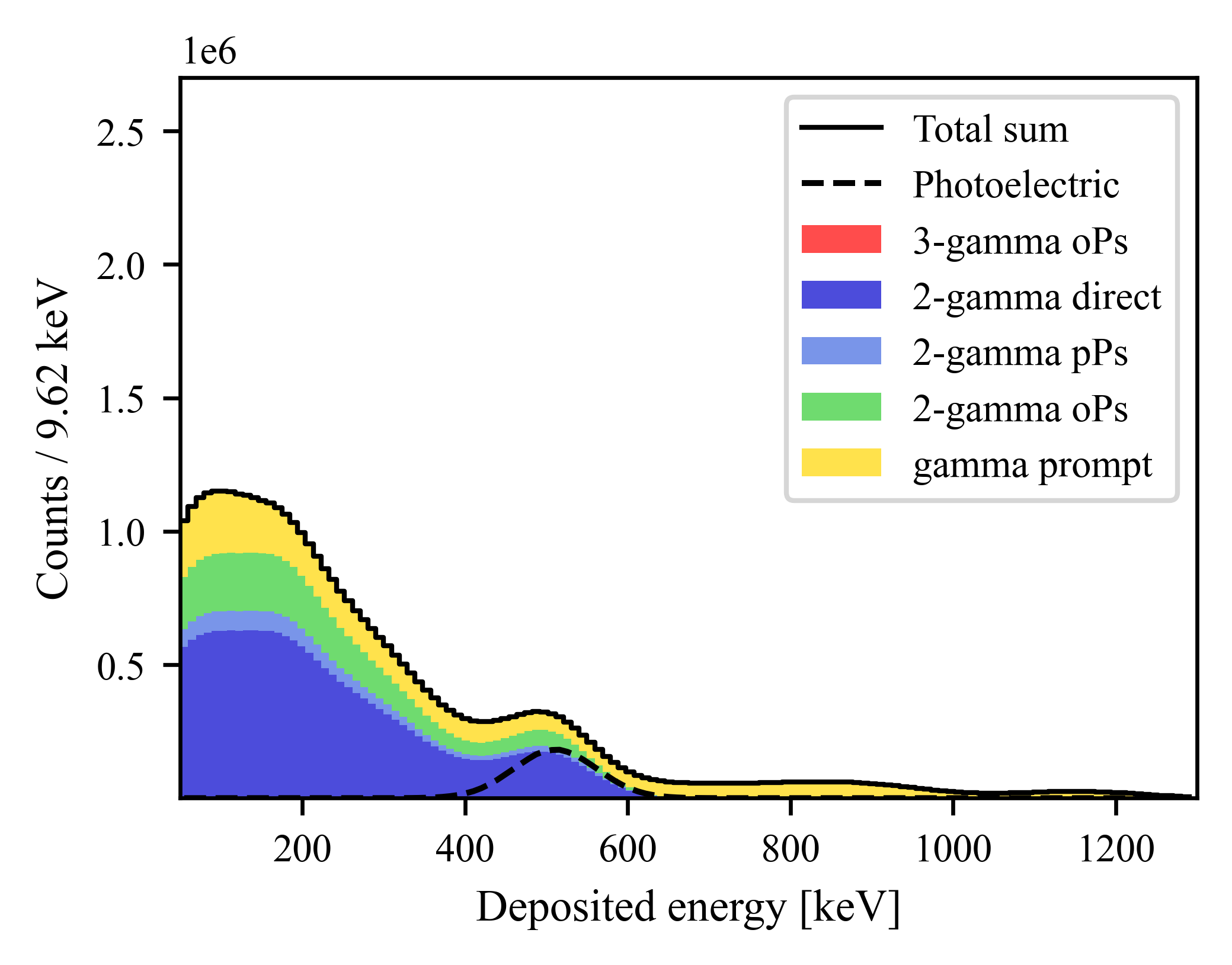}
    }
    \vspace{0.2em}

    \subfloat[Lifetime: 40 ns]{%
        \includegraphics[width=0.49\textwidth]{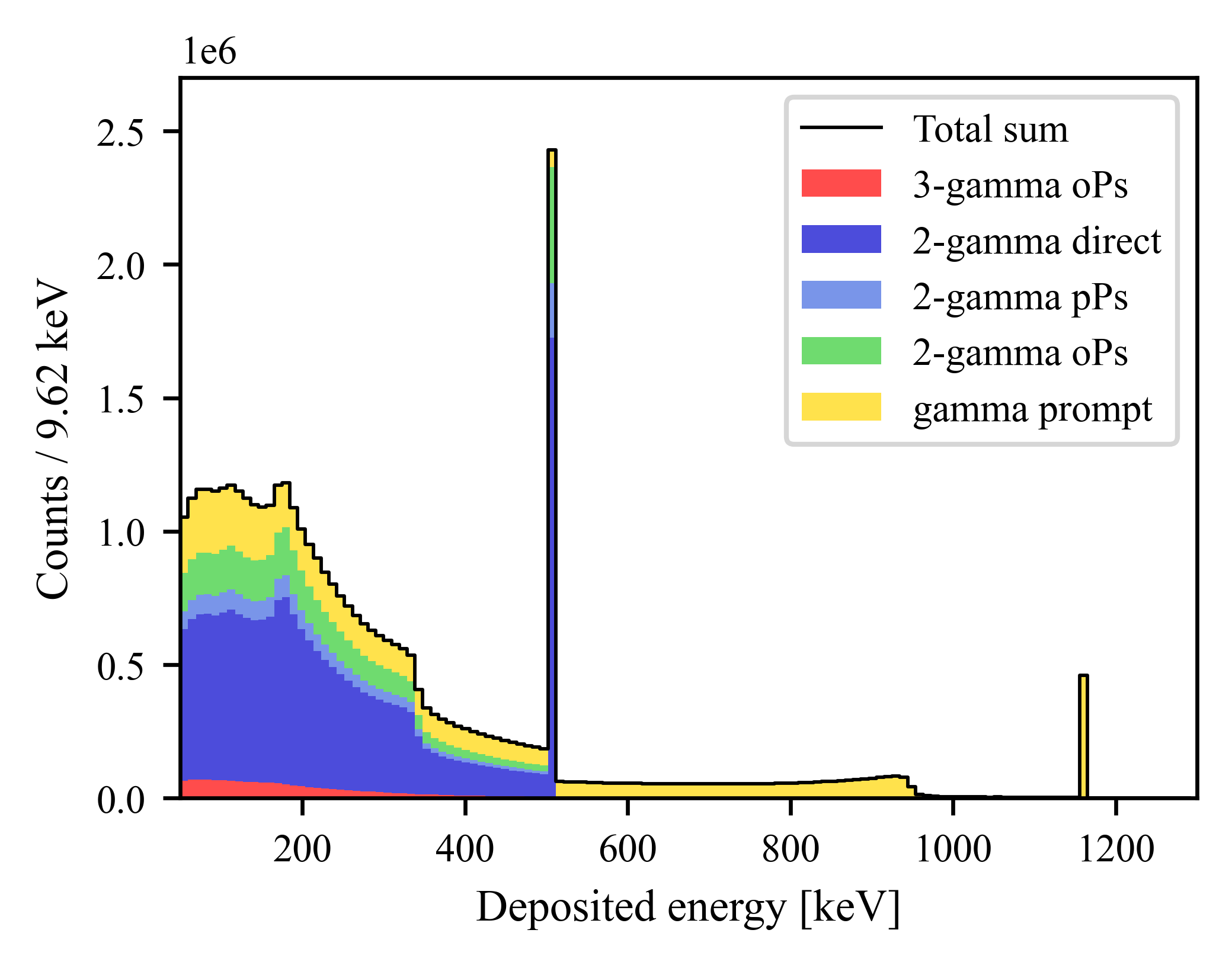}
        \includegraphics[width=0.49\textwidth]{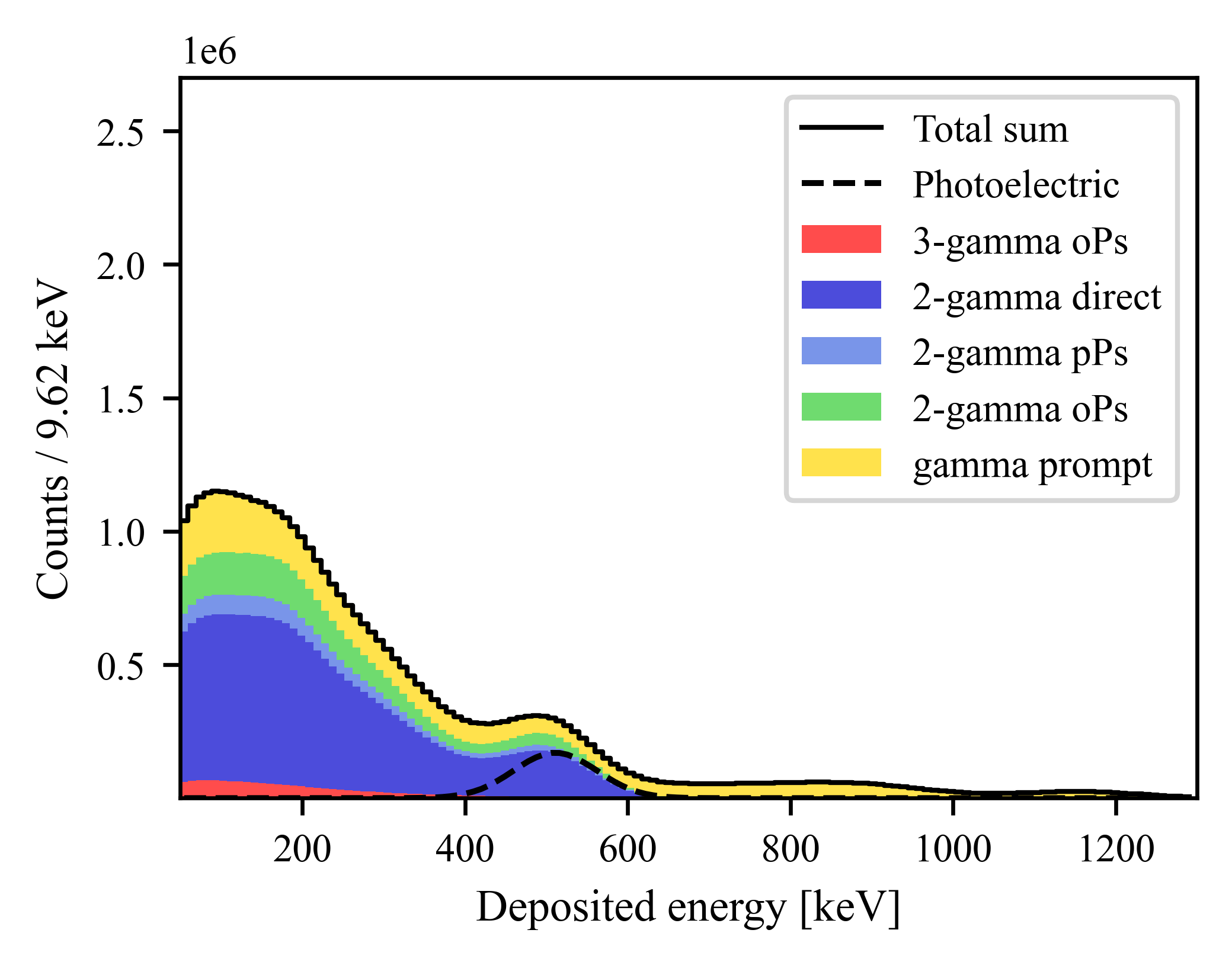}
    }
    \vspace{0.2em}

    \subfloat[Lifetime: 100 ns]{%
        \includegraphics[width=0.49\textwidth]{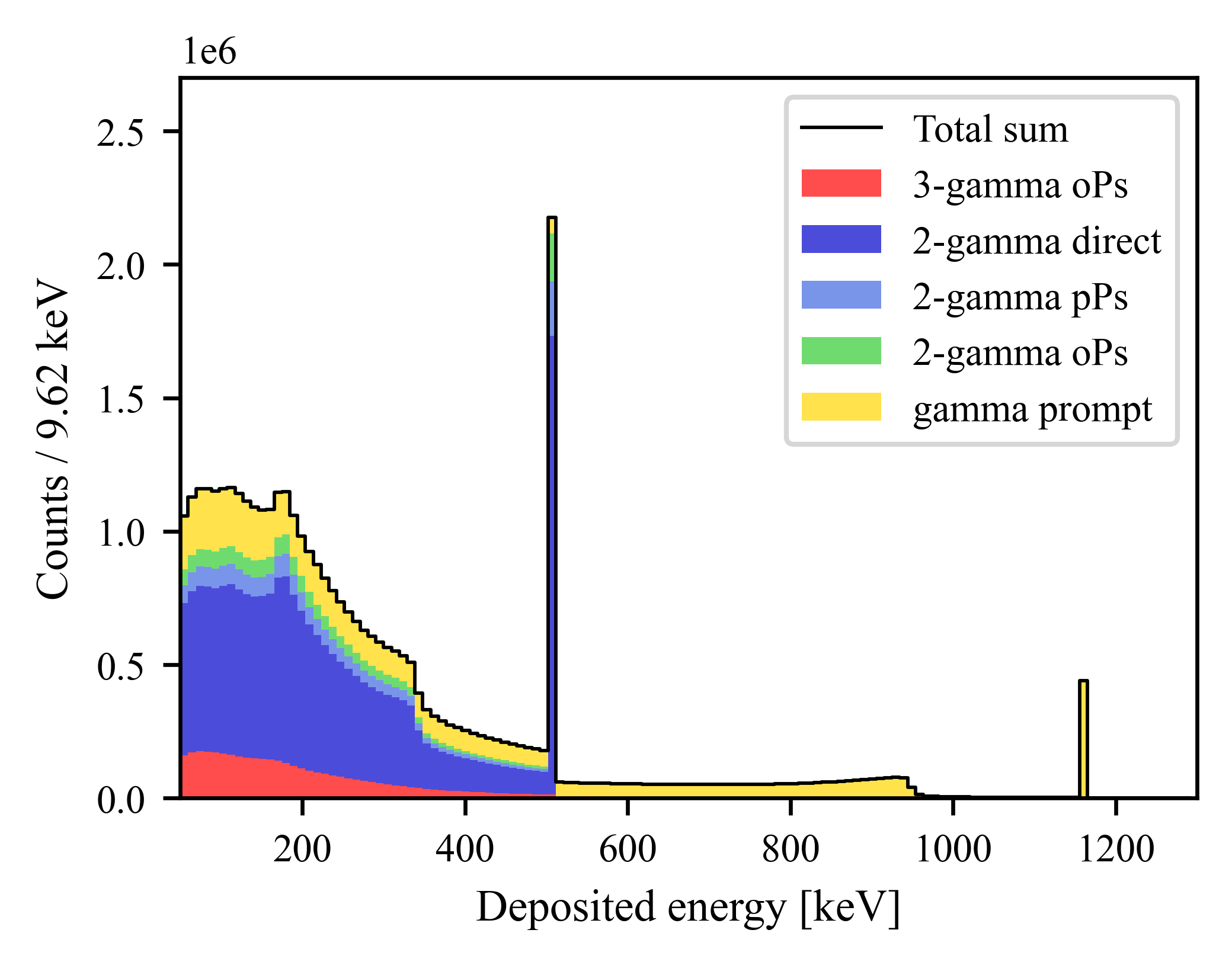}
        \includegraphics[width=0.49\textwidth]{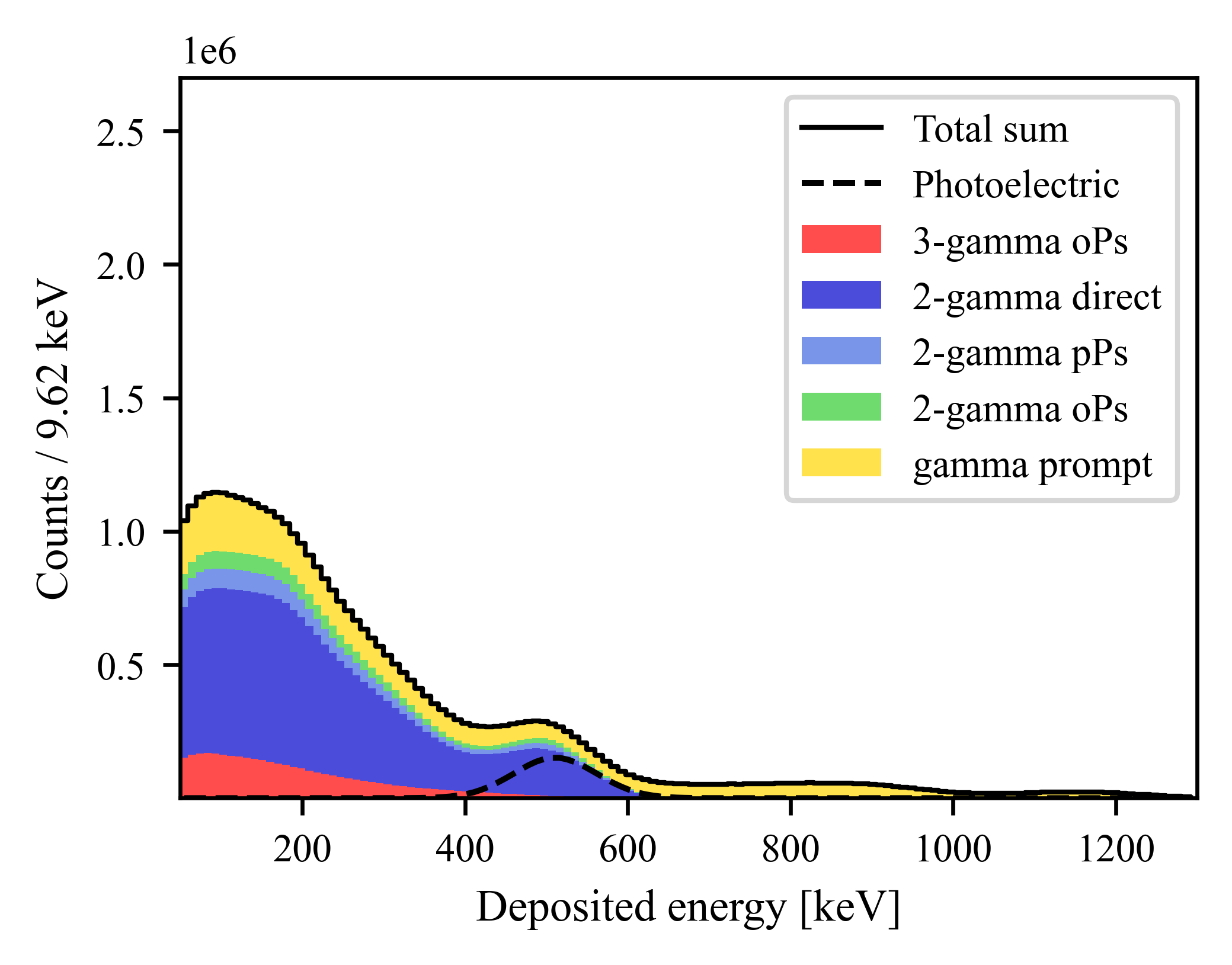}
    }

    \caption{Stacked histograms of energy deposition for the three \threetotwo simulations defined in \cref{tab:3to2_sim}, corresponding to \ac{o-Ps} lifetimes of (a) 2~ns, (b) 40~ns, and (c) 100~ns. In each panel, the left histogram shows the distribution without application of detector energy smearing; the right distribution is obtained after taking into account the detector energy smearing.}
    \label{fig:3_to_2_energy}
\end{figure}

% Energy spectra for different sources
\begin{figure}
 \centering
    \subfloat[\label{fig:deposited_energy_44Sc} Deposited energy of \nucSc source.]{
        \includegraphics[width=0.6\textwidth]{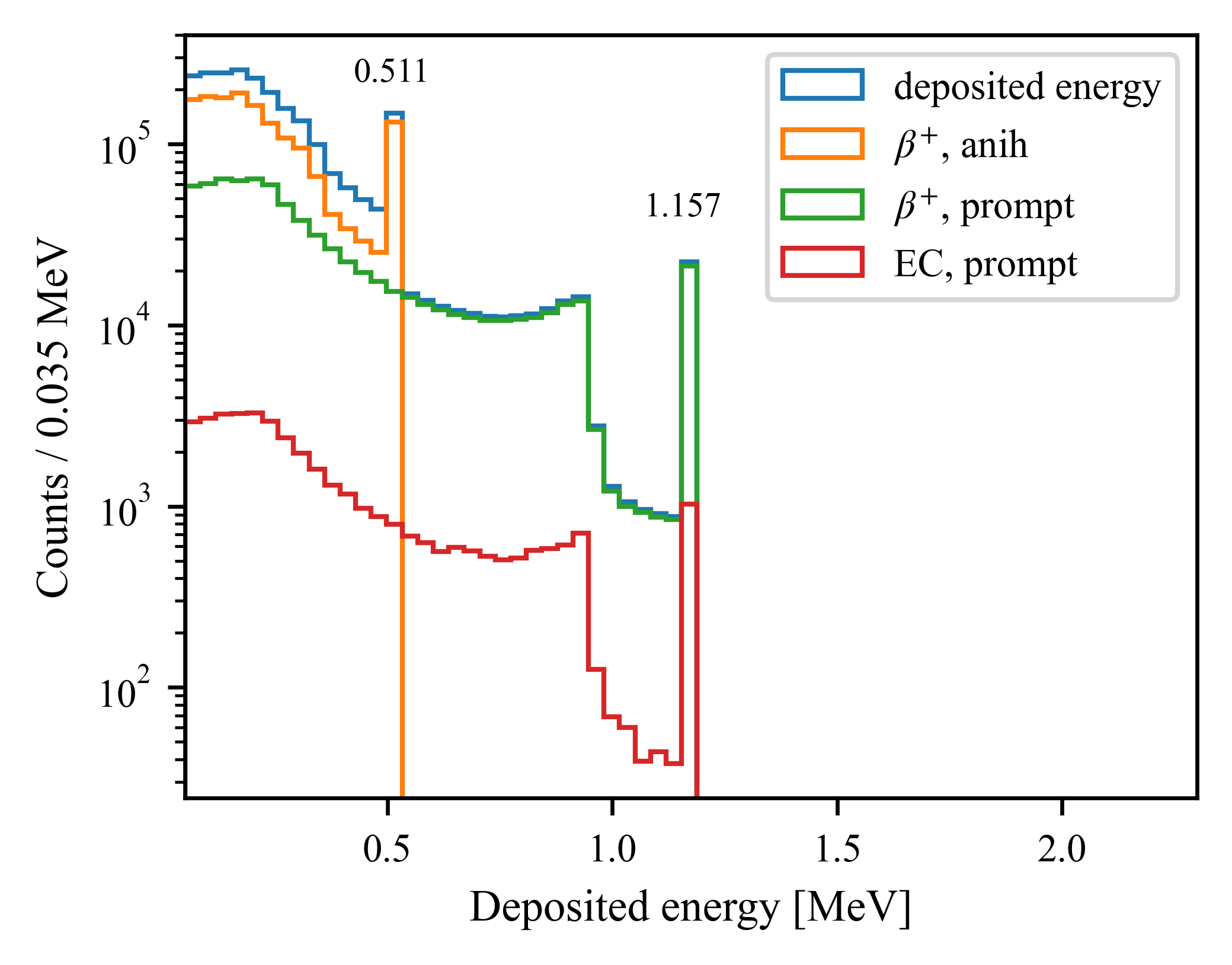}
    }

    \vspace{0.2em}

    \subfloat[\label{fig:deposited_energy_68Ga} Deposited energy of \nucGa source.]{
        \includegraphics[width=0.6\textwidth]{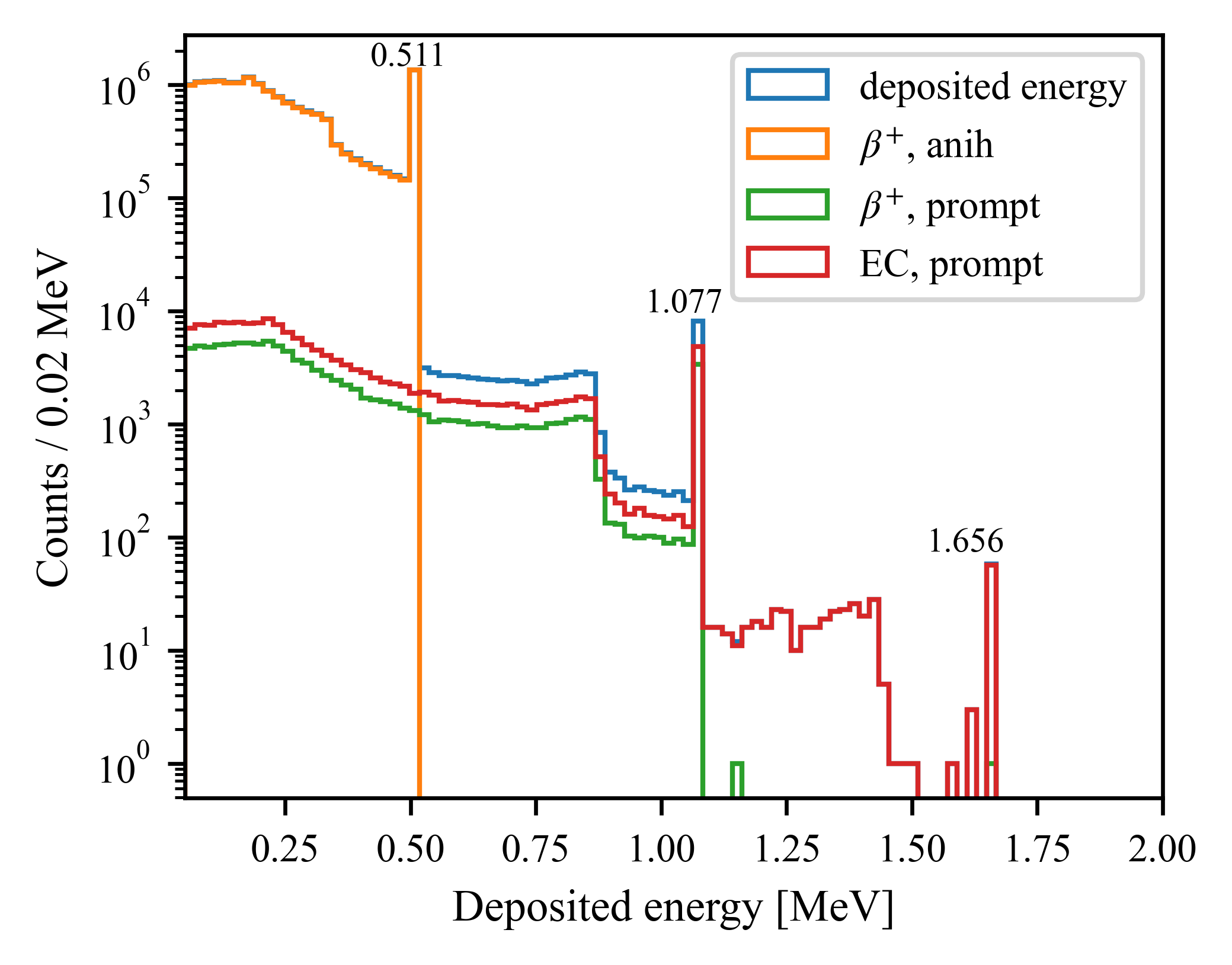}
    }

    \vspace{0.2em}

    \subfloat[\label{fig:deposited_energy_124I} Deposited energy of \nucI source.]{
        \includegraphics[width=0.6\textwidth]{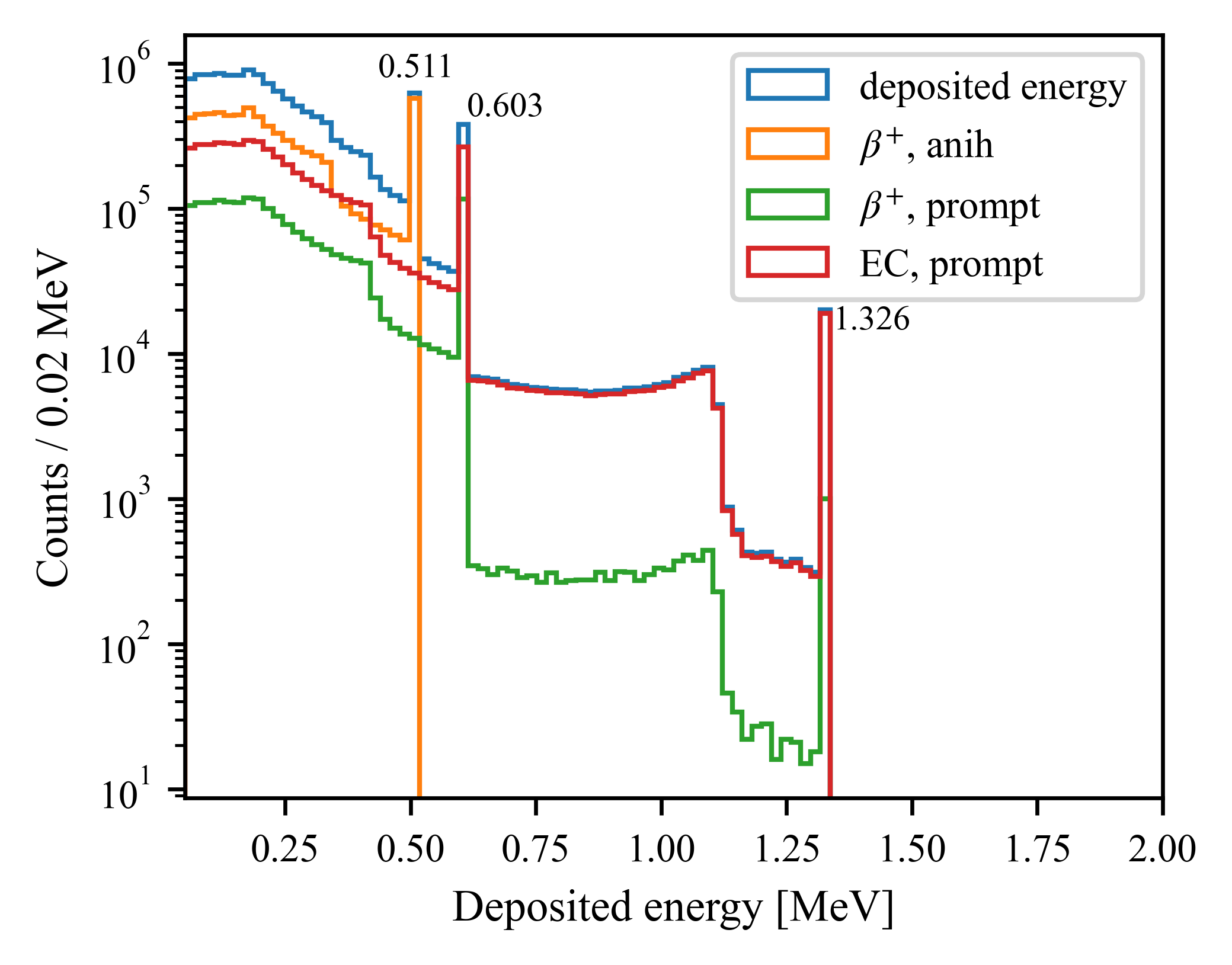}
    }

    \caption{Deposited energy distributions for different radioactive sources.}
    \label{fig:deposited_energy_sources}
\end{figure}

\newpage
\section{Example parameter sets and detailed intensity calculations}
\label{app:xad4}
In this appendix, we provide the material parameter sets used in this work, in a form that can be translated directly into a source definition. We first analyse in detail the modelling of XAD4 material (see \cref{fig:xad4}), a standard reference material in PALS,  putting emphasis on the calculation of intensities for a published component decomposition and providing an example macro setting. We then give the component structure
used for the tissue-mimicking simulations. All values are taken from published measurements, and they are intended as representative starting points.

The mean lifetimes $\tau_i$ and the total component intensities $I_i$ are the measured quantities obtained from the decomposition of the lifetime spectrum \citep{jasinskaDetermination3gammaFraction2016}. From these, the $3\gamma$
intensities are derived using \cref{eq:intensity_o_ps} for the \ac{o-Ps} components and
\cref{eq:intensity_direct} for direct annihilation, with
$I_{2\gamma} = I_i - I_{3\gamma}$. 
The \ac{p-Ps} component decays exclusively into two photons. 
The resulting configuration is summarised in \cref{tab:xad4}, where the five measured components yield nine decay channels.

\begin{table}[htbp]
\centering
\caption{Decay-channel configuration for XAD4.}
\label{tab:xad4}
    \begin{tabular}{lccccc}
        \toprule
        \textbf{Component} & $\boldsymbol{\tau_i}$ \textbf{[ns]} & $\boldsymbol{I_i}$ \textbf{[\%]}
        & \textbf{Relation} & $\boldsymbol{I_{3\gamma}}$ \textbf{[\%]} & $\boldsymbol{I_{2\gamma}}$ \textbf{[\%]} \\
        \midrule
        p-Ps      & 0.125 & 16.97  & ---                        & ---   & 16.97 \\
        direct    & 0.45  & 32.13  & \cref{eq:intensity_direct} & 0.09  & 32.04 \\
        o-Ps$_1$  & 2.5   & 3.30   & \cref{eq:intensity_o_ps}   & 0.06  & 3.24  \\
        o-Ps$_2$  & 10    & 2.80   & \cref{eq:intensity_o_ps}   & 0.20  & 2.60  \\
        o-Ps$_3$  & 90    & 44.80  & \cref{eq:intensity_o_ps}   & 28.39 & 16.41 \\
        \midrule
        \textbf{Total} & & \textbf{100.00} & & \textbf{28.74} & \textbf{71.26} \\
        \bottomrule
    \end{tabular}
\end{table}

Explicitly, for the longest-lived \ac{o-Ps} component
$I_{3\gamma} = 44.80 \times (90/142) = 28.39\,\%$ and
$I_{2\gamma} = 44.80 - 28.39 = 16.41\,\%$; for the shortest-lived one
$I_{3\gamma} = 3.30 \times (2.5/142) = 0.06\,\%$; and for direct annihilation
$I_{3\gamma} = 32.13/372 = 0.09\,\%$. The nine resulting channels are passed to the
source as the lists shown in \cref{lst:xad4} (in \ac{GATE}~9) and \cref{lst:xad4_gate10} (in \ac{GATE}~10), in the order given by \cref{tab:xad4}.
A prompt photon of 1.274~MeV is assigned to every channel, consistent with the \nucNa
decay scheme.

\begin{lstlisting}[caption={Macro definition of the XAD4 positronium source
corresponding to \cref{tab:xad4}. Fractions are given in percent and are
normalised internally.},label={lst:xad4}]
/gate/source/addSource xad4 PositroniumSource
/gate/source/xad4/setType Ps
/gate/source/xad4/setPositroniumFractions 16.97 32.04 0.09 3.24 0.06 2.60 0.20 16.41 28.39
/gate/source/xad4/setPositroniumLifetimes 0.125 0.45 0.45 2.5 2.5 10. 10. 90. 90. ns
/gate/source/xad4/setDecayKinds k2Gamma k2Gamma k3Gamma k2Gamma k3Gamma k2Gamma k3Gamma k2Gamma k3Gamma
/gate/source/xad4/setPromptPhotonProbabilites 1. 1. 1. 1. 1. 1. 1. 1. 1.
/gate/source/xad4/setPromptPhotonEnergies 1.274 1.274 1.274 1.274 1.274 1.274 1.274 1.274 1.274 MeV
\end{lstlisting}

\begin{lstlisting}[caption={\Ac{GATE}~10 definition of the positronium source defined in \cref{lst:xad4}.}, label={lst:xad4_gate10}]
source = sim.add_source("PositroniumSource", "source")
source.channels_from_fractions.fractions = [16.97, 32.04, 0.09, 3.24, 0.06, 2.60, 0.20, 16.41, 28.39]
source.channels_from_fractions.decay_kinds = ["k2Gamma", "k2Gamma","k3Gamma", "k2Gamma", "k3Gamma", "k2Gamma", "k3Gamma", "k2Gamma", "k3Gamma"]
source.positronium_lifetimes = [0.125 * ns, 0.45 * ns, 0.45 * ns, 2.5 * ns, 2.5 * ns, 10. * ns, 10. * ns, 90. * ns, 90. * ns]
source.prompt_gamma_probabilities = [1., 1., 1., 1., 1., 1., 1., 1., 1.]
source.prompt_gamma_energies = [1.274 * MeV, 1.274 * MeV, 1.274 * MeV, 1.274 * MeV, 1.274 * MeV, 1.274 * MeV, 1.274 * MeV, 1.274 * MeV, 1.274 * MeV]
\end{lstlisting}

\begin{table}[htbp]
\centering
\caption{Positronium lifetime component structures used for the tissue-mimicking
phantom simulations.}
\label{tab:tissues}
    \begin{tabular}{llcccccc}
        \toprule
        \textbf{Material} & & \textbf{p-Ps} & \textbf{direct} & \textbf{o-Ps$_1$}
        & \textbf{o-Ps$_2$} & \textbf{o-Ps$_3$} & \textbf{o-Ps$_4$} \\
        \midrule
        bone            & $\tau_i$ [ns] & 0.125 & 0.400 & 2.10 & 6.30 & 21.0 & 58.0 \\
                        & $I_i$ [\%]    & 13.0  & 76.0  & 4.8  & 2.7  & 2.2  & 1.3  \\
        \addlinespace
        fat/adipose     & $\tau_i$ [ns] & 0.152 & 0.420 & 2.54 & ---  & ---  & ---  \\
                        & $I_i$ [\%]    & 12.0  & 69.5  & 18.5 & ---  & ---  & ---  \\
        \addlinespace
        muscle          & $\tau_i$ [ns] & 0.150 & 0.440 & 2.04 & ---  & ---  & ---  \\
                        & $I_i$ [\%]    & 9.2   & 76.3  & 14.5 & ---  & ---  & ---  \\
        \addlinespace
        water           & $\tau_i$ [ns] & 0.140 & 0.400 & 1.80 & ---  & ---  & ---  \\
                        & $I_i$ [\%]    & 14.0  & 64.0  & 22.0 & ---  & ---  & ---  \\
        \bottomrule
    \end{tabular}
\end{table}

The same procedure was applied to define the tissue-mimicking sources of
\cref{subsec:NEMA_2}. The component structures assigned to the four predefined materials are collected in \cref{tab:tissues}: bone \citep{ahnHierarchicalNatureNanoscale2021}, muscle and fat/adipose tissue \citep{avachatOrthopositroniumLifetimeSofttissue2024a}, and water \citep{koteraOrthopositroniumLifetimeWater2005a}. 
Within the phantom, bone was assigned to the central cylinder, fat/adipose tissue to spheres
10 and 22, muscle tissue to spheres 13 and 28, and water to the background and the
cold spheres, as shown in \cref{fig:nemaIEC}. 
Only $\twogamma$ channels were declared for these simulations. The corresponding $3\gamma$ contributions follow from \cref{eq:intensity_o_ps,eq:intensity_direct} exactly as in
\cref{tab:xad4} and amount to at most about 1.3\,\% (bone) and below 0.6\,\% for the other materials. Each channel additionally carries a prompt photon of 1.274~MeV.

\end{document}

%% file: config.tex
\usepackage[caption=false]{subfig}

\usepackage[T1]{fontenc}
\usepackage[utf8]{inputenc}
\usepackage[english]{babel}
\usepackage{lmodern} % WK I added it

\usepackage{graphicx}
\usepackage{enumitem}
\usepackage{textcomp}
\usepackage{gensymb}
\usepackage{xparse}
\usepackage{siunitx}
\usepackage{isotope}
\usepackage{rotating}
\usepackage{float}
\usepackage{booktabs}

\usepackage{tikz}
\usetikzlibrary{shapes.geometric, arrows.meta}

\usepackage{listings}
\usepackage[sort]{natbib}
\setcitestyle{authoryear, round}

\expandafter\let\csname equation*\endcsname=\relax
\expandafter\let\csname endequation*\endcsname=\relax
\usepackage{amsmath}

\usepackage{amsfonts,amssymb,amscd}
\usepackage{xcolor, soul}
\usepackage{xspace}
\usepackage{microtype}

\usepackage{hyperref}
\usepackage{cleveref} % cleveref must be loaded after hyperref

\sethlcolor{yellow}

\hypersetup{colorlinks, linkcolor={red!50!black}, citecolor={blue!50!black}, urlcolor={blue!80!black}}
\microtypesetup{
	protrusion=alltext-nott,
	expansion=alltext-nott,
	final
}
\graphicspath{{figures/}}

\newcommand{\twogamma}{\ensuremath{2\gamma}\xspace}
\newcommand{\threegamma}{\ensuremath{3\gamma}\xspace}
\newcommand{\threetotwo}{\ensuremath{3\gamma\text{-to-}2\gamma}\xspace}

\newcommand{\nucSc}{\isotope[44]{Sc}\xspace}
\newcommand{\nucGa}{\isotope[68]{Ga}\xspace}
\newcommand{\nucNa}{\isotope[22]{Na}\xspace}
\newcommand{\nucI}{\isotope[124]{I}\xspace}

\usepackage{acro}
\acsetup{single}
\DeclareAcronym{PET}{short=PET,long=Positron Emission Tomography}
\DeclareAcronym{Ps}{short=Ps,long=positronium}
\DeclareAcronym{o-Ps}{short=o-Ps,long=ortho-positronium}
\DeclareAcronym{p-Ps}{short=p-Ps,long=para-positronium}
\DeclareAcronym{PLI}{short=PLI,long=Positronium Lifetime Imaging}
\DeclareAcronym{TOF}{short=TOF,long=time-of-flight}
\DeclareAcronym{GATE}{short=GATE,long=Geant4 Application for Tomographic Emission}
\DeclareAcronym{MC}{short=MC,long=Monte Carlo}

%% file: references.bib
@article{ahnHierarchicalNatureNanoscale2021,
  title = {Hierarchical Nature of Nanoscale Porosity in Bone Revealed by Positron Annihilation Lifetime Spectroscopy},
  author = {Ahn, Taeyong and Gidley, David W. and Thornton, Aaron W. and {Wong-Foy}, Antek G. and Orr, Bradford G. and Kozloff, Kenneth M. and Banaszak Holl, Mark M.},
  year = 2021,
  month = mar,
  journal = {ACS Nano},
  volume = {15},
  number = {3},
  pages = {4321--4334},
  publisher = {American Chemical Society},
  issn = {1936-0851},
  doi = {10.1021/acsnano.0c07478},
  urldate = {2026-01-21}
}

@article{allisonRecentDevelopmentsGeant42016,
  title = {Recent Developments in Geant4},
  author = {Allison, J. and Amako, K. and Apostolakis, J. and Arce, P. and Asai, M. and Aso, T. and Bagli, E. and Bagulya, A. and Banerjee, S. and Barrand, G. and Beck, B. R. and Bogdanov, A. G. and Brandt, D. and Brown, J. M. C. and Burkhardt, H. and Canal, {\relax Ph}. and {Cano-Ott}, D. and Chauvie, S. and Cho, K. and Cirrone, G. A. P. and Cooperman, G. and {Cort{\'e}s-Giraldo}, M. A. and Cosmo, G. and Cuttone, G. and Depaola, G. and Desorgher, L. and Dong, X. and Dotti, A. and Elvira, V. D. and Folger, G. and Francis, Z. and Galoyan, A. and Garnier, L. and Gayer, M. and Genser, K. L. and Grichine, V. M. and Guatelli, S. and Gu{\`e}ye, P. and Gumplinger, P. and Howard, A. S. and H{\v r}ivn{\'a}{\v c}ov{\'a}, I. and Hwang, S. and Incerti, S. and Ivanchenko, A. and Ivanchenko, V. N. and Jones, F. W. and Jun, S. Y. and Kaitaniemi, P. and Karakatsanis, N. and Karamitros, M. and Kelsey, M. and Kimura, A. and Koi, T. and Kurashige, H. and Lechner, A. and Lee, S. B. and Longo, F. and Maire, M. and Mancusi, D. and Mantero, A. and Mendoza, E. and Morgan, B. and Murakami, K. and Nikitina, T. and Pandola, L. and Paprocki, P. and Perl, J. and Petrovi{\'c}, I. and Pia, M. G. and Pokorski, W. and Quesada, J. M. and Raine, M. and Reis, M. A. and Ribon, A. and Risti{\'c} Fira, A. and Romano, F. and Russo, G. and Santin, G. and Sasaki, T. and Sawkey, D. and Shin, J. I. and Strakovsky, I. I. and Taborda, A. and Tanaka, S. and Tom{\'e}, B. and Toshito, T. and Tran, H. N. and Truscott, P. R. and Urban, L. and Uzhinsky, V. and Verbeke, J. M. and Verderi, M. and Wendt, B. L. and Wenzel, H. and Wright, D. H. and Wright, D. M. and Yamashita, T. and Yarba, J. and Yoshida, H.},
  year = 2016,
  month = nov,
  journal = {Nuclear Instruments and Methods in Physics Research Section A: Accelerators, Spectrometers, Detectors and Associated Equipment},
  volume = {835},
  pages = {186--225},
  issn = {0168-9002},
  doi = {10.1016/j.nima.2016.06.125},
  urldate = {2026-03-24}
}

@article{avachatOrthopositroniumLifetimeSofttissue2024a,
  title = {Ortho-Positronium Lifetime for Soft-Tissue Classification},
  author = {Avachat, Ashish V. and Mahmoud, Kholod H. and Leja, Anthony G. and Xu, Jiajie J. and Anastasio, Mark A. and Sivaguru, Mayandi and Di Fulvio, Angela},
  year = 2024,
  month = sep,
  journal = {Sci Rep},
  volume = {14},
  number = {1},
  pages = {21155},
  publisher = {Nature Publishing Group},
  issn = {2045-2322},
  doi = {10.1038/s41598-024-71695-7},
  urldate = {2026-01-21},
  copyright = {2024 The Author(s)},
  langid = {english}
}

@misc{fujimotoAdvancingPETDirect2025,
  title = {Advancing PET through Direct Imaging of Three-Photon Decay Using Pure Positron Emitters},
  author = {Fujimoto, Macoto and Shimazoe, Kenji and Sato, Riku and Hamdan, Moh and Uenomachi, Mizuki and Stephenson, Laura and Montgomery, Archie and Bordes, Julien and Watts, Daniel and Caradonna, Peter and Brown, Jamie},
  year = 2025,
  month = dec,
  publisher = {In Review},
  doi = {10.21203/rs.3.rs-7272322/v1},
  urldate = {2025-12-29},
  archiveprefix = {In Review},
  copyright = {https://creativecommons.org/licenses/by/4.0/}
}

@article{harpenPositroniumReviewSymmetry2003,
  title = {Positronium: Review of Symmetry, Conserved Quantities and Decay for the Radiological Physicist},
  shorttitle = {Positronium},
  author = {Harpen, Michael D.},
  year = 2003,
  month = dec,
  journal = {Med. Phys.},
  volume = {31},
  number = {1},
  pages = {57--61},
  issn = {00942405},
  doi = {10.1118/1.1630494},
  urldate = {2026-01-10},
  copyright = {http://doi.wiley.com/10.1002/tdm\_license\_1.1},
  langid = {english}
}

@article{hourlierExperimentalUsesPositronium2024,
  title = {Experimental Uses of Positronium and Potential for Biological Applications},
  author = {Hourlier, A. and Boisson, F. and Brasse, D.},
  year = 2024,
  month = jul,
  journal = {IEEE Transactions on Radiation and Plasma Medical Sciences},
  volume = {8},
  number = {6},
  pages = {581--594},
  issn = {2469-7303},
  doi = {10.1109/TRPMS.2024.3407981},
  urldate = {2025-07-23}
}

@article{huangFastHighresolutionLifetime2025,
  title = {Fast High-Resolution Lifetime Image Reconstruction for Positron Lifetime Tomography},
  author = {Huang, Bangyan and Wang, Zipai and Zeng, Xinjie and Goldan, Amir H. and Qi, Jinyi},
  year = 2025,
  journal = {Commun Phys},
  volume = {8},
  number = {1},
  pages = {181},
  issn = {2399-3650},
  doi = {10.1038/s42005-025-02100-6},
  langid = {english},
  pmcid = {PMC12031669},
  pmid = {40291544}
}

@article{huangHighresolutionPositroniumLifetime2024,
  title = {High-Resolution Positronium Lifetime Tomography by the Method of Moments},
  author = {Huang, Bangyan and Qi, Jinyi},
  year = 2024,
  month = dec,
  journal = {Phys Med Biol},
  volume = {69},
  number = {24},
  issn = {1361-6560},
  doi = {10.1088/1361-6560/ad9543},
  langid = {english},
  pmcid = {PMC11785412},
  pmid = {39569940}
}

@article{huangSPLITStatisticalPositronium2024,
  title = {SPLIT: Statistical Positronium Lifetime Image Reconstruction via Time-Thresholding},
  shorttitle = {SPLIT},
  author = {Huang, Bangyan and Li, Tiantian and {Arino-Estrada}, Gerard and Dulski, Kamil and Shopa, Roman Y. and Moskal, Pawel and Stepien, Ewa and Qi, Jinyi},
  year = 2024,
  month = jun,
  journal = {IEEE Trans Med Imaging},
  volume = {43},
  number = {6},
  pages = {2148--2158},
  issn = {1558-254X},
  doi = {10.1109/TMI.2024.3357659},
  langid = {english},
  pmcid = {PMC11409919},
  pmid = {38261489}
}

@article{jasinskaDetermination3gammaFraction2016,
  title = {Determination of the \$3\textbackslash gamma \$ Fraction from Positron Annihilation in Mesoporous Materials for Symmetry Violation Experiment with J-PET Scanner},
  author = {Jasi{\'n}ska, B. and Gorgol, M. and Wiertel, M. and Zaleski, R. and Alfs, D. and Bednarski, T. and Bia{\l}as, P. and Czerwi{\'n}ski, E. and Dulski, K. and Gajos, A. and G{\l}owacz, B. and Kami{\'n}ska, D. and Kap{\l}on, {\L}. and Korcyl, G. and Kowalski, P. and Kozik, T. and Krzemie{\'n}, W. and Kubicz, E. and Mohammed, M. and Nied{\'z}wiecki, {\relax Sz}. and Pa{\l}ka, M. and Raczy{\'n}ski, L. and Rudy, Z. and Rundel, O. and Sharma, N.G. and Silarski, M. and S{\l}omski, A. and Strzelecki, A. and Wieczorek, A. and Wi{\'s}licki, W. and Zgardzi{\'n}ska, B. and Zieli{\'n}ski, M. and Moskal, P.},
  year = 2016,
  journal = {Acta Phys. Pol. B},
  volume = {47},
  number = {2},
  pages = {453},
  issn = {0587-4254, 1509-5770},
  doi = {10.5506/APhysPolB.47.453},
  urldate = {2025-10-22},
  langid = {english}
}

@article{kacperskiThreegammaAnnihilationImaging2004,
  title = {Three-Gamma Annihilation Imaging in Positron Emission Tomography},
  author = {Kacperski, Krzysztof and Spyrou, Nicholas M. and Smith, F. Alan},
  year = 2004,
  month = apr,
  journal = {IEEE Trans Med Imaging},
  volume = {23},
  number = {4},
  pages = {525--529},
  issn = {0278-0062},
  doi = {10.1109/TMI.2004.824150},
  langid = {english},
  pmid = {15084078}
}

@article{krahGATE10Monte2026,
  title = {GATE 10 Monte Carlo Particle Transport Simulation: II. Architecture and Innovations},
  shorttitle = {GATE 10 Monte Carlo Particle Transport Simulation},
  author = {Krah, Nils and Arbor, Nicolas and Baudier, Thomas and Bert, Julien and Chatzipapas, Konstantinos and Favaretto, Martina and Fuchs, Hermann and Grevillot, Lo{\"i}c and Harb, Hussein and Van Hoey, Gert and Jacquet, Maxime and Jan, S{\'e}bastien and Jia, Yihan and Kagadis, George C and Kang, Han Gyu and Klever, Paul and Kochebina, Olga and Krzemien, Wojciech and Maigne, Lydia and Mohr, Philipp and Mummaneni, Guneet and Paneta, Valentina and Papadimitroulas, Panagiotis and Pereda, Alexis and Rannou, Axel and Resch, Andreas F and Roncali, Emilie and Toussaint, Maxime and Trigila, Carlotta and Tsoumpas, Charalampos and Zhang, Jing and Ziemons, Karl and Sarrut, David},
  year = 2026,
  month = jan,
  journal = {Phys. Med. Biol.},
  volume = {71},
  number = {1},
  pages = {015043},
  issn = {0031-9155, 1361-6560},
  doi = {10.1088/1361-6560/ae237c},
  urldate = {2026-01-15}
}

@article{mercolliPhantomImagingDemonstration2025,
  title = {Phantom Imaging Demonstration of Positronium Lifetime with a Long Axial Field-of-View PET/CT and 124I},
  author = {Mercolli, Lorenzo and Steinberger, William M. and Rathod, Narendra and Conti, Maurizio and Moskal, Pawe{\l} and Rominger, Axel and Seifert, Robert and Shi, Kuangyu and St{\k e}pie{\'n}, Ewa {\L} and Sari, Hasan},
  year = 2025,
  journal = {EJNMMI Phys},
  volume = {12},
  number = {1},
  pages = {80},
  issn = {2197-7364},
  doi = {10.1186/s40658-025-00790-z},
  langid = {english},
  pmcid = {PMC12379202},
  pmid = {40855031}
}

@misc{mercolliVivoPositroniumLifetime2024,
  title = {In Vivo Positronium Lifetime Measurements with Intravenous Tracer Administration and a Long Axial Field-of-View PET/CT},
  author = {Mercolli, Lorenzo and Steinberger, William M. and Sari, Hasan and {Afshar-Oromieh}, Ali and Caobelli, Federico and Conti, Maurizio and Felgosa Cardoso, {\^A}ngelo R. and Mingels, Clemens and Moskal, Pawe{\l} and Pyka, Thomas and Rathod, Narendra and Schepers, Robin and St{\k e}pie{\'n}, Ewa {\L}. and Viscione, Marco and Rominger, Axel and Shi, Kuangyu and Seifert, Robert},
  year = 2024,
  month = oct,
  publisher = {{Radiology and Imaging}},
  doi = {10.1101/2024.10.19.24315509},
  urldate = {2026-01-21},
  archiveprefix = {Radiology and Imaging},
  copyright = {http://creativecommons.org/licenses/by/4.0/},
  langid = {english}
}

@article{mercolliVivoVoxelwisePositronium2026,
  title = {In Vivo Voxel-Wise Positronium Lifetime Imaging of Thyroid Cancer Using Clinically Routine I-124 PET/CT},
  author = {Mercolli, Lorenzo and Steinberger, William M. and L{\"a}ppchen, Tilman and Amon, Michelle and Bregenzer, Carola and Conti, Maurizio and Cardoso, {\^A}ngelo R. Felgosa and Mingels, Clemens and Moskal, Pawe{\l} and Rathod, Narendra and Sari, Hasan and St{\k e}pie{\'n}, Ewa {\L}. and Weidner, Sabine and Rominger, Axel and Shi, Kuangyu and Seifert, Robert},
  year = 2026,
  month = mar,
  journal = {EANM Innovation},
  volume = {2},
  pages = {100017},
  issn = {3051-2913},
  doi = {10.1016/j.eanmi.2025.100017},
  urldate = {2026-01-09}
}

@article{moskalPositroniumImageHuman2024a,
  title = {Positronium Image of the Human Brain in Vivo},
  author = {Moskal, Pawe{\l} and Baran, Jakub and Bass, Steven and Choi{\'n}ski, Jaros{\l}aw and Chug, Neha and Curceanu, Catalina and Czerwi{\'n}ski, Eryk and Dadgar, Meysam and Das, Manish and Dulski, Kamil and Eliyan, Kavya V. and Fronczewska, Katarzyna and Gajos, Aleksander and Kacprzak, Krzysztof and Kajetanowicz, Marcin and Kaplanoglu, Tevfik and Kap{\l}on, {\L}ukasz and Klimaszewski, Konrad and Kobylecka, Ma{\l}gorzata and Korcyl, Grzegorz and Kozik, Tomasz and Krzemie{\'n}, Wojciech and Kubat, Karol and Kumar, Deepak and Kunikowska, Jolanta and M{\k a}czewska, Joanna and Migda{\l}, Wojciech and Moskal, Gabriel and Mryka, Wiktor and Nied{\'z}wiecki, Szymon and Parzych, Szymon and {del Rio}, Elena P. and Raczy{\'n}ski, Lech and Sharma, Sushil and Shivani, Shivani and Shopa, Roman Y. and Silarski, Micha{\l} and Skurzok, Magdalena and Tayefi, Faranak and Ardebili, Keyvan T. and Tanty, Pooja and Wi{\'s}licki, Wojciech and Kr{\'o}licki, Leszek and St{\k e}pie{\'n}, Ewa {\L}.},
  year = 2024,
  month = sep,
  journal = {Science Advances},
  volume = {10},
  number = {37},
  pages = {eadp2840},
  publisher = {American Association for the Advancement of Science},
  doi = {10.1126/sciadv.adp2840},
  urldate = {2025-10-22}
}

@article{oreThreePhotonAnnihilationElectronPositron1949,
  title = {Three-Photon Annihilation of an Electron-Positron Pair},
  author = {Ore, A. and Powell, J. L.},
  year = 1949,
  month = jun,
  journal = {Phys. Rev.},
  volume = {75},
  number = {11},
  pages = {1696--1699},
  issn = {0031-899X},
  doi = {10.1103/PhysRev.75.1696},
  urldate = {2026-01-09},
  copyright = {http://link.aps.org/licenses/aps-default-license},
  langid = {english}
}

@article{pevovarRatioPositronAnnihilation2007,
  title = {Ratio of Positron Annihilation into Three Photons versus Two},
  author = {Pevovar, S. C. and Weber, M. H. and Lynn, K. G.},
  year = 2007,
  journal = {physica status solidi c},
  volume = {4},
  number = {10},
  pages = {3447--3450},
  issn = {1610-1642},
  doi = {10.1002/pssc.200675786},
  urldate = {2025-10-22},
  copyright = {Copyright \copyright{} 2007 WILEY-VCH Verlag GmbH \& Co. KGaA, Weinheim},
  langid = {english}
}

@article{qiPositroniumLifetimeImage2022,
  title = {Positronium Lifetime Image Reconstruction for TOF PET},
  author = {Qi, Jinyi and Huang, Bangyan},
  year = 2022,
  month = oct,
  journal = {IEEE Trans Med Imaging},
  volume = {41},
  number = {10},
  pages = {2848--2855},
  issn = {0278-0062},
  doi = {10.1109/TMI.2022.3174561},
  urldate = {2025-10-16},
  pmcid = {PMC9829407},
  pmid = {35584079}
}

@article{sarrutAdvancedMonteCarlo2021a,
  title = {Advanced Monte Carlo Simulations of Emission Tomography Imaging Systems with GATE},
  author = {Sarrut, David and Ba{\l}a, Mateusz and Bardi{\`e}s, Manuel and Bert, Julien and Chauvin, Maxime and Chatzipapas, Konstantinos and Dupont, Mathieu and Etxebeste, Ane and M Fanchon, Louise and Jan, S{\'e}bastien and Kayal, Gunjan and S Kirov, Assen and Kowalski, Pawe{\l} and Krzemien, Wojciech and Labour, Joey and Lenz, Mirjam and Loudos, George and Mehadji, Brahim and M{\'e}nard, Laurent and Morel, Christian and Papadimitroulas, Panagiotis and Rafecas, Magdalena and Salvadori, Julien and Seiter, Daniel and Stockhoff, Mariele and Testa, Etienne and Trigila, Carlotta and Pietrzyk, Uwe and Vandenberghe, Stefaan and Verdier, Marc-Antoine and Visvikis, Dimitris and Ziemons, Karl and Zvolsk{\'y}, Milan and Roncali, Emilie},
  year = 2021,
  month = may,
  journal = {Phys. Med. Biol.},
  volume = {66},
  number = {10},
  pages = {10TR03},
  issn = {0031-9155, 1361-6560},
  doi = {10.1088/1361-6560/abf276},
  urldate = {2026-01-15}
}

@article{sarrutGATE10Monte2026,
  title = {GATE 10 Monte Carlo Particle Transport Simulation: I. Development and New Features},
  shorttitle = {GATE 10 Monte Carlo Particle Transport Simulation},
  author = {Sarrut, David and Arbor, Nicolas and Baudier, Thomas and Bert, Julien and Chatzipapas, Konstantinos and Favaretto, Martina and Fuchs, Hermann and Grevillot, Lo{\"i}c and Harb, Hussein and Van Hoey, Gert and Jacquet, Maxime and Jan, S{\'e}bastien and Jia, Yihan and Kagadis, George C and Gyu Kang, Han and Klever, Paul and Kochebina, Olga and Krzemien, Wojciech and Maigne, Lydia and Mohr, Philipp and Mummaneni, Guneet and Paneta, Valentina and Papadimitroulas, Panagiotis and Pereda, Alexis and Rannou, Axel and Resch, Andreas F and Roncali, Emilie and Toussaint, Maxime and Trigila, Carlotta and Tsoumpas, Charalampos and Zhang, Jing and Ziemons, Karl and Krah, Nils},
  year = 2026,
  month = jan,
  journal = {Phys. Med. Biol.},
  volume = {71},
  number = {1},
  pages = {015042},
  issn = {0031-9155, 1361-6560},
  doi = {10.1088/1361-6560/ae237b},
  urldate = {2026-01-15}
}

@article{sarrutOpenGATEEcosystemMonte2022b,
  title = {The OpenGATE Ecosystem for Monte Carlo Simulation in Medical Physics},
  author = {Sarrut, David and Arbor, Nicolas and Baudier, Thomas and Borys, Damian and Etxebeste, Ane and Fuchs, Hermann and Gajewski, Jan and Grevillot, Lo{\"i}c and Jan, S{\'e}bastien and Kagadis, George C and Kang, Han Gyu and Kirov, Assen and Kochebina, Olga and Krzemien, Wojciech and Lomax, Antony and Papadimitroulas, Panagiotis and Pommranz, Christian and Roncali, Emilie and Rucinski, Antoni and Winterhalter, Carla and Maigne, Lydia},
  year = 2022,
  month = sep,
  journal = {Phys. Med. Biol.},
  volume = {67},
  number = {18},
  pages = {184001},
  issn = {0031-9155, 1361-6560},
  doi = {10.1088/1361-6560/ac8c83},
  urldate = {2026-01-15}
}

@article{shibuyaOxygenSensingAbility2020a,
  title = {Oxygen Sensing Ability of Positronium Atom for Tumor Hypoxia Imaging},
  author = {Shibuya, Kengo and Saito, Haruo and Nishikido, Fumihiko and Takahashi, Miwako and Yamaya, Taiga},
  year = 2020,
  month = oct,
  journal = {Commun Phys},
  volume = {3},
  number = {1},
  pages = {173},
  issn = {2399-3650},
  doi = {10.1038/s42005-020-00440-z},
  urldate = {2026-01-09},
  langid = {english}
}

@article{steinbergerPositroniumLifetimeValidation2024a,
  title = {Positronium Lifetime Validation Measurements Using a Long-Axial Field-of-View Positron Emission Tomography Scanner},
  author = {Steinberger, William M. and Mercolli, Lorenzo and Breuer, Johannes and Sari, Hasan and Parzych, Szymon and Niedzwiecki, Szymon and Lapkiewicz, Gabriela and Moskal, Pawel and Stepien, Ewa and Rominger, Axel and Shi, Kuangyu and Conti, Maurizio},
  year = 2024,
  month = aug,
  journal = {EJNMMI Physics},
  volume = {11},
  number = {1},
  pages = {76},
  issn = {2197-7364},
  doi = {10.1186/s40658-024-00678-4},
  urldate = {2025-10-16}
}

@article{kacperskiPerformanceThreephotonPET2005,
  title = {Performance of Three-Photon PET Imaging: Monte Carlo Simulations},
  shorttitle = {Performance of Three-Photon PET Imaging},
  author = {Kacperski, Krzysztof and Spyrou, Nicholas M.},
  year = 2005,
  month = dec,
  journal = {Physics in Medicine and Biology},
  volume = {50},
  number = {23},
  pages = {5679--5695},
  issn = {0031-9155},
  doi = {10.1088/0031-9155/50/23/019},
  langid = {english},
  pmid = {16306661}
}

@article{koteraOrthopositroniumLifetimeWater2005a,
  title = {Measurement of positron lifetime to probe the mixed molecular states of liquid water},
  author = {Kotera, Katsushige and Saito, Tadashi and Yamanaka, Taku},
  year = 2005,
  journal = {Physics Letters A},
  volume = {345},
  number = {1},
  pages = {184-190},
  issn = {0375-9601},
  doi = {https://doi.org/10.1016/j.physleta.2005.07.018},
  urldate = {2026-08-06},
}

@article{mercolliFirstPositroniumLifetime2025a,
  title = {First Positronium Lifetime Imaging with Scandium-44 on a Long Axial Field-of-View PET/CT},
  author = {Mercolli, Lorenzo and Steinberger, William M. and Grundler, Pascal V. and Moiseeva, Anzhelika and Braccini, Saverio and Conti, Maurizio and Moskal, Pawe{\l} and Rathod, Narendra and Rominger, Axel and Sari, Hasan and Schibli, Roger and Seifert, Robert and Shi, Kuangyu and Stepie{\'n}, Ewa {\L}. and Van Der Meulen, Nicholas P.},
  year = 2025,
  month = nov,
  journal = {Front. Nucl. Med.},
  volume = {5},
  pages = {1648621},
  issn = {2673-8880},
  doi = {10.3389/fnume.2025.1648621},
  urldate = {2026-01-26},
  langid = {english}
}

@article{tashimaThreeGammaImagingNuclear2024a,
  title = {Three-Gamma Imaging in Nuclear Medicine: A Review},
  shorttitle = {Three-Gamma Imaging in Nuclear Medicine},
  author = {Tashima, Hideaki and Yamaya, Taiga},
  year = 2024,
  month = nov,
  journal = {IEEE Trans. Radiat. Plasma Med. Sci.},
  volume = {8},
  number = {8},
  pages = {853--866},
  issn = {2469-7311, 2469-7303},
  doi = {10.1109/TRPMS.2024.3470836},
  urldate = {2026-03-24},
  copyright = {https://creativecommons.org/licenses/by/4.0/legalcode}
}

@inproceedings{tashimaModelingPositroniumLifetime2023,
  title = {Modeling Positronium Lifetime Distribution in Geant4 Monte Carlo Simulation},
  booktitle = {2023 IEEE Nuclear Science Symposium, Medical Imaging Conference and International Symposium on Room-Temperature Semiconductor Detectors (NSS MIC RTSD)},
  author = {Tashima, H. and Takyu, S. and Nishikido, F. and Takahashi, M. and Yamaya, T.},
  year = 2023,
  month = nov,
  pages = {1--1},
  issn = {2577-0829},
  doi = {10.1109/NSSMICRTSD49126.2023.10338264},
  urldate = {2026-03-06}
}

@misc{NEMA:2018,
    author = {NEMA},
    note    = {National Electrical Manufacturers Association (NEMA NU 2-2018)},
    title     = {NEMA Standards Publication NU 2-2018: Performance Measurements of Positron Emission Tomographs (PET)},
    year      = "2018"
}

@misc{LNHB_database,
  author = {{Laboratoire National Henri Becquerel (LNHB)}},
  title = {Table of Radionuclides / Nuclear Data Tables},
  year = {2026},
  url = {http://www.lnhb.fr/accueil/donnees-nucleaires/donnees-nucleaires-tableau/},
  note = {Accessed: 2026-04-24}
}

@article{moskalFeasibilityStudyPositronium2019c,
  title = {Feasibility Study of the Positronium Imaging with the J-PET Tomograph},
  author = {Moskal, P and Kisielewska, D and Curceanu, C and Czerwi{\'n}ski, E and Dulski, K and Gajos, A and Gorgol, M and Hiesmayr, B and Jasi{\'n}ska, B and Kacprzak, K and Kap{\l}on, {\L} and Korcyl, G and Kowalski, P and Krzemie{\'n}, W and Kozik, T and Kubicz, E and Mohammed, M and Nied{\'z}wiecki, Sz and Pa{\l}ka, M and {Pawlik-Nied{\'z}wiecka}, M and Raczy{\'n}ski, L and Raj, J and Sharma, S and {Shivani} and Shopa, R Y and Silarski, M and Skurzok, M and St{\k e}pie{\'n}, E and Wi{\'s}licki, W and Zgardzi{\'n}ska, B},
  year = 2019,
  month = mar,
  journal = {Phys. Med. Biol.},
  volume = {64},
  number = {5},
  pages = {055017},
  issn = {1361-6560},
  doi = {10.1088/1361-6560/aafe20},
  urldate = {2026-04-24}
}

@article{PALSAV1,
  title = {Commissioning of the J-PET detector in view of the positron annihilation lifetime spectroscopy},
  author = {Dulski, K and Curceanu, C and Czerwi{\'n}ski, E and Gajos, A and Gorgol, M and Gupta-Sharma, N and Hiesmayr, B C and Jasi{\'n}ska, B and Kacprzak, K and Kap{\l}on, {\L} and Kisielewska, D and Klimaszewski, K and Korcyl, G and Kowalski, P and Krawczyk, N and Krzemie{\'n}, W and Kozik, T and Kubicz, E and Mohammed, M and Nied{\'z}wiecki, Sz and Pa{\l}ka, M and Pawlik-Nied{\'z}wiecka, M and Raczy{\'n}ski, L and Raj, J and Rakoczy, K and Rudy, Z and Sharma, S and Shivani and Shopa, R Y and Silarski, M and Skurzok, M andWi{\'s}licki, W and Zgardzi{\'n}ska, B and Moskal P},
  year = 2018,
  journal = {Hyperfine Interact},
  volume = {239},
  number = {40},
  pages = {1},
  doi = {10.1007/s10751-018-1517-z},
  urldate = {2026-04-26}
}

@article{PALSAV2,
  title = {PALS Avalanche — A New PAL Spectra Analysis Software},
  author = {Dulski, K},
  year = 2020,
  journal = {Acta Phys. Pol. A},
  volume = {137},
  number = {2},
  pages = {167-170},
  doi = {10.12693/APhysPolA.137.167},
  urldate = {2026-04-26}
}

@article{ADKINS2002136,
title = {Two-Loop Correction to the Orthopositronium Decay Rate},
journal = {Annals of Physics},
volume = {295},
number = {2},
pages = {136-193},
year = {2002},
issn = {0003-4916},
doi = {https://doi.org/10.1006/aphy.2001.6219},
url = {https://www.sciencedirect.com/science/article/pii/S0003491601962190},
author = {Gregory S. Adkins and Richard N. Fell and J. Sapirstein}
}

@article{PhysRevA.68.032512,
  title = {Two-loop corrections to the decay rate of parapositronium},
  author = {Adkins, Gregory S. and McGovern, Nathan M. and Fell, Richard N. and Sapirstein, J.},
  journal = {Phys. Rev. A},
  volume = {68},
  issue = {3},
  pages = {032512},
  numpages = {11},
  year = {2003},
  month = {Sep},
  publisher = {American Physical Society},
  doi = {10.1103/PhysRevA.68.032512},
  url = {https://link.aps.org/doi/10.1103/PhysRevA.68.032512}
}

@manual{SiemensBiographVision2022,
  author  = {{Siemens Healthineers}},
  title   = {Biograph Vision PET/CT},
  year    = {2022},
  month   = dec,
  url     = {https://marketing.webassets.siemens-healthineers.com/29135ac8e7d52de3/8b64a2477e30/siemens-healthineers_MI_biograph-vision-petct-brochure.pdf},
  urldate = {2026-05-11}
}

@article{janGATESimulationToolkit2004,
  title = {GATE: A Simulation Toolkit for PET and SPECT},
  shorttitle = {GATE},
  author = {Jan, S and Santin, G and Strul, D and Staelens, S and Assi{\'e}, K and Autret, D and Avner, S and Barbier, R and Bardi{\`e}s, M and Bloomfield, P M and Brasse, D and Breton, V and Bruyndonckx, P and Buvat, I and Chatziioannou, A F and Choi, Y and Chung, Y H and Comtat, C and Donnarieix, D and Ferrer, L and Glick, S J and Groiselle, C J and Guez, D and Honore, P-F and {Kerhoas-Cavata}, S and Kirov, A S and Kohli, V and Koole, M and Krieguer, M and Laan, D J Van Der and Lamare, F and Largeron, G and Lartizien, C and Lazaro, D and Maas, M C and Maigne, L and Mayet, F and Melot, F and Merheb, C and Pennacchio, E and Perez, J and Pietrzyk, U and Rannou, F R and Rey, M and Schaart, D R and Schmidtlein, C R and Simon, L and Song, T Y and Vieira, J-M and Visvikis, D and Walle, R Van De and Wie{\"e}rs, E and Morel, C},
  year = 2004,
  month = oct,
  journal = {Phys. Med. Biol.},
  volume = {49},
  number = {19},
  pages = {4543--4561},
  issn = {0031-9155, 1361-6560},
  doi = {10.1088/0031-9155/49/19/007},
  urldate = {2026-05-11}
}

@article{raczynskiVertex2026,
  author  = {Raczyński, Lech and Krzemień, Wojciech and Coussat, Aurelien and Bała, Mateusz and Hiesmayr, Beatrix C. and Klimaszewski, Konrad and Obara, Michał and Shopa, Roman Y.},
  doi     = {https://doi.org/10.48550/arXiv.2605.09174},
  journal = {pre-print},
  title   = {Ortho-Positronium Three-Photon Decays: Physics Constraints and a Closed-Form Energy Method for Annihilation Vertex Reconstruction},
  url     = {https://arxiv.org/abs/2605.09174},
  year    = 2026
}

@article{arce2014,
  title = {Gamos: {{A}} Framework to Do {{Geant4}} Simulations in Different Physics Fields with an User-Friendly Interface},
  author = {Arce, Pedro and Lagares, Juan Ignacio and Harkness, Laura and {P{\'e}rez-Astudillo}, Daniel and Ca{\~n}adas, Mario and Rato, Pedro and {de Prado}, Mar{\'i}a and Abreu, Yamiel and {de Lorenzo}, Gianluca and Kolstein, Machiel and others},
  year = {2014},
  journal = {NIMA},
  volume = {735},
  pages = {304--313}
}

@article{faddegon2020,
  title = {The {{TOPAS}} Tool for Particle Simulation, a {{Monte Carlo}} Simulation Tool for Physics, Biology and Clinical Research},
  author = {Faddegon, Bruce and {Ramos-M{\'e}ndez}, Jos{\'e} and Schuemann, Jan and McNamara, Aimee and Shin, Jungwook and Perl, Joseph and Paganetti, Harald},
  year = {2020},
  month = apr,
  journal = {Physica Medica},
  volume = {72},
  pages = {114--121},
  issn = {1724-191X},
  doi = {10.1016/j.ejmp.2020.03.019},
  langid = {english},
  pmcid = {PMC7192305},
  pmid = {32247964}
}
